\documentclass{ar-1col-S2O}
\usepackage[numbers]{natbib}
\usepackage{amsmath}
\usepackage{url}
\usepackage{caption}
\usepackage{subcaption}
\usepackage{bm}
\usepackage[utf8]{inputenc}

\newcommand{\orcidlink}[1]{}

\usepackage{hyperref}
\hypersetup{breaklinks=true,colorlinks=true,linkcolor=blue,citecolor=blue,filecolor=blue,urlcolor=blue}

\jname{Xxxx. Xxx. Xxx. Xxx.}
\jvol{AA}
\jyear{YYYY}
\doi{10.1146/((please add article doi))}

\newcommand{\bfr}{\bm{r}}
\newcommand{\bfs}{\bm{s}}

\newcommand{\bfj}{\bm{j}}

\newcommand{\caN}{\mathcal N}

\newcommand{\rmd}{\mbox{d}}

\newcommand{\citemany}[2]{\protect\cite{#1}--\protect\cite{#2}}

\AtBeginDocument{
    \hypersetup{
        linkcolor=blue,
    }
}

\begin{document}

\markboth{J. Dobaczewski {\it et al.}}{Nuclear DFT moments}

\title{Electromagnetic and Exotic Moments in Nuclear DFT}

\author{%
J. Dobaczewski\orcidlink{0000-0002-4158-3770},$^{1,2}$
B. C. Backes\orcidlink{0000-0002-4490-2802},$^1$
R.P. de Groote\orcidlink{0000-0003-4942-1220},$^3$
A. Restrepo-Giraldo\orcidlink{0000-0002-7811-6252},$^1$
X. Sun\orcidlink{0000-0002-0130-6269}$^1$ and
H. Wibowo\orcidlink{0000-0003-4093-0600}$^1$
\affil{$^1$School of  Physics, Engineering and Technology, University of York, \\ Heslington, York YO10 5DD, United Kingdom}
\affil{$^2$Institute of Theoretical Physics, Faculty of Physics, University of Warsaw, \\ ul. Pasteura 5, PL-02-093 Warsaw, Poland}
\affil{$^3$KU Leuven, Instituut voor Kern-en Stralingsfysica, B-3001 Leuven, Belgium}}

\begin{abstract}
Electromagnetic interactions serve as essential probes for studying and testing our understanding of the atomic nucleus, as they reveal emergent properties across the nuclear chart. We analyse their corresponding observables, which relate to charge and current distributions in nuclei expressed through their multipole components. We focus on theoretical results obtained within nuclear density functional theory (DFT) to derive self-consistent, symmetry-restored nuclear wave functions along with their spectroscopic multipole moments. We demonstrate how these compare with experimental data. We also discuss potential improvements in the formulation of magnetic dipole operators by including two-body meson-exchange contributions. Discussions of exotic symmetry-breaking moments emphasise their importance for understanding fine details of fundamental nuclear interactions. Detailed derivations are provided in the accompanying Supplemental Material.
\end{abstract}

\maketitle

\tableofcontents

\section{INTRODUCTION}

Nuclear electromagnetic moments have been measured and calculated for many years, References~\cite{(Kop58b),(Cas90c)}. It was early recognised that they serve as sensitive probes of nuclear properties such as the collective or single-particle character of ground and excited states. Initial descriptions of moments were almost exclusively centred on shell-model applications, including those based on phenomenological deformed mean-field studies introduced by Nilsson, Reference~\cite{(Nil55a)}. Later, numerous studies within nuclear density functional theory (DFT) focused on properties of even-even nuclei. Only recently have DFT analyses of odd nuclei been carried out, where electromagnetic moments are directly measurable. This area has recently gained momentum, which we intend to review in this article.

We also examine the subject from a wider perspective. We start by connecting low-energy nuclear electromagnetic observables to both classical and quantum models of electromagnetism (Section~\ref{QED}). Then, in Section~\ref{one-body}, we review the classical properties of the magnetic dipole and electric quadrupole moments, before discussing the less-known magnetic octupole moments. We only include a summary of experimental results, which were thoroughly reviewed in the recent Reference~\cite{(Geo25)}. Sections~\ref{fundamental} and~\ref{MEC} discuss current topics about exotic moments for studying fundamental interactions and two-body contributions to magnetic dipole moments, respectively. Followed by Sections~\ref{Pairing}--\ref{Results}, which explain modern DFT methodology for determining nuclear moments and summarise recent results. The review is complemented by the Supplemental Material (Reference~\cite{ARNPS-suppl}), providing additional details.

\section{NUCLEAR ELECTROMAGNETIC MOMENTS WITHIN QED\label{QED}}

The relativistic Lagrangian of the nuclear system reads as follows,
\begin{equation}\label{eq:lag_nucl}
    \mathcal{L} = \mathcal{L}_{\textrm{nucl}} + \mathcal{L}_{\textrm{int}} - \frac{1}{4}F_{\mu\nu}F^{\mu\nu},
\end{equation}
where $\mathcal{L}_{\textrm{nucl}}$ refers the nuclear part, and the electromagnetic interaction $\mathcal{L}_{\textrm{int}}$ is mediated by the photons through the following coupling,
\begin{equation}\label{eq:lag_int_EM}
    \mathcal{L}_{\textrm{int}} = -e\bar{\psi}(x)\gamma^{\mu}\frac{ 1-\tau_3}{2}\psi(x) A_{\mu}(x),
\end{equation}
where $\psi(x)$ is the nucleon field with $\bar{\psi}(x)$ its Dirac adjoint, $\tau_3=1$ and $\tau_3=-1$ refers to neutrons and protons, respectively, $A_{\mu}(x)$ is the photon field, and $F^{\mu\nu}=\partial^{\mu}A^{\nu}-\partial^{\nu}A^{\mu}$ is the anti-symmetric electromagnetic tensor. The equation of motion of the photon satisfies the following equation,
\begin{equation}\label{eq:eos_photon}
        \partial_\mu F^{\mu\nu} = j^\nu ,
\end{equation}
with the electromagnetic current $j^\nu=-e\bar{\psi}\gamma^{\nu}\frac{ 1-\tau_3}{2}\psi$.

In relativistic quantum mechanics, the Dirac spinor and the corresponding four-vector current naturally incorporate spin.
From the Gordon decomposition (Reference~\cite{Gordon1928ZFP}),
\begin{equation}
    \bar\psi \gamma^\mu\psi =\frac{i}{2m} \left(\bar \psi \partial^\mu\psi -(\partial^\mu\bar \psi) \psi\right)+\frac{1}{m} \partial_\nu\left(\bar\psi \Sigma^{\mu\nu}\psi\right),
\end{equation}
with $\Sigma^{\mu\nu} = \frac {i}{4} [\gamma^\mu,\gamma^\nu]$, the four-vector current can be separated into two terms. The first term is a convective current relating to the motion of the particles, and the second term is a spin current depending on the curvature of the spin density.

The Lagrangian in Equation~\ref{eq:lag_int_EM}, however, does not contain complete information about nuclear magnetic moments. This is because a nucleon is a composite object made of more fundamental particles and possesses an additional magnetic moment arising from its spin. Therefore, to address the magnetic moments,
an anomalous current (Reference~\cite{Furnstahl1989PRC}),
\begin{equation}
    \tilde{j}^{\mu} = \frac{e}{m}\partial_{\nu}\left[\bar{\psi} \left(\kappa_n\frac{1+\tau_3}{2}+\kappa_p\frac{1-\tau_3}{2}\right)\Sigma^{\mu\nu}\psi \right],
\end{equation}
with the anomalous gyromagnetic ratios of the neutron and proton, $\kappa_n=-1.913$ and $\kappa_p=1.793$,
should be added to the electromagnetic current, namely, $j^{\mu}_{\text{tot}} = j^{\mu} + \tilde{j}^{\mu}$.
Then, the time- and space-components are respectively $j^0_{\text{tot}} = \rho_p$ and $\bm{j}_{\text{tot}} = \bm{j}_L^p + \bm{j}_S^p + \bm{j}_S^n$, with
\begin{equation}\label{Eq:rel_4cur}
    \begin{aligned}
        \rho_p &= e\psi^{\dagger}\frac{1-\tau_3}{2}\psi,\\
        \bm{j}_L^p&= e\frac{1}{2m}\left(i\nabla{\psi}^{\dagger}\frac{1-\tau_3}{2}\psi -{\psi}^{\dagger}\frac{1-\tau_3}{2}i\nabla\psi\right),\\
        \bm{j}_S^p&= e\frac{1+\kappa_p}{m}\nabla\times({\psi}^{\dagger}\frac{1-\tau_3}{2}\bm{S}\psi),\\
        \bm{j}_S^n&= e\frac{\kappa_n}{m}\nabla\times({\psi}^{\dagger}\frac{1+\tau_3}{2}\bm{S}\psi),\\
    \end{aligned}
\end{equation}
where
\begin{equation}
\bm{S}=(\Sigma^{23},\Sigma^{31},\Sigma^{12})=\frac{1}{2}
\begin{pmatrix}
    \bm{\sigma} & 0 \\
    0 & \bm{\sigma}
\end{pmatrix}.
\end{equation}

Equation~\ref{eq:eos_photon} corresponds to Maxwell's equations with the charge density $\rho(\bfr)$ and the current density $\bm{j}(\bfr)$ as the time and space components of $j^{\nu}=(c\rho,\bm{j})$, and the electric field $\phi$ and the magnetic vector potential $\bm{A}$ are the time  and space components of $A_{\nu}=(\phi/c,-\bm{A})$.
Under the Coulomb gauge $\nabla\cdot\bm{A}=0$ and requesting $\phi=0$, the solutions of $\bm{A}$ satisfy equation (Reference~\cite{(Rin80)}),
\begin{equation}
    \bm{A}(\bm{r},t) = \sum_{\alpha k}\sum_{\lambda\mu}\left\{ e^{-ickt}\bm{A}_{\alpha k \lambda \mu}(\bm{r})\hat{a}^{\dagger}_{\alpha k \lambda \mu}+h.c.\right\},
\end{equation}
with $\alpha$ standing for $E$ (electric) and $M$ (magnetic) radiation,
$k$ denoting the wave number, $\mu$ being the projection of multipolarity $\lambda$, $\hat{a}^{\dagger}_{\lambda k I m}$ denoting the creation operators of photons, and
\begin{equation}
    \begin{aligned}
    \bm{A}_{Ek\lambda \mu}(\bm{r}) &=\frac{\sqrt{4\pi \hbar ck/R}}{k\sqrt{\lambda(\lambda+1)}}\bm{\nabla}\times(\bm{r}\times\bm{\nabla})[j_\lambda(kr)Y_{\lambda \mu}(\theta,\phi)], \\
    \bm{A}_{Mk\lambda \mu}(\bm{r}) &= \frac{-i\sqrt{4\pi \hbar ck/R}}{\sqrt{\lambda(\lambda+1)}}(\bm{r}\times\bm{\nabla})[j_\lambda(kr)Y_{\lambda \mu}(\theta,\phi)].
    \end{aligned}
\end{equation}
The parameter $R$ defines a spherical standing wave boundary, hence discretizes the wave number $k$ as the zeros of either the spherical Bessel function $j_\lambda(kR)$ or $\partial_r(rj_{\lambda}(kr)|_{r=R}$, which also determines the density of the state $g(k)=R/\hbar c\pi$ at the limit $R\to\infty$.

In classical electrodynamics~\cite{(Jac75b)}, in a given reference frame, the charge density $\rho(\bfr)$ generates the electric field and the current density  $\bfj_{\text{tot}}(\bfr)$ generates the magnetic field. When observing a given system of localised charge and current densities from a distance, it is practical to decompose the densities and fields into a series of components, as defined by the so-called multipole expansion,~\cite{(Jac75b)}~Equation~16.127.

In non-relativistic quantum mechanics~\cite{(Mes99)}, the electromagnetic charge, orbital current, and spin densities, $\rho(\bfr)$, $\bfj_L(\bfr)$, and $\bfs_\tau(\bfr)$, are replaced by the average values of particular one-body operators, that is,
\begin{eqnarray}
\label{density1}
\rho(\bfr)&=&e\langle\Psi|\sum_{i=1}^{Z}\delta(\bfr-\bfr_i)|\Psi\rangle,
\quad
\bfj_L(\bfr)=e\langle\Psi|\sum_{i=1}^{Z}\Big[\tfrac{-i\hbar}{2m}{\bm{\nabla}_i}\delta(\bfr-\bfr_i)\Big]|\Psi\rangle,
\\
\label{density2}
\bfs_\tau(\bfr)&=&\langle\Psi|\sum_{i=1}^{N(\tau)}\tfrac{1}{2}\bm{\sigma}_i\delta(\bfr-\bfr_i)|\Psi\rangle,
\end{eqnarray}
where the sums in Equation~\ref{density1} run over $Z$ charged particles of a given quantum system (protons) and $|\Psi\rangle$ is its many-body wave function\footnote{In the literature discussing the Skyrme density functionals, see, e.g., References~\nocite{(Eng75c),(Ben03e),(Per04c)}\citemany{(Eng75c)}{(Per04c)}, the current density in Equation~\ref{density1} is defined without the factor $\hbar/m$.}.
In addition, the magnetic fields are generated not only by the orbital current but also by spin densities in units of $\hbar$ (Equation~\ref{density2}), where the sums run over $N(n)=N$ neutrons or $N(p)=Z$ protons.
The total current density,
$\bm{j}_{\text{tot}}=\bm{j}_L+e\kappa_n\nabla\times\bfs_n+e(1+\kappa_p)\nabla\times\bfs_p$
is the sum of the orbital and spin currents.
Moreover, in quantum field theory, the effects of charged virtual fields must also, in principle, be taken into account, cf.~Section~\ref{MEC}.

The classical static electric potential $\Phi(\bm{r})$ and
magnetic field $\bm{B}(\bm{r})$ are determined in terms of the charge $\rho(\bm{r})$ and current $\bm{j}_{\text{tot}}(\bm{r})$ densities by the Poisson and vector Poisson equations, respectively, that is,
\begin{equation}
    \begin{aligned}
        \nabla^2 \Phi(\bm{r}) = -\frac{\rho(\bm{r})}{4\pi\epsilon_0},\quad
        \nabla^2 \bm{B}(\bm{r}) = -\frac{4\pi}{c}\bm{\nabla}\times\bm{j}_{\text{tot}}(\bm{r}).
    \end{aligned}
\end{equation}
The multipole expansion of the Green function of the Poisson equation can be obtained using the following identity:
\begin{equation}\label{Eq:inverse_rel}
    \frac{1}{|\bm{r}-\bm{r}'|}=\sum_{\lambda=0}^{\infty}\sum_{\mu=-\lambda}^{\lambda}
    \frac{4\pi}{2\lambda+1}\frac{r_<^{\lambda}}{r^{\lambda+1}_>}
    Y_{\lambda\mu}^*(\theta',\phi')Y_{\lambda\mu}(\theta,\phi),
\end{equation}
where $r_<$ ($r_>$) is the smaller (larger) of $|\bm{r}|$ and $|\bm{r}'|$.
At the distance $|\bm{r}|>|\bm{r}'|$, the static electric potential can be expressed as follows,
\begin{equation}
\begin{aligned}
    \Phi(\bm{r}) = \frac{1}{4\pi\varepsilon_0} \int \rmd^3\bm{r}' \frac{\rho(\bm{r}')}{|\bm{r}-\bm{r}'|}
    = \frac{1}{4\pi\varepsilon_0}\sum_{\lambda=0}^{\infty}\sum_{\mu=-\lambda}^{\lambda}
        \frac{4\pi}{2\lambda+1} Q^*_{\lambda\mu}\frac{Y_{\lambda\mu}(\theta,\phi)}{r^{\lambda+1}},
\end{aligned}
\end{equation}
with the electric multipole moment $Q_{\lambda\mu}$ defined as
\begin{equation}
\label{Electric}
    Q_{\lambda\mu} = \int \rmd^3\bm{r} \rho(\bm{r}) r^{\lambda} Y_{\lambda\mu}(\theta,\phi).
\end{equation}
The multipole expansion of the static magnetic field can be obtained similarly by using the scalar potential method (References~\cite{Bronzan1971AJP,Gray1978AJP}). Expanding the radial component of the magnetic field gives,
\begin{equation}
\begin{aligned}
    \bm{r}\cdot\bm{B} &= \frac{1}{c}\int\rmd^3\bm{r}' \frac{\bm{r}'\cdot\nabla'\times\bm{j}_{\text{tot}}(\bm{r}')}{|\bm{r}-\bm{r}'|} =\sum_{\lambda=0}^{\infty}\sum_{\mu=-\lambda}^{\lambda}(\lambda+1)\frac{4\pi}{2\lambda+1}M^*_{\lambda\mu}\frac{Y_{\lambda\mu}(\theta,\phi)}{r^{\lambda+1}},
\end{aligned}
\end{equation}
with the magnetic multipole moment $M_{\lambda\mu}$ defined as
\begin{equation}
\label{Magnetic}
    M_{\lambda\mu} = \frac{1}{c}\left(\frac{1}{\lambda+1}\right)
    \int\rmd^3\bm{r}\, r^{\lambda} Y_{\lambda\mu}(\theta,\phi)\bm{r}\cdot\left(\bm{\nabla}\times\bm{j}_{\text{tot}}\right).
\end{equation}

Additionally, the time-dependent charge and current densities generate a time-dependent electromagnetic field, and, in particular, electromagnetic radiation. For completeness, the corresponding expressions are summarised in the Supplemental Material (Reference~\cite{ARNPS-suppl}, Section~\ref{Radiation}).

\section{ONE-BODY MOMENTS\label{one-body}}

The one-body electric and magnetic moments, $Q_{\lambda\mu}$ and $M_{\lambda\mu}$, Equations~\ref{Electric} and~\ref{Magnetic}, respectively, can be expressed as follows, Reference~\cite{(Dob00e)},
   \begin{eqnarray}\label{eq701}
        Q_{\lambda\mu} = \langle\Psi |\hat{Q}_{\lambda\mu}|\Psi\rangle
            =  \int\rmd^3\bfr\, q_{\lambda\mu}(\bfr)
                                              , \label{eq701a} \quad
        M_{\lambda\mu}=\langle\Psi |\hat{M}_{\lambda\mu}|\Psi\rangle
            =  \int\rmd^3\bfr\, m_{\lambda\mu}(\bfr)
                                              , \label{eq701b}
   \end{eqnarray}%
where $|\Psi\rangle$ is a many-body wave function and the one-body electric and magnetic operators are defined as,
\begin{eqnarray}
    \hat{Q}_{\lambda\mu}&=&e\sum_{i=1}^Z\,{Q}_{\lambda\mu}(\bfr_i)\label{eq:electric}, \\
    \hat{M}_{\lambda\mu}&=&\mu_{\text{N}}\sum_{\tau=n,p}\sum_{i=1}^{N(\tau)}\left[\bm{\nabla}_iQ_{\lambda\mu}(\bfr_i)\right]\cdot\left(
    g_{s}^{(\tau)}\hat{\bm{S}}_i^{(\tau)}+
    g_{l}^{(\tau)}\tfrac{2}{\lambda+1}\hat{\bm{L}}_i^{(\tau)}\right), \quad\mbox{or}\label{eq:magnetic}
 \\
\hat{M}_{\lambda\mu}&=&\mu_{\text{N}} \sqrt{\lambda(2\lambda+1)}\sum_{\tau=n,p}\sum_{i=1}^{N(\tau)}\left[Q_{\lambda-1}(\bfr_i)\otimes\left(
    g_{s}^{(\tau)}\hat{S}_{1,i}^{(\tau)}+
    g_{l}^{(\tau)}\tfrac{2}{\lambda+1}\hat{L}_{1,i}^{(\tau)}\right)\right]_{\lambda\mu},
\label{eq:magnetic2}
\end{eqnarray}
where in Equation~\ref{eq:magnetic2}, $\hat{S}_{1,i}^{(\tau)}$ and $\hat{L}_{1,i}^{(\tau)}$ are the $i$th particle's rank-1 spherical tensors of the spin and orbital angular momentum operators, respectively, which in the Cartesian coordinates of Equation~\ref{eq:magnetic} are given by
 $\hat{\bm{S}}_i^{(\tau)}=\tfrac{1}{2}\hat{\bm{\sigma}}_i$ and
$\hat{\bm{L}}_i^{(\tau)}=-i\bfr_i\times\bm{\nabla}_i$ (with units of $\hbar$ omitted),

In Equations~\ref{eq:electric}--\ref{eq:magnetic2}, the multipole functions (solid harmonics), Reference~\cite{(Var88)}, Equation~5.1.7(16), have the standard form, $Q_{\lambda\mu}(\bfr_i)$=$r_i^\lambda Y_{\lambda\mu}(\theta_i,\phi_i)$, $N(n)=N$ and $N(p)=Z$ denote the numbers of neutrons and protons, and the gyromagnetic factors (the $g$ factors) and nuclear magneton $\mu_{\text{N}}$ are equal to
\begin{equation}
    g_l^{(p)}=1,\hspace{3mm}
    g_l^{(n)}=0,\hspace{3mm}
    g_s^{(p)}=2(1+\kappa_p)=5.587,\hspace{3mm}
    g_s^{(n)}=2\kappa_n=-3.826,\hspace{3mm}
    \mu_{\text{N}}=\frac{e\hbar}{2m_p c}.
\label{eq:g_factors}
\end{equation}
Derivations of the connection between the magnetic multipole moments given by the multipole expansion, Equations~\ref{Magnetic}, and the two forms of the magnetic multipole moment operator, Equations~\ref{eq:magnetic} and~\ref{eq:magnetic2}, are given in the Supplemental Material (Reference~\cite{ARNPS-suppl}, Sections~\ref{One-body_magnetic} and~\ref{Gradient}, respectively).

In Equation~\ref{eq701}, $q_{\lambda\mu}(\bfr)$ is the electric-moment density and $m_{\lambda\mu}(\bfr)$ is the magnetic-moment density (magnetisation), that is,
   \begin{eqnarray}
   \label{eq702a}
        q_{\lambda\mu}(\bfr) &=&
                \rho^{(p)}(\bfr)Q_{\lambda\mu}(\bfr), \\
   \label{eq702b}
m_{\lambda\mu}(\bfr) &=&
\mu_{\text{N}}\hspace*{-1mm}\sum_{\tau=n,p}\sum_{k=x,y,z}\hspace*{-1mm}\left(
	    g_s^{(\tau)}{s_k^{(\tau)}}(\bfr)\nabla_k Q_{\lambda\mu}(\bfr)
          -  {\tfrac{2}{\lambda+1}}
                     g_l^{(\tau)}{j_k^{(\tau)}}(\bfr)\big(\bfr\times
                 \bm{\nabla}{}Q_{\lambda\mu}(\bfr)\big)_k\right),
\end{eqnarray}%
where $\rho(\bfr)$, $\bfj(\bfr)$, and $\bfs(\bfr)$ are the charge, current, and spin densities, respectively, Equations~\ref{density1} and~\ref{density2}.

\subsection{Magnetic dipole operator and moment\label{Magnetic_dipole}}

The magnetic properties of a quantum many-body system derive from the dynamics of its constituent particles' charges and spins. In the case of the atomic nucleus, protons ($\tau=p$) contribute through both their orbital motion and their intrinsic spin, while neutrons ($\tau=n$) contribute solely via their spin. The magnetic dipole operator, representing the lowest-order magnetic multipole of Equation~\ref{eq:magnetic} for $\lambda=1$, is typically expressed by its Cartesian $z$ component.
\begin{equation}
\begin{aligned}
    \hat{\mu}_z = \sqrt{\tfrac{4\pi}{3}}\hat{M}_{10} =\mu_{\text{N}}\sum_{\tau=n,p}\Big[g_s^{(\tau)}\hat{S}_z^{(\tau)}+g_l^{(\tau)}\hat{L}_z^{(\tau)}\Big] .
\label{eq:mu_qm_operator_total}
\end{aligned}
\end{equation}
Equation~\ref{eq:mu_qm_operator_total} follows directly from the standard normalisation of the solid harmonics, $Q_{10} = \sqrt{\tfrac{3}{4\pi}}z$, where the gradient in Equation~\ref{eq:magnetic} then reduces to a constant value.

\begin{marginnote}[]
\entry{Magnetic moment}{Characterises the distribution of nuclear spin and current and defines the magnetic field inside and outside of the nucleus.}
\end{marginnote}
We note here that for any spherical tensor operator $\hat{O}_{\lambda\mu}$, such as the electric $\hat{Q}_{\lambda\mu}$ and magnetic $\hat{M}_{\lambda\mu}$ moments, the Wigner-Eckart theorem (Reference~\cite{(Var88)}, Equation~13.1.1(2) and Section~\ref{Symmetry}), defines its matrix elements $\langle{I'M'}|\hat{O}_{\lambda\mu}|IM\rangle$ for any projections $M',M,\mu$ of angular momenta $I',I,\lambda$ through one constant called the reduced matrix element $\langle{I'}\|\hat{O}_{\lambda}\|I\rangle$.
Therefore, one set of projections $M',M,\mu$ can be freely selected to characterise all the projections. Traditionally, for the diagonal matrix elements $I=I'$ of the electromagnetic moments, one chooses $M'=M=I$,
which defines the so-called spectroscopic moments. In particular, the standard definition of the spectroscopic magnetic dipole moment $\mu$ used in the literature is as follows,
\begin{equation}
    \mu =  \sqrt{\tfrac{4\pi}{3}}\langle II|\hat{M}_{10}|II\rangle
        =  \langle II |\hat{\mu}_{z}|II\rangle,
\label{eq:mm_definition}
\end{equation}
and is reported for all measurements.

\subsubsection{Single-particle estimates, the Schmidt moments\label{Schmidt_moments}}

The single-particle estimates offer a crucial perspective on analysing the electromagnetic interactions of the nucleus. These estimates are also significant for electromagnetic transitions, especially for evaluating transition strengths that define the so-called Weisskopf units (W.u.) (Reference~\cite{PhysRev.83.1073}), which are often utilised in experimental reports and databases.

The mathematical form of the magnetic dipole operator in Equation~\ref{eq:mu_qm_operator_total} does not involve radial coordinates of a single particle; thus, calculations of Equation~\ref{eq:mm_definition} are independent of the radial profile of the potential and radial quantum number $n$ of the single-particle wave function. If we consider the coupled total angular momentum single-particle wave functions characterised by the radial ($n$) orbital-angular-momentum $l$, total-angular-momentum $j$, and its projection $m$ quantum numbers, we obtain the following expressions, see derivation in the Supplemental Material (Reference~\cite{ARNPS-suppl}, Section~\ref{Schmidt}),
\begin{equation}
\begin{aligned}
     \mu_{\text{s.p.}}^{(\tau)} = \begin{cases}
       \mu_{\text{N}}\Big[\Big(j-\frac{1}{2}\Big)g^{(\tau)}_l+\frac{1}{2}g_s^{(\tau)}\Big], \hspace{17mm}j = l+1/2,\\
       \mu_{\text{N}}{\textstyle{\frac{j}{j+1}}}\Big[ \Big(j+\frac{3}{2}\Big)g^{(\tau)}_l-\frac{1}{2}g_s^{(\tau)} \Big], \hspace{11mm}j = l-1/2.
     \end{cases}
\label{eq:mu_sp}
\end{aligned}
\end{equation}
These are called the Schmidt moments and represent extreme cases where a single valence nucleon accounts for the entire magnetic moment of the nucleus. In Section~\ref{Experimental}, these estimates are compared with the experimental data.

\subsection{Electric quadrupole operator and moment}

An arbitrary localised charge distribution can be expressed as a sum of multipole moments (Equation~\ref{Electric}), a convenient approach for studying its properties. The lowest-order multipole moments dominate the nuclear electric field. For systems composed of a single type of charge, such as the atomic nucleus, the distribution's geometry can be directly associated with it. Observations show that the nuclear interactions preserve parity, so the lowest experimentally accessible multipole is the monopole, equivalent to the number of protons, and the quadrupole, which defines the lowest-order shape of the nucleus.

Traditionally, the spherical tensor of the quadrupole moment used in the literature, $\bar{Q}_{2\mu}$, is defined by introducing a specific factor, for which its axial component $\mu=0$ is given as follows,
\begin{equation}
    \bar{Q}_{20} =\sqrt{\tfrac{16\pi}{5}}{Q}_{20}
    =e\sqrt{\tfrac{16\pi}{5}}\sum_{i=1}^Z r^{2}_iY_{20}(\theta_i,\phi_i)
    =e\sum_{i=1}^Z (2z^2_i-x^2_i-y^2_i).
    \label{eq:Q_multipole}
\end{equation}
\begin{marginnote}[]
\entry{Electric moment}{Characterises the distribution of nuclear charge and defines the electric field inside and outside of the nucleus.}
\end{marginnote}
Similarly to Equation~\ref{eq:mm_definition}, the spectroscopic quadrupole moment reported for all measurements is defined by
\begin{equation}
\begin{aligned}
    Q = \sqrt{\tfrac{16\pi}{5}}\langle II|\hat{Q}_{20}|II\rangle,
    \label{eq:qadrupole_moment}
\end{aligned}
\end{equation}
and quantifies the axially symmetric elongation or flattening of the charge distribution. Three regimes can be identified in the possible values of $Q$, Equation~\ref{eq:qadrupole_moment}; a spherically symmetric distribution is characterised by $Q=0$, a prolate (elongated) shape by $Q>0$, and an oblate (flattened) shape by $Q<0$, see also Section~\ref{Symmetry} for the definition and discussion of the intrinsic multipole moments.

\subsubsection{Single-particle estimates\label{Qsp}}

The observable in Equation~\ref{eq:qadrupole_moment} can be calculated for single-particle states to quantify the deviation from spherical shape of individual nuclear orbitals. For a nucleon in a single-particle state characterised by the total angular momentum $j$, its quadrupole moment is given by (see derivation in the Supplemental Material, Reference~\cite{ARNPS-suppl}, Section~\ref{Single_particle_quadrupole}),
\begin{equation}
\begin{aligned}
    Q_{\text{s.p.}} &= -e\frac{2j-1}{2(j+1)}\langle r^2\rangle,
    \label{eq:q_sp}
\end{aligned}
\end{equation}
where $\langle r^2\rangle$ represents the mean square radius of that state, which depends on its radial wave function. Note that $Q_{\text{s.p.}}=0$ for $j=1/2$ states, $Q_{\text{s.p.}}=-\tfrac{2}{5}e\langle r^2\rangle$ for $j=3/2$ states, and for large-$j$ states it gradually increases in magnitude towards $Q_{\text{s.p.}}=-e\langle r^2\rangle$. The latter value is therefore the maximum quadrupole moment for any $j$ in a single-particle state.  In Section~\ref{Experimental}, this maximum estimate is compared with the experimental data.

\subsection{Magnetic octupole operator and moment}

Following Reference~\cite{(Sch1955)}, the spectroscopic magnetic octupole moment $\Omega$, which is reported for all measurements, is defined for $\lambda\mu=30$ in Equation~\ref{eq:magnetic} using a specific prefactor,
\begin{equation}
    \Omega=-\sqrt{\frac{4\pi}{7}}\langle II|\hat{M}_{30}|II\rangle,
    \label{eq:octupole_moment}
\end{equation}
where the magnetic octupole operator $\hat{M}_{30}$ reads explicitly as follows,
\begin{equation}
    \begin{aligned}
\hat{M}_{30}&=-3\mu_{\text{N}}\sqrt{\frac{7}{16\pi}}\sum_{\tau=n,p}\sum_{i=1}^{N(\tau)}\Big[xz\left(g_{s}^{(\tau)}\hat{S}^{(\tau)}_{i,x}+g_{l}^{(\tau)}\hat{L}^{(\tau)}_{i,x}\right)+yz\left(g_{s}^{(\tau)}\hat{S}^{(\tau)}_{i,y}+g_{l}^{(\tau)}\hat{L}^{(\tau)}_{i,y}\right)\\&\hspace{50mm}+(2z^2-x^2-y^2)\left(g_{s}^{(\tau)}\hat{S}^{(\tau)}_{i,z}+g_{l}^{(\tau)}\hat{L}^{(\tau)}_{i,z}\right)\Big].
\label{eq:octupole_moment2}
    \end{aligned}
\end{equation}

\subsubsection{Single-particle estimates, the Schwartz moments\label{Schwartz_moments}}

The extreme single-particle limits for the magnetic octupole moment are called the Schwartz moments (References~\cite{(Sch1955),BOFOS2024101672}, see the derivation in the Supplemental Material, Reference~\cite{ARNPS-suppl}, Section~\ref{Schwartz}),
\begin{align}
    \Omega^{(\tau)}_{\text{s.p.}}=+\mu_{\text{N}}\frac{3}{2}\frac{(2j-1)}{(2j+4)(2j+2)}\langle r^{2}\rangle\times\left\{\begin{array}{c}
         (j+2)\left[(j-\tfrac{3}{2})g^{(\tau)}_{l}+g^{(\tau)}_{s}\right],\;\;\;j=l+\tfrac{1}{2},  \\[4pt]
         (j-1)\left[(j+\tfrac{5}{2})g^{(\tau)}_{l}-g^{(\tau)}_{s}\right],\;\;\;j=l-\tfrac{1}{2}.
    \end{array}\right.
    \label{eq:schwartz_limit}
\end{align}

\subsection{Shapes of densities and currents\label{Shapes}}

The intrinsic nuclear densities $\rho(\bfr)$ are 3D space distributions, which can be difficult to analyse. Therefore, they are often parameterised and visualised in terms of the sharp-edge density distribution $\bar{\rho}(\bfr)$, which is equal to a constant value $\rho_0$ inside a predefined surface and vanishes outside. One of several popular definitions of the surface is given by the so-called Bohr deformation parameters $\beta_{\lambda\mu}$, and in spherical coordinates $(r,\theta,\phi)$ it reads as follows,
\begin{equation}
\label{Bohr}
R(\theta,\phi)=R_0\sum_{\lambda=0,1,2,3\ldots}^{\lambda_{\text{max}}}\sum_{\mu=-\lambda}^{\mu=\lambda}\beta_{\lambda\mu}\,Y_{\lambda\mu}(\theta,\phi),\quad\bar{\rho}(\bfr)=\left\{
\begin{array}{rl}
\rho_0&\mbox{for}~r\leq{}R(\theta,\phi) \\
     0&\mbox{for}~r>R(\theta,\phi)
\end{array}\right.,
\end{equation}
where $Y_{\lambda\mu}(\theta,\phi)$ are spherical harmonic functions, Reference~\cite{(Var88)}, Equation~5.2(1). For every $\lambda_{\text{max}}$, the multipole moments $Q_{\lambda\mu}$ (Equation~\ref{eq701}) for $\lambda=0,1,2,3,\ldots,\lambda_{\text{max}}$ can be determined as functions of $\beta_{\lambda\mu}$ and inverted, so a $Q$-equivalent sharp-edge shape can be visualised and examined. Although the exact functions $\beta_{\lambda\mu}(Q_{\lambda'\mu'})$ are difficult to derive, this is not really necessary as a very efficient numerical algorithm exists, see Reference~\cite{(Dob09g)}. The main point here is that the principal shape that generates a neutron, proton, or nucleon multipole moment $Q_{\lambda\mu}$ is given by $\beta_{\lambda\mu}=\tfrac{4\pi{}Q_{\lambda\mu}}{3NR_0^\lambda}$, where $N$ is the number of neutrons, protons, or nucleons, and $R_0=\sqrt{\tfrac{5}3}R_{\text{rms}}$ is the geometric radius of neutrons, protons, or nucleons, respectively (Supplemental Material, Reference~\cite{ARNPS-suppl}, Section~\ref{Bohr_deformation}).

In this Section, we also aim to visualise the geometric properties of the nuclear current densities and to illustrate the geometry of the ground-state magnetic multipole moments of odd nuclei. We assume that the signature-breaking states correspond to the angular momenta aligned with the axial symmetry axis.  For simplicity, we consider only the axial shapes and currents, where the Bohr surface (Equation~\ref{Bohr}) is independent of the azimuthal angle $\phi$, and reads as follows,
\begin{equation}
\label{Legendre}
R(\theta)=R_0\sum_{\lambda=0,1,2,3\ldots}^{\lambda_{\text{max}}}\sqrt{\tfrac{2\lambda+1}{4\pi}}\beta_{\lambda}\,P_{\lambda}(\cos\theta),\quad\bar{\rho}(\bfr)=\left\{
\begin{array}{rl}
\rho_0&\mbox{for}~r\leq{}R(\theta) \\
     0&\mbox{for}~r>R(\theta)
\end{array}\right.,
\end{equation}
and where $\beta_\lambda\equiv\beta_{\lambda0}$, and $P_{\lambda}(\cos\theta)$ are the standard Legendre polynomials, Reference~\cite{(Gra2007)}, Subsection~8.91.

\begin{figure}[h]
\centering\includegraphics[width=0.98\textwidth]{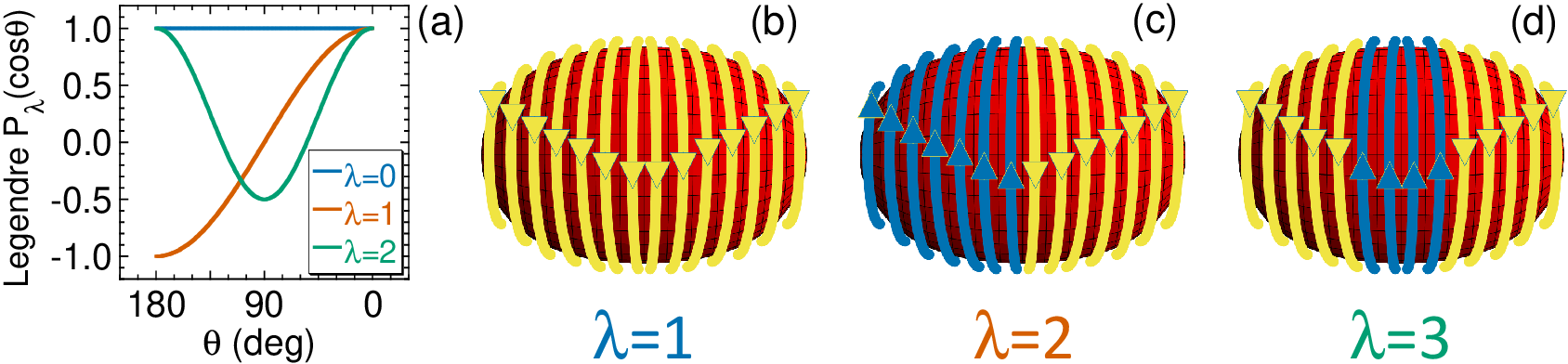}
\caption{Legendre polynomials, $P_\lambda(\cos\theta)$ for $\lambda=0,1,2$, (a), and the patterns of the current flows for the magnetic dipole (b), quadrupole (c), and octupole (d) moments of symmetry-broken aligned ground states of odd nuclei.}
\label{fig:Axial_shapes}
\end{figure}

In the sharp-edge approximation (Equation~\ref{Legendre}), the Legendre polynomials, shown in {\bf Figure}~\ref{fig:Axial_shapes}(a) generate standard nuclear shapes. For example, polynomial $P_{\lambda=2}$ increases the nuclear size both in the direction along the $z$-axis ($\theta=0$) and opposite to it ($\theta=180^\circ$) and thus corresponds to the prolate shape.

For nuclear currents in axial symmetry, let us assume their simplest geometry, namely the flow in the form of rings perpendicular to the $z$-axis. Then, in cylindrical coordinates $(z,\eta,\phi)$, the currents have only azimuthal components $j_\phi(z,\eta)$ and every ring generates the corresponding $z$ component of the magnetisation $m_z(z,\eta)$, Equation~\ref{eq702b}. Inspired by the form of the magnetic operator in Equation~\ref{eq:magnetic2}, we then postulate (Supplemental Material (Reference~\cite{ARNPS-suppl}, Section~\ref{Gradient}),
\begin{equation}
\begin{aligned}
\label{current}
j_\phi(z,\eta)&=\sum_{\lambda=1,2,3\ldots}^{\lambda_{\text{max}}}j_{\lambda}(r)\,
P_{\lambda-1}(\cos{\theta}),\quad\mbox{with}~r=\sqrt{\eta^2+z^2},\quad \cos{\theta}=\frac{z}{r},\quad\mbox{and}\\
j_{\lambda}(r) &= \frac{2\lambda-1}{2}\int_{0}^{\pi}\sin\theta\rmd{\theta}\, j_{\phi}(z,\eta) P_{\lambda-1}(\cos{\theta}),
\end{aligned}
\end{equation}
where in the first-order approximation, the term with $P_{\lambda-1}$ generates the magnetisation of multipolarity $\lambda$. This is illustrated in {\bf Figure}~\ref{fig:Axial_shapes}, where the dipole magnetisation is generated by $P_0$, that is, all rings of current flowing in the same direction, panel (a), the quadrupole parity-odd magnetisation, by currents at $z>0$ and $z<0$ flowing in opposite directions, panel (b), and the octupole magnetisation, with currents at small $|z|$ flowing in opposite directions than those at large $|z|$, panel (c).

\subsection{Experimental data\label{Experimental}}

Collecting nuclear data and ensuring its integrity are crucial for advancing the field, as they help structure our understanding of the nucleus and provide a solid foundation for theory and applications. Electromagnetic interactions are significant probes of the nucleus and are very well controlled by contemporary laboratory techniques, making them the interaction of choice for nuclear physics studies. Over the years, many measurements of static multipole moments, electromagnetic decays, and charge radii have been compiled in evaluated nuclear databases, providing a reliable and accessible source of information on the nucleus, of great interest to the scientific community (see References~\nocite{nuclear_data_sheets,(Sto19a),(Sto20),(Sto21)}\citemany{nuclear_data_sheets}{(Sto21)} for compilations of electromagnetic moments).

\begin{figure}[h]
     \centering
\begin{subfigure}[b]{0.4005
\textwidth}\includegraphics[width=\textwidth]{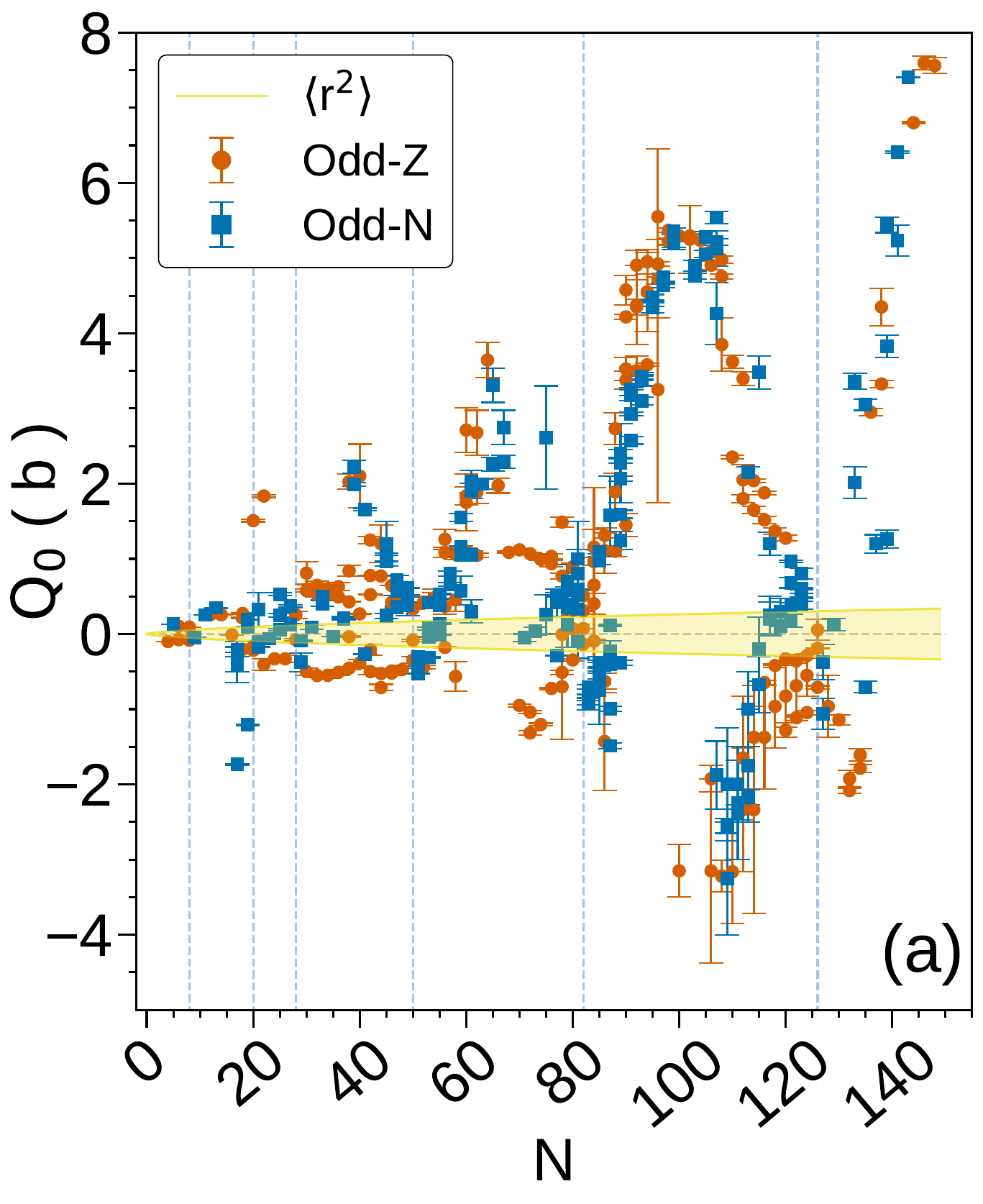}
     \end{subfigure}
\begin{subfigure}[b]{0.343
\textwidth} \vspace{3pt}\includegraphics[width=\textwidth]{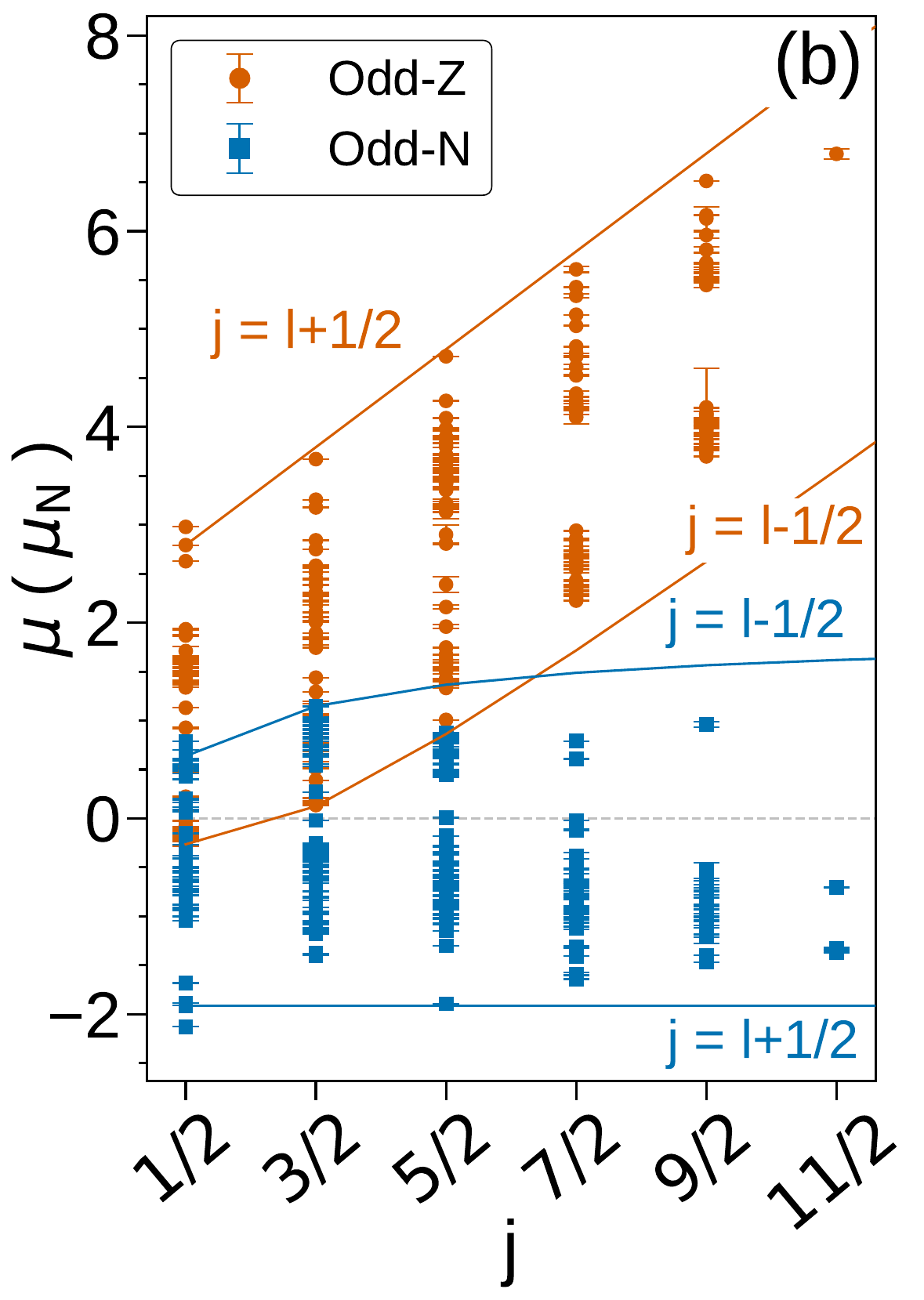}
     \end{subfigure}
     \begin{subfigure}[b]
{0.235
\textwidth}\includegraphics[width=\textwidth]{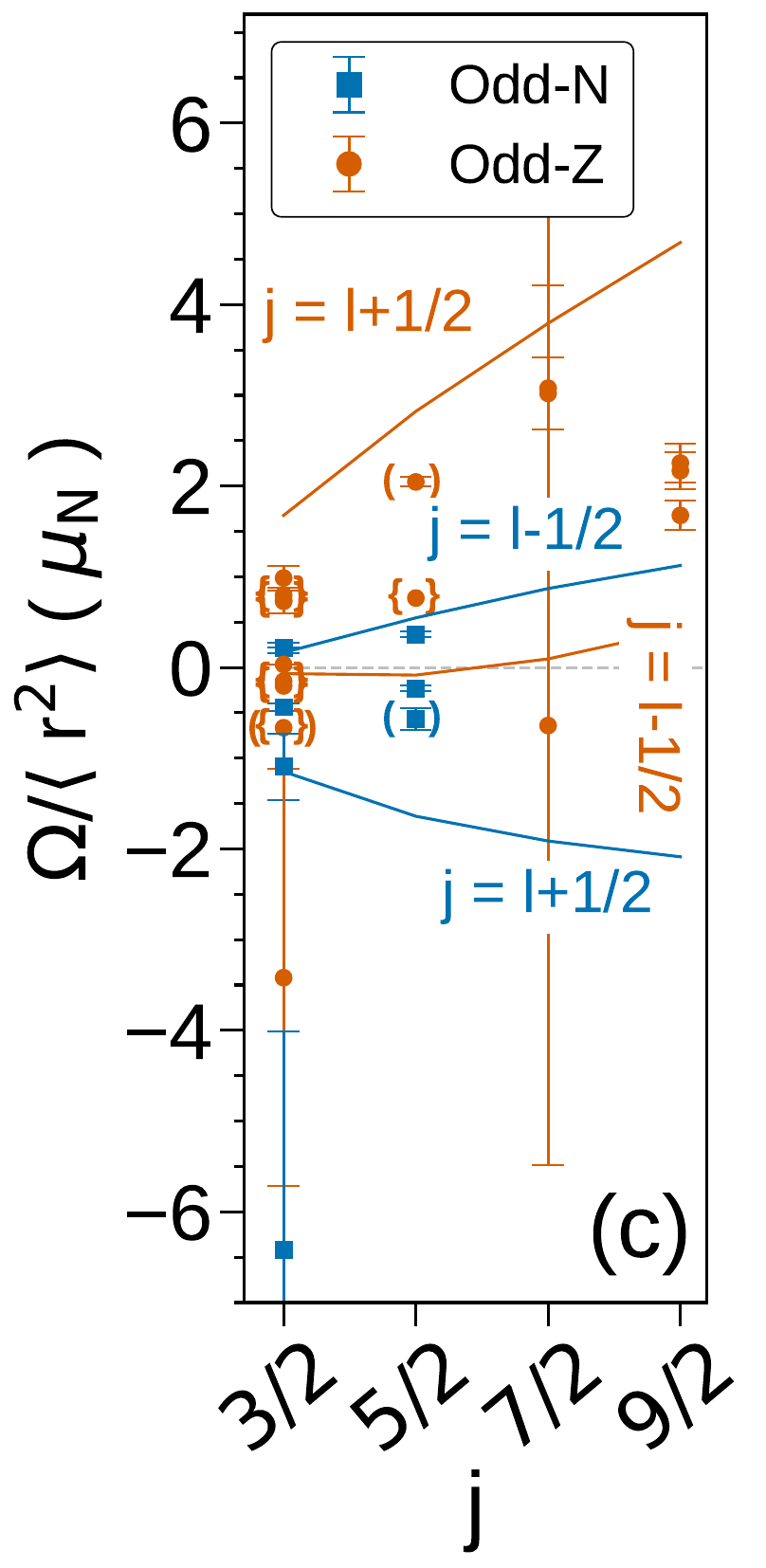}\\
     \end{subfigure}
\caption{\\
(a) Effective intrinsic electric quadrupole moments $Q_0\equiv{Q}^{\text{intr}}_{\text{eff}}(2II)$ (in barn) of ground states of odd-$A$ nuclei obtained through the large-axial-deformation approximation, Equation~\protect\ref{spectroscopic_final3}, from experimental spectroscopic moments (Reference~\protect\cite{(Sto21)}, values with measured signs only). The shaded band illustrates the maximum magnitude of single-particle estimates (Section~\protect\ref{Qsp}) calculated for $\langle r^2\rangle\simeq\frac{3}{5}R_0^2A^{2/3}$, with $R_0=1.2$\,fm and the stability line parametrised as $A(N)\approx N+Z(N)\approx\lfloor 1.2N + 1.1N^{0.8}\rfloor$. Vertical dashed lines indicate the neutron magic numbers. The value of $Q_0(^{233}\text{Pa})=-7.5\pm1 $\,b is outside plot's scale.\\
(b) Same as (a), but for the experimental spectroscopic magnetic dipole moments (in $\mu_{\text{N}}$) (Reference~\protect\cite{(Sto19a)}, values with measured signs only), compared with Schmidt moments (Section~\protect\ref{Schmidt_moments}) shown with lines. \\
(c) Same as (a), but for the experimental spectroscopic magnetic octupole moments listed in Table~\protect\ref{tab:MO} of the Supplemental Material (Reference~\cite{ARNPS-suppl}, Section~\ref{Magnetic_octupole}) divided by $\langle r^2\rangle\simeq\frac{3}{5}R_0^2A^{2/3}$ (in $\mu_{\text{N}}$), and Schwartz moments (Section~\protect\ref{Schwartz_moments}). Braces denote measurements for which the experimental uncertainties were not reported. Parentheses denote experimental values where unmeasured signs are replaced by the signs of Schwartz moments. 
\label{fig:ElectromagneticMoments}}
\end{figure}

Among the multipole moments that can be directly accessed experimentally today are the magnetic dipole, electric quadrupole, and magnetic octupole moments of ground and excited states. The magnetic dipole and electric quadrupole moments are determined using several model-independent techniques, such as laser spectroscopy, muonic-atom transitions, magnetic resonance, \textit{etc.} (References~\cite{Gerda_Neyens_2003, Blaum_2013, YANG2023104005,(Geo25)}).
Characterising quadrupole $Q$ and magnetic dipole $\mu$ moments is essential for studying emergent properties in the nuclear chart. Its main technical difficulty lies in properly setting up an external field and coupling it to the nucleus. In the case of $\mu$, the interaction energy is proportional to the external magnetic field $\bm{B}$. In contrast, for $Q$ it is proportional to the gradient of the external electric field $\nabla\bm{E}$, making its direct measurement more challenging at the nuclear scale. These observables are also the primary source of evidence for collectivity and single-particle dynamics in the nucleus, thus making them crucial for testing theories. Determining the exact values of the multipole moments remains a challenge across various theoretical approaches due to the numerous interactions involved and their varying importance across the nuclear chart.

Early attempts to use simple models could reproduce limiting cases or some qualitative features, but remained far from accounting for the actual data. In {\bf Figure}~\ref{fig:ElectromagneticMoments}(a), we display the ground-state ($\Omega=I$) effective intrinsic quadrupole moments, Equation~\ref{spectroscopic_final3},  of odd-$A$ nuclei, which represent measured spectroscopic electric quadrupole moments and illustrate nuclear quadrupole shapes. It is clear that between the magic numbers, the single-particle estimates (Section~\protect\ref{Qsp}) underestimate the experimental results by a significant factor. This systematic result constitutes the most direct evidence that open-shell nuclei are deformed, that is, that not just one but all occupied single-particle states contribute constructively to the system's total deformation.

For the magnetic dipole moments, {\bf Figure}~\ref{fig:ElectromagneticMoments}(b), the situation is entirely different. Here, almost all experimental data lie between the single-particle estimates given by the Schmidt moments (Section~\ref{Schmidt_moments}). Nevertheless, nearly none of the data exactly matches those estimates. Therefore, fine details of how the odd particle and the rest of the particles share the total magnetic moment are of primary importance.

\begin{textbox}[h]\section{Symmetry-breaking moments inferred from transition probabilities}
The ground state of the ammonia molecule conserves parity, and therefore its electric dipole moment, which breaks parity, must be equal to zero. On the other hand, we ``know'' that the ammonia molecule has the parity-breaking shape of a pyramid and thus should have a non-zero electric dipole moment. This apparent contradiction is the essence of the spontaneous symmetry breaking phenomenon in finite systems~\protect\cite{(And72),(She21)}. In a nutshell, the spontaneous symmetry breaking manifests itself through increased transition probabilities between states belonging to {\em different} representations of the broken-symmetry group. For example,
\begin{itemize}
    \item Large electric dipole moment of the pyramid-shaped ammonia molecule leads to a strong E1 transition between its lowest {\em opposite-parity} states~\protect\cite{(Fey05)}.
    \item Large electric quadrupole moments of the deformed even-even nuclei lead to strong E2 transitions between their lowest excited $J^\pi=2^+$ and ground $J^\pi=0^+$ states~\protect\cite{(Ram01a)}.
    \item Large electric octupole moments of the pear-shaped even-even nuclei lead to strong E3 transitions between their lowest excited $J^\pi=3^-$ and ground $J^\pi=0^+$ states~\protect\cite{(But96)}.
\end{itemize}
\end{textbox}

Higher-order multipole moments are considerably more challenging to measure due to their weaker coupling to electromagnetic probes, pushing current experimental techniques to their limits. Few magnetic octupole moments have been measured and evaluated, see the recent review in Reference~\cite{BOFOS2024101672}. The importance of the higher-order moments lies in their hypothetical ability to constrain better interactions in the isoscalar time-odd mean-field channels, cf.~References~\cite{SENKOV2002351, (deG22b)}. The observable of the magnetic octupole moment remains largely unexplored, requiring both experimental and theoretical efforts to advance, as proposed in Reference~\cite{(deG26)}. In {\bf Figure}~\ref{fig:ElectromagneticMoments}(c), we display the currently available data for magnetic octupole moments compared with the single-particle estimates given by the Schwartz moments (Section~\protect\ref{Schwartz_moments}). See also preliminary DFT results presented in Section~\protect\ref{DFT_Magnetic_octupole} and Supplemental Material (Reference~\cite{ARNPS-suppl}, Section~\ref{Magnetic_octupole}).


\section{PARITY-BREAKING MOMENTS FOR FUNDAMENTAL INTERACTIONS\label{fundamental}}

Recently, increasing attention has been given to exotic parity-breaking moments, as seen in References~\cite{GarciaRuiz2020,Arrowsmith-Kron_2024}, namely the electric dipole, anapole, electric octupole, and magnetic quadrupole moments, for their significance in explaining the fundamental symmetries of nature and in advancing ultra-precise laboratory techniques.

\subsection{Electric dipole moments, the Schiff moments\label{Schiff}}

The subject matter has already been presented in the recent review article (Reference~\cite{(Eng25)}). Here, we summarise the basic principles and present selected DFT results.
\begin{marginnote}[]
\entry{Nuclei, atoms, and molecules}{Sensitive probes for new parity-breaking fundamental interactions that may induce otherwise impossible exotic non-zero parity-breaking moments}
\end{marginnote}
An atom can acquire an EDM from the nucleus. In turn, a nucleus acquires an EDM from either nucleon EDM or parity- and time-reversal symmetry-violating interaction between nucleons and pions. The nuclear quantity that induces the atomic EDM is not the nuclear dipole moment itself due to the Schiff theorem (Reference~\cite{(Sch63a)}), which states that, in a homogeneous external field, the nuclear EDM causes the rearrangement of electrons, which, in turn, generates an opposite electric field. The so-called nuclear Schiff moment is the relevant nuclear quantity that induces the atomic EDM. The definitions and derivations of the Schiff operator
    $\hat{S}_{z}=\frac{e}{10}\sum_{p}\left(r^{2}_{p}-\frac{5}{3}\langle r^{2}\rangle_{\text{ch}}\right)z_{p}$
can be found in References~\cite{(Spevak1997),(Aue08a),(Eng25a)}. The sum runs only over protons, and $\langle r^{2}\rangle_{\text{ch}}$ stands for the mean-square charge radius. In the recent publication, Reference~\cite{(Eng25a)}, Engel added to the Schiff operator, $\hat{S}_{z}$, the nucleon contribution, which is not discussed here.

The laboratory Schiff moment, $S_z^{\text{lab}}$,
is determined using second-order perturbation theory,
    $S_z^{\text{lab}}\approx\sum_{i\neq 0}\frac{\langle\Psi_{0}|\hat{S}_{z}|\Psi_{i}\rangle\langle\Psi_{i}|\hat{V}_{\text{PT}}|\Psi_{0}\rangle}{E_{0}-E_{i}}+\text{c.c.}$,
where $|\Psi_{0}\rangle$ is the ground state and the sum is over excited states.
The $P,T$-violating $NN$ interaction, $\hat{V}_{\text{PT}}$, is given in References~\nocite{Maekawa2011,(Hax83a), (Her88)}\citemany{Maekawa2011}{(Her88)},
\begin{align}
    \hat{V}_{\text{PT}}(\bm{r}_{1}-\bm{r}_{2})&=-\frac{gm^{2}_{\pi}}{8\pi m_{N}}\left\{(\hat{\bm{\sigma}}_{1}-\hat{\bm{\sigma}}_{2})\cdot(\bm{r}_{1}-\bm{r}_{2})\Big[\bar{g}_{0}\hat{\Vec{\tau}}_{1}\cdot\hat{\Vec{\tau}}_{2}-\frac{\bar{g}_{1}}{2}(\hat{\tau}_{1z}+\hat{\tau}_{2z})+\right.\nonumber\\
    &+\left.\bar{g}_{2}(3\hat{\tau}_{1z}\hat{\tau}_{2z}-\hat{\Vec{\tau}}_{1}\cdot\hat{\Vec{\tau}}_{2})\Big]-\frac{\bar{g}_{1}}{2}(\hat{\bm{\sigma}}_{1}+\hat{\bm{\sigma}}_{2})\cdot(\bm{r}_{1}-\bm{r}_{2})(\hat{\tau}_{1z}-\hat{\tau}_{2z})\right\}\nonumber\\
    &\times\frac{e^{-m_{\pi}r}}{m_{\pi}r^{2}}\left[1+\frac{1}{m_{\pi}r}\right]+\frac{1}{2m^{3}_{N}}\left[\bar{c}_{1}+\bar{c}_{2}\hat{\Vec{\tau}}_{1}\cdot\hat{\Vec{\tau}}_{2}\right](\hat{\bm{\sigma}}_{1}-\hat{\bm{\sigma}}_{2})\cdot\bm{\nabla}\delta^{3}(\bm{r}_{1}-\bm{r}_{2}).
    \label{VPT}
\end{align}
Note that a time-even Schiff operator $\hat{S}_{z}$ probes a time-odd interaction $\hat{V}_{\text{PT}}$. This is possible because the nuclear anti-linear time-reversal operator does not generate any selection rules.
In Equation~\ref{VPT}, $r=|\bm{r}_{1}-\bm{r}_{2}|$ is the relative distance between two nucleons, $m_{\pi}$ is pion mass, $m_{N}$ is nucleon mass, $g$ is the $\pi NN$ coupling constant, and the convention $\hbar=c=1$ is used. The isoscalar, isovector, and isotensor $T$-violating pion-nucleon coupling constants are denoted by $\bar{g}_{0}$, $\bar{g}_{1}$, and $\bar{g}_{2}$, respectively. The coupling constants $\bar{c}_{1}$ and $\bar{c}_{2}$ belong to the short-range interaction. The linearity of $\hat{V}_{\text{PT}}$ allows us to express the laboratory Schiff moment, $S_z^{\text{lab}}$, as in Reference~\cite{(Eng25a)},
    $S_z^{\text{lab}}=a_{0}g\bar{g}_{0}+a_{1}g\bar{g}_{1}+a_{2}g\bar{g}_{2}+b_{1}\bar{c}_{1}+b_{2}\bar{c}_{2}$,
where
coefficients $a$ and $b$ are the results of the calculation.


The matrix elements $\langle\Psi_{0}|\hat{S}_{z}|\bar{\Psi}_{0}\rangle$ and $\langle\Bar{\Psi}_{0}|\hat{V}_{\text{PT}}|\Psi_{0}\rangle$ can, in principle, be calculated for parity and angular-momentum symmetry-restored states, Section~\ref{Symmetry}, but in deformed nuclei, precise results can be obtained by employing the large-deformation approximation, Equation~\ref{spectroscopic_final2}.
In terms of the coupling constants $g$, $\bar{g}_{0}$, $\bar{g}_{1}$, $\bar{g}_{2}$, $\bar{c}_{1}$, and $\bar{c}_{2}$, we have
    $\langle\hat{V}_{\text{PT}}\rangle_{\text{intr}}=v_{0}g\bar{g}_{0}+v_{1}g\bar{g}_{1}+v_{2}g\bar{g}_{2}+w_{1}\bar{c}_{1}+w_{2}\bar{c}_{2}$,
which leads to the laboratory Schiff moment
given by the following constants,
    $a_{i}=-\frac{2J}{J+1}\frac{v_{i}\langle\hat{S}_{z}\rangle_{\text{intr}}}{\Delta E}\;\;\;\text{and}\;\;\;b_{j}=-\frac{2J}{J+1}\frac{w_{j}\langle\hat{S}_{z}\rangle_{\text{intr}}}{\Delta E}$,
with $i=0,1,2$ and $j=1,2$.

\begin{figure}[h]
    \centering
     \includegraphics[width=\textwidth]{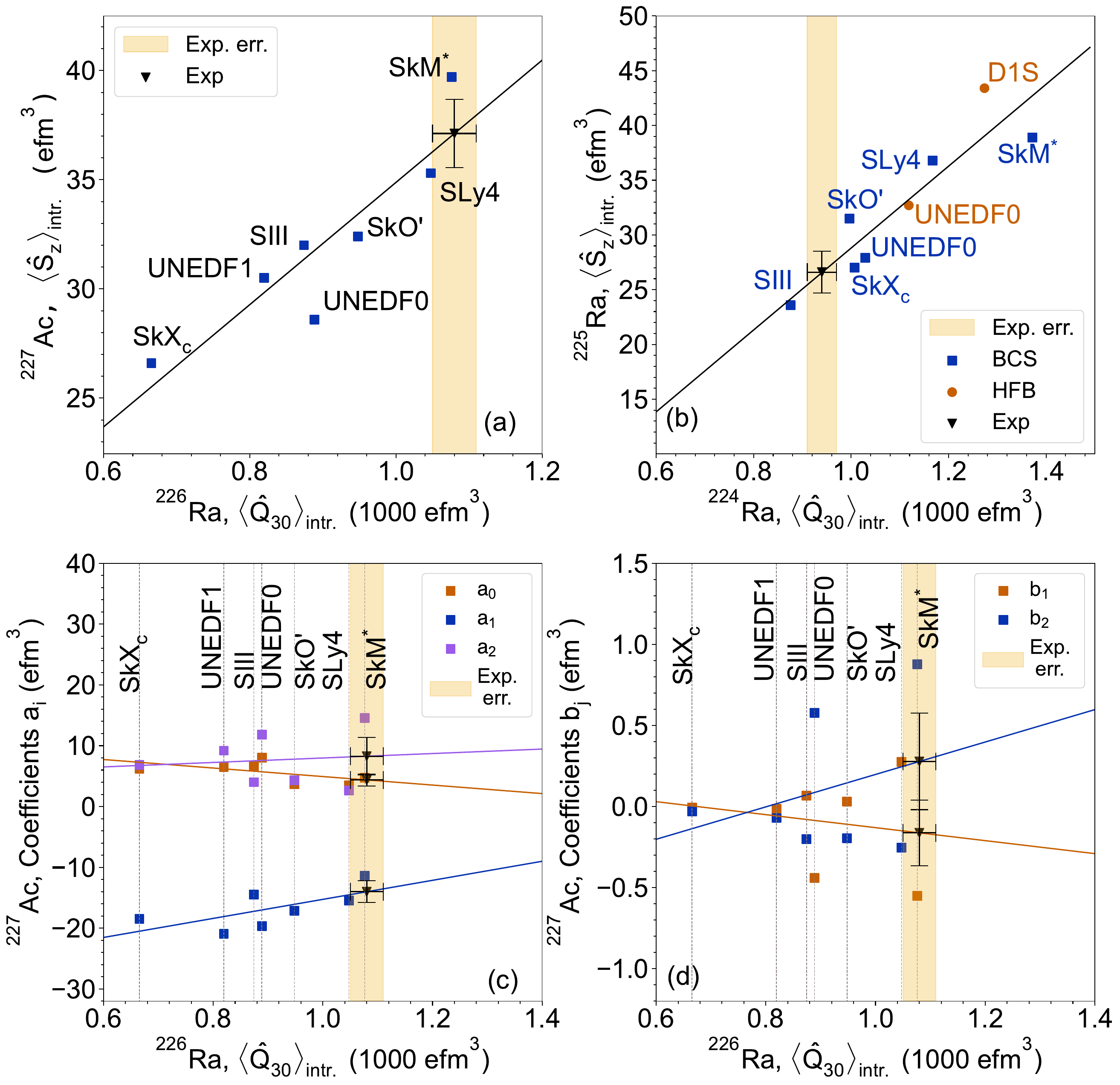}
    \caption{The DFT Schiff and octupole moments calculated in $^{227}$Ac and $^{225}$Ra, see text. Panel (a) is adapted from~\protect\cite{(Ath25a)} and panel (b) is adapted from~\protect\cite{(Dob18a)}.\label{fig: Ac227 Schiff moment}}
\end{figure}
Following earlier DFT calculations of the coefficients $a_i$ and $b_i$ performed for several actinides, Reference~\cite{(Dob18a)}, they have recently been performed also for $^{227}$Ac, Reference~\cite{(Ath25a)}. The $^{227}$Ac results obtained for seven Skyrme energy density functionals are summarised in {\bf Figure}~\ref{fig: Ac227 Schiff moment}. Panel (a) demonstrates a strong correlation between intrinsic Schiff moment, $\langle\hat{S}_{z}\rangle_{\text{intr}}$, of $^{227}$Ac and the intrinsic electric octupole moment, $\langle\hat{Q}_{30}\rangle_{\text{intr}}=e\sum_{i=1}^Z\langle{}r_i^{3}Y_{30}(\theta_i,\phi_i)\rangle_{\text{intr}}$, of $^{226}$Ra. As illustrated in panel (b), in the previous work (Reference~\cite{(Dob18a)}), the intrinsic Schiff moment of $^{225}$Ra was also correlated with the intrinsic octupole moment of $^{224}$Ra. The shaded bands illustrate the measured intrinsic octupole moments, 940(30) efm$^3$ in $^{224}$Ra and 1080(30) efm$^{3}$ in $^{226}$Ra, References~\cite{(Gaf13)} and~\cite{(Wol93)}, respectively. The propagated intrinsic Schiff moments of $^{227}$Ac and $^{225}$Ra and their uncertainties at the experimental intrinsic octupole moments of $^{226}$Ra and $^{224}$Ra, respectively, are 37.1(1.6) and 26.6(1.9) efm$^3$. The $^{227}$Ac coefficients $a_{0}$, $a_{1}$, $a_{2}$, $b_{1}$, and $b_{2}$, determined by regression analysis, are shown in panels (c) and (d) of {\bf Figure}~\ref{fig: Ac227 Schiff moment}. The uncertainties of coefficients $a_{i}$ are considerably smaller than those of $b_{j}$, implying that coefficients $a_{i}$ are better correlated with the intrinsic octupole moment of $^{226}$Ra.

\subsection{Anapole moments}

The nuclear anapole operator is the lowest-order parity non-conservation (PNC), time-reversal-odd\footnote{Here we define the time-reversal symmetry with respect to the time reversal of the nuclear wave function only, see  Reference~\protect\cite{(Hax02a)} for another convention.} multipole operator. Following Reference~\cite{(Wib26a)}, the spherical-vector ($\lambda=1$) anapole operator $\hat{a}_{\eta}$ with $\eta=0,\pm 1$ is defined as in Reference~\cite{(Hax02a)},
\begin{equation}
    \hat{a}_{\eta}=-\frac{M^{2}}{9}\int \rmd^{3}\bfr\;r^{2}\left[\hat{j}_{\text{tot};\eta}(\bm{r})+\sqrt{2\pi}\left[Y_{2}(\theta,\phi)\otimes\hat{\bm{j}}_{\text{tot}}(\bm{r})\right]_{1\eta}\right],
\end{equation}
where $M$ is the nucleon mass. The coupling between quadrupole spherical harmonics ${Y}_{2\eta}(\theta,\phi)$ and current operator $\hat{\bm{j}}_{\text{tot}}(\bm{r})$ is defined as follows,
    $\left[Y_{2}(\theta,\phi)\otimes\hat{\bm{j}}_{\text{tot}}(\bm{r})\right]_{1\eta}=\sum_{t}C^{1\eta}_{2(\eta-t)1t}Y_{2(\eta-t)}(\theta,\phi)\hat{j}_{\text{tot};t}(\bm{r})$.
Another definition of the anapole operator is more commonly used, Reference~\cite{(Fla1984)},
    $\hat{a}_{\eta}=-\pi\int \rmd^{3}\bfr\;r^{2}\;\hat{j}_{\text{tot};\eta}(\bm{r})$.

Analogous to the laboratory Schiff moment, $S_z^{\text{lab}}$, the laboratory anapole moment is
    $a_z^{\text{lab}}\approx\sum_{i\neq 0}\frac{\langle\Psi_{0}|\hat{a}_{z}|\Psi_{i}\rangle\langle\Psi_{i}|\hat{V}_{\text{PNC}}|\Psi_{0}\rangle}{E_{0}-E_{i}}+\text{c.c.}$,
where $\hat{V}_{\text{PNC}}$ is used in the form of the Desplanques Donoghue, and Holstein (DDH) potential, Reference~\cite{(Des80)}, and has the form as follows,
\begin{align}
    \hat{V}^{\text{DDH}}(\bm{r})&=\tfrac{i}{2\sqrt{2}}h^{1}_{\pi}g_{\pi NN}\left(\hat{\bm{\tau}}_{1}\times\hat{\bm{\tau}}_{2}\right)_{z}(\hat{\bm{\sigma}}_{1}+\hat{\bm{\sigma}}_{2})\cdot\tfrac{1}{2M}\big[\hat{\bm{p}}_{1}-\hat{\bm{p}}_{2},\omega_{\pi}(r)\big]\nonumber\\
    &-g_{\rho}\Big(h^{0}_{\rho}\hat{\bm{\tau}}_{1}\cdot\hat{\bm{\tau}}_{2}+\tfrac{1}{2}h^{1}_{\rho}(\hat{\bm{\tau}}_{1}+\hat{\bm{\tau}}_{2}\big)_{z}+\tfrac{1}{2\sqrt{6}}h^{2}_{\rho}(3\hat{\tau}_{1z}\hat{\tau}_{2z}-\hat{\bm{\tau}}_{1}\cdot\hat{
    \bm{\tau}}_{2})\Big)\nonumber\\
    &~~~\times\tfrac{1}{2M}\Big((\hat{\bm{\sigma}}_{1}-\hat{\bm{\sigma}}_{2})\cdot\big\{\hat{\bm{p}}_{1}-\hat{\bm{p}}_{2},\omega_{\rho}(r)\big\}+i(1+\chi_{V})(\hat{\bm{\sigma}}_{1}\times\hat{\bm{\sigma}}_{2})\cdot\big[\hat{\bm{p}}_{1}-\hat{\bm{p}}_{2},\omega_{\rho}(r)\big]\Big)\nonumber\\
    &-g_{\omega}\Big(h^{0}_{\omega}+\tfrac{1}{2}h^{1}_{\omega}\left(\hat{\bm{\tau}}_{1}+\hat{\bm{\tau}}_{2}\right)_{z}\Big)\nonumber\\
    &~~~\times\tfrac{1}{2M}\Big((\hat{\bm{\sigma}}_{1}-\hat{\bm{\sigma}}_{2})\cdot\big\{\hat{\bm{p}}_{1}-\hat{\bm{p}}_{2},\omega_{\omega}(r)\big\}+i(1+\chi_{S})(\hat{\bm{\sigma}}_{1}\times\hat{\bm{\sigma}}_{2})\cdot\big[\hat{\bm{p}}_{1}-\hat{\bm{p}}_{2},\omega_{\omega}(r)\big]\Big)\nonumber\\
    &+\tfrac{1}{2}\left(\hat{\bm{\tau}}_{1}-\hat{\bm{\tau}}_{2}\right)_{z}(\hat{\bm{\sigma}}_{1}+\hat{\bm{\sigma}}_{2})\cdot\tfrac{1}{2M}\Big(g_{\rho}h^{1}_{\rho}\big\{\hat{\bm{p}}_{1}-\hat{\bm{p}}_{2},\omega_{\rho}(r)\big\}-g_{\omega}h^{1}_{\omega}\big\{\hat{\bm{p}}_{1}-\hat{\bm{p}}_{2},\omega_{\omega}(r)\big\}\Big)\nonumber\\
    &-\tfrac{i}{2}g_{\rho}h^{1'}_{\rho}\left(\hat{\bm{\tau}}_{1}\times\hat{\bm{\tau}}_{2}\right)_{z}(\hat{\bm{\sigma}}_{1}+\hat{\bm{\sigma}}_{2})\cdot\tfrac{1}{2M}\big[\hat{\bm{p}}_{1}-\hat{\bm{p}}_{2},\omega_{\rho}(r)\big].\label{eq: DDH potential from Wick's paper}
\end{align}
 Here $\omega_{i}(r)=\exp(-m_{i}r)/4\pi r$ ($i=\pi,\;\omega,\;\rho$) are the Yukawa potentials, the momentum operator is $\hat{\bm{p}}_{j}=-i{\bm{\nabla}}_{j}$ ($j=1,\;2$), and the brackets and braces denote commutators and anticommutators, respectively. Similarly, as we noted for the Schiff moment in Section~\ref{Schiff}, here a time-odd anapole operator probes a time-even DDH interaction. This is again possible because the nuclear anti-linear time-reversal operator does not generate any selection rules. The DFT implementation of the DDH potential follows that of the $\hat{V}_{\text{PT}}$ in Equation~\ref{VPT} (Reference~\cite{(Dob09g)}).

\section{MESON EXCHANGE CURRENTS\label{MEC}}

For many years, discrepancies existed between the calculated and experimental magnetic dipole moments in heavy nuclei, which were tentatively attributed to the fact that the one-body magnetic dipole moment operator is not the most general, as it does not account for the interaction between nucleons mediated by pions. Only very recently, this conjecture was tested in the {\it ab initio} calculations of References~\cite{(Seu23),(Miy24)} and is also available in nuclear DFT, Reference~\cite{(Wib26b)}.

\begin{marginnote}[]
\entry{Meson exchange currents}{Induce corrections to nuclear moments via virtual meson fields}
\end{marginnote}
A more general expression of the magnetic dipole operator is, $
    \hat{\bm{\mu}}=\sum_{k}^{A}\hat{\bm{\mu}}_{1\text{b},k}+\sum_{k<\ell}^{A}\hat{\bm{\mu}}_{2\text{b},k\ell},$
where the first term is the one-body (1b) term given by Equation~(\ref{eq:mu_qm_operator_total}) and the second takes into account meson-exchange corrections (MECs),
\begin{equation}
    \hat{\bm{\mu}}_{2\text{b}}(\bm{r}_{1},\bm{r}_{2})=\frac{1}{2}\int \rmd^{3}\bm{x}\;\bm{x}\times\hat{\bm{j}}_{2\text{b}}(\bm{x},\bm{r}_{1},\bm{r}_{2}).\label{eq: two-body operator}
\end{equation}
The two-body current $\hat{\bm{j}}_{2\text{b}}$ is determined from two diagrams: seagull and pion-in-flight (Reference~\cite{(Cas90c)}).
They represent the next-to-leading order (NLO) contributions, as per the chiral effective field theory ($\chi$EFT), References~\cite{(Seu23),(Miy24)}. It is worth mentioning that, although the explicit pion term is absent in the Skyrme forces, the study of MEC is still possible since the
pions are virtual (Reference~\cite{MORSE1990}).

The two-body operator (\ref{eq: two-body operator}) has two terms, called the intrinsic term,
\begin{equation}
    \hat{\bm{\mu}}^{\text{ int}}_{2\text{b}}(\bm{r}_{1},\bm{r}_{2})=-\frac{g^{2}_{A}m_{\pi}}{32\pi F^{2}_{\pi}}(\hat{\bm{\tau}}_{1}\times\hat{\bm{\tau}}_{2})_{z}\left\{\left(1+\frac{1}{u}\right)\left[(\hat{\bm{\sigma}}_{1}\times\hat{\bm{\sigma}}_{2})\cdot\hat{\bm{r}}\right]\hat{\bm{r}}-(\hat{\bm{\sigma}}_{1}\times\hat{\bm{\sigma}}_{2})\right\}e^{-u},
    \label{intinsic}
\end{equation}
and the Sachs term,
\begin{equation}
    \hat{\bm{\mu}}_{2\text{b}}^{\text{Sachs}}(\bm{r}_{1},\bm{r}_{2})=-\frac{m^{3}_{\pi}g^{2}_{A}}{96\pi F^{2}_{\pi}}(\hat{\bm{\tau}}_{1}\times\hat{\bm{\tau}}_{2})_{z}(\bm{R}\times\bm{r})\left[\hat{S}_{12}\left(1+\frac{3}{u}+\frac{3}{u^{2}}\right)+\hat{\bm{\sigma}}_{1}\cdot\hat{\bm{\sigma}}_{2}\right]\frac{e^{-u}}{u}.
    \label{Sachs}
\end{equation}
Here the axial coupling constant $g_{A}=1.27$, pion decay constant $F_{\pi}=92.3$ MeV, $m_{\pi}=0.7045\;\text{fm}^{-1}$ is the average pion mass, and the scaled distance $u=m_{\pi}r$ is used. The intrinsic term depends only on the relative coordinate $\bm{r}\equiv\bm{r}_{1}-\bm{r}_{2}=r\hat{\bm{r}}$, whereas the Sachs term also depends on the center-of-mass coordinate $\bm{R}\equiv\tfrac{1}{2}(\bm{r}_{1}+\bm{r}_{2})$ between two nucleons. The tensor operator $\hat{S}_{12}$ is defined as $\hat{S}_{12}=3(\hat{\bm{r}}\cdot\hat{\bm{\sigma}}_{1})(\hat{\bm{r}}\cdot\hat{\bm{\sigma}}_{2})-\hat{\bm{\sigma}}_{1}\cdot\hat{\bm{\sigma}}_{2}$.


Within nuclear DFT, the MEC contributions are calculated as the expectation values of two two-body operators (Equations~\ref{intinsic} and ~\ref{Sachs}), that is
\begin{equation}
    \langle\hat{\bm{\mu}}_{2\text{b}}(\bm{r}_{1},\bm{r}_{2})\rangle=\frac{1}{2}\sum_{ij}\langle ij|\hat{\bm{\mu}}_{2\text{b}}(\bm{r}_{1},\bm{r}_{2})|ij\rangle-\frac{1}{2}\sum_{ij}\langle ij|\hat{\bm{\mu}}_{2\text{b}}(\bm{r}_{1},\bm{r}_{2})\hat{P}^{M}\hat{P}^{\sigma}\hat{P}^{\tau}|ij\rangle.\label{eq: MEC correction}
\end{equation}
The second (exchange) term differs from the first (direct) term by the negative sign and presence of the spatial-exchange (Majorana) $\hat{P}^{M}$, the spin-exchange $\hat{P}^{\sigma}$, and the isospin-exchange $\hat{P}^{\tau}$ operators. As it turns out, in the particle-hole (ph) channel, the direct isospin matrix elements of the operator $(\hat{\bm{\tau}}_{1}\times\hat{\bm{\tau}}_{2})_{z}$
vanish. Furthermore, following Reference~\cite{(Per04c)}, in the particle-particle (pp) channel, matrix elements of $(\hat{\bm{\tau}}_{1}\times\hat{\bm{\tau}}_{2})_{z}$ also vanish. Hence, the sole contribution to $\langle\hat{\bm{\mu}}_{2\text{b}}(\bm{r}_{1},\bm{r}_{2})\rangle$ comes from the exchange term of Equation~\ref{eq: MEC correction} in the ph channel. These facts greatly simplify the DFT calculations of the MEC contributions, which will be presented in a forthcoming publication (Reference~\cite{(Wib26b)}).

\section{NUCLEAR-DFT SELF-CONSISTENT STATES\label{Pairing}}

In this section, for completeness, we provide a brief overview of the DFT theory with pairing, which is extensively discussed in textbooks, see, for instance, References~\cite{(Rin80),(Sch19b)}. In particular, we focus on describing odd-particle-number states, which is essential for understanding magnetic moments.

\subsection{Even paired fermion states\label{Even-Pairing}}

To account for pairing, we use the quasiparticle product states as given by the Hartree-Fock-Bogoliubov (HFB) formalism, References~\cite{(Rin80),(Sch19b)}. The Bogoliubov transformation defines the quasiparticle creation and annihilation operators,
\begin{eqnarray}
    \alpha_\alpha^+ &=& \sum_\nu \left(A_{\nu\alpha} a_\nu^+ + B_{\nu\alpha}a_\nu\right),\quad
    \alpha_\alpha = \sum_\nu \left(A^*_{\nu\alpha} a_\nu + B^*_{\nu\alpha}a^+_\nu\right),
\end{eqnarray}
respectively. The HFB vacuum is a zero-quasiparticle state,
$|\Phi\rangle = \mathcal{N}\prod_\nu \alpha_\nu|0\rangle, \quad\alpha_\nu|\Phi\rangle=0$,
where the product runs over the so-called essential quasiparticles $\alpha_\nu$, that is, those that do not annihilate the particle vacuum, $\alpha_\nu|0\rangle\neq0$, and $\mathcal{N}$ is a normalisation constant. We note that: (i) if the number of essential quasiparticles is even (odd), the state $|\Phi\rangle$ is a linear combination of states with even (odd) numbers of particles, and (ii) the phase of state $|\Phi\rangle$ depends on the product of phases of all essential quasiparticles.

Another useful form of the HFB vacuum for even numbers of particles is given by the pair-condensate state, defined by the Thouless theorem, which reads as follows,
\begin{equation}
    |\Phi\rangle_{\mathrm{even}} = \mathcal{N}_{\mathrm{even}} \exp\Big(-\tfrac{1}{2} \sum_{\mu\nu} Z^*_{\mu\nu} a^+_\mu a^+_\nu\Big) |0\rangle,
    \label{eq-thouless}
\end{equation}
where $Z_{\mu\nu}$ is the anti-symmetric Thouless matrix, which is linked to the Bogoliubov transformation via $Z=BA^{-1}$, and $a^+$ are the particle creation operators. Equation~\ref{eq-thouless} holds only for $\langle0|\Phi\rangle_{\mathrm{even}}\neq0$, which is equivalent to $A$ being non-singular. All information about the state $|\Phi\rangle_{\mathrm{even}}$ is contained in the generalised density matrix $\mathcal{R}$
\begin{equation}
    \mathcal{R} = \begin{pmatrix}
        \rho &\kappa\\-\kappa^* &1-\rho^*
    \end{pmatrix} = \begin{pmatrix}
        B^*B^T & B^*A^T\\A^*B^T & B^*B^T,
    \end{pmatrix} = \begin{pmatrix}
        B^* \\A^*
    \end{pmatrix} \begin{pmatrix}
        B^T & A^T
    \end{pmatrix}, \quad \mathcal{R}^2= \mathcal{R}.
    \label{QpR}
\end{equation}
 After using the pair-condensate state, Equation~\ref{eq-thouless}, as a trial wave function parameterised by $Z^*_{\mu\nu}$, one obtains the HFB equations $\mathcal{HA=AE}$, or explicitly,
\begin{equation}
    \begin{pmatrix}h^\prime-\lambda & \Delta\\ -\Delta^* & -h^{\prime *} +\lambda \end{pmatrix}  =  \begin{pmatrix}A & B^* \\ B & A^*\end{pmatrix}\begin{pmatrix}E & 0\\0 & -E\end{pmatrix}, ~~\mbox{for}~~\chi\equiv\begin{pmatrix}A \\ B \end{pmatrix}, \varphi\equiv\begin{pmatrix} B^* \\  A^*\end{pmatrix},
    \label{HFB}
\end{equation}
where $\mathcal{A}$ is the matrix Bogoliubov transformation, $E$ is a diagonal matrix containing the quasiparticle energies, $\mathcal{H}$ is the  quasiparticle Hamiltonian, $h^\prime$ is the single-particle Routhian matrix, $\lambda$ is the Fermi energy, and $\Delta$ is the pairing (or particle-particle) mean-field. Since $h^\prime$ depends on the density matrix $\rho$ and $\Delta$ depends on the pairing tensor $\kappa$, defined in Equation~\ref{QpR}, the HFB quasiparticle Hamiltonian $\mathcal{H}$ has to be diagonalised iteratively until the self-consistent solution $[\mathcal{H},\mathcal{R}]=0$ is found, that is, the columns of the Bogoliubov transformation $\mathcal{A}$ (the quasiparticle wave functions) become eigenstates of $\mathcal{H}$. Equation~\ref{QpR} then shows that $\mathcal{R}$ projects the space of the quasiparticle states onto the subspace of occupied quasiparticles, which have negative quasiparticle energies, and leaves the remaining quasiparticle states, which have positive quasiparticle energies, empty.

\subsection{Odd paired fermion states and blocking\label{Odd-Pairing}}

The blocking approximation is the standard tool for describing odd-$A$ and odd-odd nuclei. Within this approach, a quasiparticle $\beta$ is created atop the even-even state of equation \ref{eq-thouless}

\begin{equation}
|\Phi_\beta \rangle _{\mathrm{odd}} = \mathcal{N}_{\text{odd}} \alpha^+_\beta \exp\left(-\frac{1}{2} \sum_{\mu\nu} Z^*_{\mu\nu} a^+_\mu a^+_\nu\right) |0\rangle.
\label{blocking}
\end{equation}
This corresponds to replacing in the generalised density matrix $\mathcal{R}$ the negative-energy $\beta$-th column $\varphi_\beta$, Equation~\ref{HFB}, which represents an occupied quasiparticle state, with the positive-energy column $\chi_\beta$, which represents an empty quasiparticle state. After such a replacement, one of the single-particle states becomes fully occupied and thus does not contribute to pairing correlations. The effects of blocking, therefore, include an increase in the total energy (decrease in the binding energy) of the system due to the corresponding pair-breaking contribution, as well as a change in deformation due to polarisation.

\begin{marginnote}[]
\entry{Blocking}{Defines characteristics of the shape and angular-momentum polarisation of the nucleus induced by the odd nucleon.
}
\end{marginnote}
An essential aspect of the odd-particle-number quasiparticle states, defined in Equation~\ref{blocking}, is that those corresponding to different blocked states $\beta$ and $\beta'$ can be, barring conserved symmetries, smoothly connected to one another by a suitable path in the space of parameters $Z^*_{\mu\nu}$. This non-trivial fact can be best explained in the language of group theory, whereby the even and odd states belong to two different spinor representations of the special orthogonal group in even dimensions (see, for instance, Reference~\cite{(Dob81d)}) and can thus be parameterised within two disconnected compact manifolds.

In practice, this mathematical fact translates into a specific variational property of odd states, where an unconstrained variation only yields the lowest odd state of a given symmetry. Such a variational method, therefore, does not allow access to the full spectrum of odd quasiparticle states. To achieve this, a constrained variation is necessary, in which the blocked state is fixed at each self-consistent iteration using a specific method. This involves making consistent decisions at each iteration about which eigenstate of $\mathcal{H}$ to block.

To address this issue, the tagging mechanism, References~\cite{(Dob09g),(Wib25d)}, allows us to make those decisions in the following way: In each iteration of the self-consistent loop, one looks for the largest overlap between a fixed single-particle wave function, $\phi_{\text{tag}}$, and all quasiparticles $\varphi_\beta$. This is done by determining the overlaps with the upper $B_\beta^*$ and the time-reversed lower $\hat{T}A^*_\beta\hat{T}$ quasiparticle component, $\mathcal{O}_\beta=\mathrm{max}\left( (\phi_{\text{tag}}|B^*_\beta),(\phi_{\text{tag}}|\hat{T}A^*_\beta\hat{T}) \right)$ keeping the greater of the two. The quasiparticle with the greatest overlap $\mathcal{O}_\beta$ is then chosen to be blocked during that iteration. This procedure is insensitive to whether the quasiparticles are of the particle or hole type, as it follows the desired quasiparticle despite eventual crossing of the Fermi energy.

Traditionally, the odd-particle-number states are described in the language of the so-called particle-core coupling approach (Reference~\cite{(Suh07a)}), in which the states of its even-particle-number subsystem (the core) are coupled with the states of the odd single-particle. In this approach, the core polarisation is understood as taking into account not only the ground state but also a suitable linear combination of the excited core states.

In nuclear DFT, the approach differs because of the self-consistent angular-momentum and shape polarisation exerted by the blocked quasiparticle $\alpha^+_\beta$ on the even-particle-number subsystem in Equation~\ref{blocking}, which means the latter is always distinct from the genuine even system in Equation~\ref{eq-thouless}. Such polarisation is not described solely by the spectrum of eigenstates of the even system. Instead, it is non-perturbatively embedded into a single, modified state, which is different for every blocked quasiparticle $\alpha^+_\beta$. This warrants coining a new name, c{\oe}r, to the even state in Equation~\ref{blocking} (Reference~\cite{(Bon26)}) instead of the standard core. The analysis of the particle-c{\oe}r coupling is particularly fruitful for one-body operators, such as nuclear moments, whose averages can be exactly represented as the sums of products of contributions coming from the c{\oe}r and the odd particle, see the first application in Reference~\cite{(Ver22b)}.

\subsection{Convergence}

The convergence of the self-consistent iterative process is never guaranteed, and the speed at which the results converge heavily depends on the structure of the quasiparticle states and their energies. In particular, with crossings or a high density of quasiparticle energies, states may shift towards triaxial deformations despite being axial at self-consistency. This can significantly slow down or prevent convergence completely. Projecting the wave functions onto the axial shapes, Reference~\cite{(Dob21f)}, helps stabilise convergence if the system is genuinely axial, but the problem may persist such that no energy minima are found anyway.

\section{SYMMETRY RESTORATION\label{Symmetry}}

Depending on the physical situation one wishes to describe, the DFT solutions can be pre-configured to break or conserve various symmetries. In addition, for some observables, specific symmetries must be restored before comparing the DFT results with data, while for others, such a restoration yields nearly identical results and is not worth the effort. The decision to break and restore symmetries is, therefore, a matter of defining the research strategy and can impact both the meaningfulness of the obtained results and the time required to get them.

\begin{marginnote}[]
\entry{Symmetry breaking and restoration}{Essential elements to describe emerging low-energy features of heavy nuclei.}
\end{marginnote}
As discussed in Section~\ref{Results}, the results presented there were obtained by conserving the parity and axial symmetry and breaking the signature, rotational, and particle-number symmetries. As it turned out (Reference~\cite{(Wib25d)}), the particle-number symmetry restoration did not affect the results obtained for nuclear moments, whereas the restoration of rotational symmetry was crucial. Therefore, we discuss the latter in more detail.

General features of the rotational-symmetry restoration are amply discussed in the literature, see, for instance, References~\cite{(Rin80),(Dob09g),(She21),(Sch19b),(Bal21)}. Below, we focus on the specific case of restoring rotational symmetry for axial states, see, for instance, References~\cite{(Egido04),(Tan2023)}. We start by presenting the general case of the projected wave function $|\Phi_{IM}\rangle$, which, with well-defined angular momentum quantum numbers $I$ and $M$, is given by,
\begin{eqnarray}
\label{projected}
|\Phi_{IM}\rangle&=&\sum_{K=-I}^I g_{IK}|\Phi_{IMK}\rangle~~\mbox{for}~~
|\Phi_{IMK}\rangle=\hat{P}_{MK}^{I}|\Phi\rangle,~~\mbox{and}\\
\label{projector}
    \hat{P}_{MK}^{I}&=&\frac{2I+1}{8\pi^{2}}{{\int_0^{2\pi}} \rmd\alpha\int_0^{\pi}\sin\beta\,\rmd\beta\int_0^{2\pi}\rmd\gamma}\;D^{I\ast}_{MK}(\alpha,\beta,\gamma)\hat{R}(\alpha,\beta,\gamma),
\end{eqnarray}
where the projection operators $\big(\hat{P}_{KM}^{I}\big)^+=\hat{P}_{MK}^{I}$ fulfil the multiplication rule $\hat{P}_{K'M'}^{I'}\hat{P}_{MK}^{I}=\delta_{II'}\delta_{MM'}\hat{P}_{K'K}^{I}$, that is, $\hat{P}_{KK}^{I}$ are Hermitian and  projective, $\big(\hat{P}_{KK}^{I}\big)^2=\hat{P}_{KK}^{I}$. $|\Phi\rangle$ is a normalised, $\langle\Phi|\Phi\rangle$=1, rotational-symmetry-breaking wave function. ${D}^{I}_{MK}(\alpha,\beta,\gamma)$ and $\hat{R}(\alpha,\beta,\gamma)$ are the Wigner $D$-function (Reference~\cite{(Var88)}, Equation~4.1(1)) and rotation operator, respectively,
\begin{equation}
    {D}^{I}_{MK}(\alpha,\beta,\gamma)=e^{-i\alpha{}M}{d}^{I}_{MK}(\beta)e^{-i\gamma{}K}, \quad
    \hat{R}(\alpha,\beta,\gamma)=e^{-i\alpha\hat{I}_{z}}e^{-i\beta\hat{I}_{y}}e^{-i\gamma\hat{I}_{z}},
\label{Wigner}
\end{equation}
$\alpha,\beta,\gamma$ are the three Euler angles, and $\hat{\bm{I}}$ is the angular momentum operator. We also note that the symmetry-breaking states of even or odd numbers of nucleons can be expressed through the unity resolution, see Supplemental Material, Reference~\cite{ARNPS-suppl}, Section~\ref{unity_resolution}, as the sums of, respectively, integer-$I$ or half-integer-$I$ projected states,
\begin{equation}
    \label{complete}
|\Phi\rangle=\sum_{I}^\infty\sum_{K=-I}^I \hat{P}_{KK}^{I}|\Phi\rangle=\sum_{I}^\infty\sum_{K=-I}^I |\Phi_{IKK}\rangle,
\end{equation}
where, for $K\neq{}K'$, states $|\Phi_{IMK}\rangle$ and $|\Phi_{IMK'}\rangle$ are neither normalised nor orthogonal.

For any spherical-tensor operator $\hat{O}_{\lambda\mu}$, which can represent in particular the quadrupole or magnetic multipole operators of Equations~\ref{eq:electric}--\ref{eq:magnetic2}, the projection operators and their matrix elements have two fundamental properties (Supplemental Material (Reference~\cite{ARNPS-suppl}, Section~\ref{3D}),
\begin{equation}
    \hat{P}^{I'}_{K'M'}\hat{O}_{\lambda\mu}\hat{P}^{I}_{MK}=C^{I'M'}_{IM\lambda\mu}\sum_{\nu\mu'}C^{I'K'}_{I\nu\lambda\mu'}\hat{O}_{\lambda\mu'}\hat{P}^{I}_{\nu K},
    \label{Shift}
\end{equation}
and the Wigner-Eckart theorem,
\begin{equation}
    \langle\Phi_{I'M'}|\hat{O}_{\lambda\mu}|\Phi_{IM}\rangle=\frac{1}{\sqrt{2I'+1}}C^{I'M'}_{IM\lambda\mu}\langle\Phi_{I'}\|\hat{O}_{\lambda}\|\Phi_{I}\rangle,\label{Wigner-Eckart}
\end{equation}
where the so-called reduced matrix element is defined as,
\begin{align}\label{reduced_matrix_element}
    \langle\Phi_{I'}\|\hat{O}_{\lambda}\|\Phi_{I}\rangle&=\sqrt{2I'+1}\sum_{\mu{}M}C^{I'M'}_{IM\lambda\mu}\langle\Phi_{I'M'}|\hat{O}_{\lambda\mu}|\Phi_{IM}\rangle.
\end{align}

In the case of conserved axial symmetry and broken signature, which is of interest here, the normalised, $\langle\Phi_\Omega|\Phi_\Omega\rangle$=1,  rotational-symmetry-breaking aligned wave function, $|\Phi_\Omega\rangle$, of an odd-$A$ nucleus is an eigenstate of the projection of the angular momentum on the axial-symmetry $z$-axis with a half-integer eigenvalue $\Omega$, that is,
$\hat{I}_{z}|\Phi_\Omega\rangle=\Omega|\Phi_\Omega\rangle$. Then, integration over the angle $\gamma$ in Equation~\ref{projector} gives the factor of $2\pi\delta_{K\Omega}$, which reduces the sum in Equation~\ref{projected} to one term, and gives $|\Phi_{IM}\rangle=|\widetilde{\Phi}_{IM\Omega}\rangle={\caN}_{I\Omega}|\Phi_{IM\Omega}\rangle$ for $|\Phi_{IM\Omega}\rangle\equiv\hat{P}^I_{M\Omega}|\Phi_\Omega\rangle$. In this way, $g_{I\Omega}\equiv{\caN}_{I\Omega}$ becomes a normalisation factor of $|\Phi_{IM\Omega}\rangle$ and the normalised projected wave function $|\Phi_{IM}\rangle=|\widetilde{\Phi}_{IM\Omega}\rangle$ inherits the quantum number $\Omega$ of the broken-symmetry state $|\Phi_\Omega\rangle$.

By the same argument, the matrix element of Equation~\ref{Shift} between two different aligned states $\langle\Phi_{\Omega_1}|$ and $|\Phi_{\Omega_2}\rangle$ is given by,
\begin{equation}
\begin{aligned}
    \langle\Phi_{\Omega_1}|\hat{P}^{I'}_{\Omega_1M'}\hat{O}_{\lambda\mu}\hat{P}^{I}_{M\Omega_2}|\Phi_{\Omega_2}\rangle&=
    \langle\Phi_{I'M'\Omega_1}|\hat{O}_{\lambda\mu}|\Phi_{IM\Omega_2}\rangle\\&=
    C^{I'M'}_{IM\lambda\mu}\sum_{\nu\mu'}C^{I'\Omega_{1}}_{I\nu\lambda\mu'}\langle\Phi_{\Omega_1}|\hat{O}_{\lambda\mu'}\hat{P}^{I}_{\nu \Omega_{2}}|\Phi_{\Omega_2}\rangle.
    \label{Shift_element}
\end{aligned}
\end{equation}
By using the fact that a spherical tensor $\hat{O}^+_{\lambda\mu'}=(-1)^{\lambda+\mu'}\hat{O}_{\lambda,-\mu'}$ subtracts the angular-momentum projection $\mu'$ from state $|\Omega_1\rangle$, that is, $\hat{I}_z\hat{O}^+_{\lambda\mu'}|\Omega_1\rangle=(\Omega_1-\mu')\hat{O}^+_{\lambda\mu'}|\Omega_1\rangle$, or equivalently, $\langle\Phi_{\Omega_1}|\hat{O}_{\lambda\mu'}\hat{I}_z=\langle\Phi_{\Omega_1}|\hat{O}_{\lambda\mu'}(\Omega_1-\mu')$, the integration over angle $\alpha$ in Equation~\ref{projector} reduces again to $2\pi\delta_{\Omega_1-\mu',\nu}$ and thus only the integration over $\beta$ remains. Then, from Equation~\ref{Shift_element} we derive the final result,
\begin{eqnarray}
\langle\widetilde{\Phi}_{I'M'\Omega_1}|\hat{O}_{\lambda}|\widetilde{\Phi}_{IM\Omega_2}\rangle
&=&\tfrac{\caN_{I_1\Omega_1}\caN_{I_2\Omega_2}(2I+1)}{2}C^{I'M'}_{IM\lambda\mu}\sum_{\mu'}C^{I'\Omega_{1}}_{I\Omega_1-\mu',\lambda\mu'}
\nonumber \\ &&\times
\int_0^{\pi}\sin\beta\,\rmd\beta\;d^{I}_{\Omega_1-\mu',\Omega_2}(\beta)\langle\Phi_{\Omega_1}|\hat{O}_{\lambda\mu'}e^{-i\beta\hat{I}_{y}}|\Phi_{\Omega_2}\rangle,
    \label{reduced_final}
\end{eqnarray}
where the integral over $\beta$ must be calculated numerically.
In particular, the spectroscopic moments of Equations~\ref{eq:mm_definition}, \ref{eq:qadrupole_moment}, and~\ref{eq:octupole_moment} are given by the diagonal terms, $I_1M_1\Omega_1\equiv{}I_2M_2\Omega_2\equiv{}II\Omega$.
 We also note that in axial odd nuclei with half-integer $\Omega$, Equation~\ref{complete} reduces to
\begin{equation}
|\Phi_\Omega\rangle=\sum_{I=\Omega}^\infty\hat{P}_{\Omega\Omega}^{I}|\Phi_\Omega\rangle=\sum_{I=\Omega}^\infty|\Phi_{I\Omega\Omega}\rangle=\sum_{I=\Omega}^\infty{\caN}_{I\Omega}^{-1}|\widetilde{\Phi}_{I\Omega\Omega}\rangle.
    \label{complete2}
\end{equation}

As the last result of this section, we define the so-called large-axial-deformation approximation of spectroscopic moments, which assumes that rapidly-varying functions of $\beta$ $\langle\Phi_{\Omega}|\hat{O}_{\lambda\mu'}e^{-i\beta\hat{I}_{y}}|\Phi_{\Omega}\rangle$ and $\langle\Phi_{\Omega}|e^{-i\beta\hat{I}_{y}}|\Phi_{\Omega}\rangle$ are such that their ratio, called the reduced kernel, is a slowly-varying function of $\beta$, which can be expanded in the Taylor series as follows,
\begin{equation}
\frac{\langle\Phi_{\Omega}|\hat{O}_{\lambda\mu'}e^{-i\beta\hat{I}_{y}}|\Phi_{\Omega}\rangle}{\langle\Phi_{\Omega}|e^{-i\beta\hat{I}_{y}}|\Phi_{\Omega}\rangle}=\sum_{n=0}n_{\lambda\mu'}^{(n)}\beta^n
=\delta_{\mu'0}\langle\Phi_{\Omega}|\hat{O}_{\lambda0}|\Phi_{\Omega}\rangle+\sum_{n=1}n_{\lambda\mu'}^{(n)}\beta^n,    \label{reduced_kernel}
\end{equation}
where the $n=0$ term is derived by setting $\beta=0$.
In particular, keeping only that term, we obtain the standard approximate relationship between the spectroscopic ${O}^{\text{spec}}_{\lambda{I}\Omega}$ and intrinsic ${O}^{\text{intr}}_{\lambda\Omega}$ moments (Supplemental Material, Reference~\cite{ARNPS-suppl}, Section~\ref{large-axial-deformation approximation}),
\begin{equation}
{O}^{\text{spec}}_{\lambda{I}\Omega}\equiv\langle\widetilde{\Phi}_{II\Omega}|\hat{O}_{\lambda0}|\widetilde{\Phi}_{II\Omega}\rangle
\simeq C^{II}_{II,\lambda0}C^{I\Omega}_{I\Omega,\lambda0}\langle\Phi_{\Omega}|\hat{O}_{\lambda0}|\Phi_{\Omega}\rangle \equiv C^{II}_{II,\lambda0}C^{I\Omega}_{I\Omega,\lambda0}{O}^{\text{intr}}_{\lambda\Omega} ,
    \label{spectroscopic_final2}
\end{equation}
or a parameterisation of the spectroscopic moment ${O}^{\text{spec}}_{\lambda{I}\Omega}$ in terms of the effective intrinsic moment ${O}^{\text{intr}}_{\text{eff}}(\lambda{I}\Omega)$,
\begin{equation}
{O}^{\text{spec}}_{\lambda{I}\Omega} = C^{II}_{II,\lambda0}C^{I\Omega}_{I\Omega,\lambda0} {O}^{\text{intr}}_{\text{eff}}(\lambda{I}\Omega) ,
    \label{spectroscopic_final3}
\end{equation}
see Reference~\cite{(Dob26e)}.

\section{DFT RESULTS\label{Results}}

\subsection{Generic features}


Along with several DFT studies of nuclear electromagnetic moments, see References~\nocite{(Bal14b),(Bon15),(Co15),(Bor16),(Bor17),(Li18),(Per21a),(Bal22a),(Rys22c),(Bal23),(Zho24),(Zho25b),(Zho25a),(Nak26)}\citemany{(Bal14b)}{(Nak26)}, reviewed in Reference~\cite{(Dob26e)}, a novel approach was introduced by Sassarini {\it et al.} in Reference~\cite{(Sas22c)}. This latter approach was employed in various experimental studies, References~\cite{(deG22b),(Ver22b)} and \nocite{(Nie23),(Gra23),(deG24),(Kar24a)}\citemany{(Nie23)}{(Kar24a)}, and applied across long chains of isotopes of several heavy elements, References~\cite{(Bon23c),(Wib25d),(Dob26e)}. The novelty of this approach did not lie in reinventing the wheel; all necessary components were previously known and used in dedicated research. However, combining them was essential for gaining a comprehensive understanding of the physics problem. We now present its main features.

We begin by discussing a key simplification of the method, specifically limiting it to axial shapes. Several previous results were obtained by analysing the collectivity within triaxial shapes, References~\cite{(Bal14b),(Bor16),(Bor17)} and~\nocite{(Bal22a),(Rys22c),(Zho25a)}\citemany{(Bal22a)}{(Zho25a)}. This has been the most advanced approach to date; however, when combined with full angular-momentum and particle-number symmetry restoration, Section~\ref{Symmetry}, it demands a computational effort of over four orders of magnitude greater than the axial approach used in References~\cite{(Bon23c),(Wib25d),(Dob26e)}. Consequently, triaxial results could only be obtained for a few nuclides, while the axial method enabled systematic studies of thousands of nuclides and excited states. Strategically, the latter enabled the creation of a reference set of results against which triaxial calculations can subsequently be performed to assess the impact of this additional symmetry breaking on the description of the experimental data. Furthermore, tests of the particle-number symmetry restoration indicated that it does not alter the results by more than about 1\%, and therefore can be safely omitted (Reference~\cite{(Wib25d)}). We also note that the availability of results across a wide range of neutron and proton numbers is essential for drawing meaningful conclusions about the general properties of a theoretical description, rather than spotlighting calculations in isolated nuclides.

A comprehensive account of the c{\oe}r's angular-momentum polarisation (Section~\ref{Odd-Pairing}) is essential in determining the nuclear magnetic moments. This involves two aspects: (i) the need to consider the time-odd mean fields and (ii) the importance of the angular momentum's direction relative to the shape of the charge distribution. Without the time-odd mean fields, the time-reversed states of the odd nucleon are perfectly Kramers-degenerate, and any linear combination of them can be blocked with the same total energy, see Section~\ref{Odd-Pairing}. This introduces the concept of alispin~\cite{(Sch10b)}, which becomes a symmetry related to those arbitrary linear combinations. Additionally, the absence of the c{\oe}r's angular-momentum polarisation induces the signature symmetry\footnote{The signature operator rotates the nucleus by 180$^\circ$ about the axis perpendicular to the axial-symmetry axis.} of the c{\oe}r. Consequently, it eliminates odd-spin states from its set of angular-momentum-projected states (References~\cite{(Bon26),(Ver22b)}). An essential aspect of particle-c{\oe}r coupling disappears along with it.

The second aspect of the polarisation of angular momentum concerns its direction within the nucleus. In Reference~\cite{(Sat12a)}, it has already been recognised that for the time-odd mean fields acting in a triaxial (axial) nucleus, three (two) independent self-consistent solutions are possible, with angular momenta aligned along one of the density's principal axes. The impact of these different orientations on the magnetic moments remains unknown. Two choices have been independently explored in various works, related to the conserved, References~\cite{(Bal14b),(Bal22a),(Bal23),(Bor16),(Bor17),(Per21a),(Rys22c)}, or broken, References~\cite{(Bon15),(Sas22c),(Bon23c),(Wib25d),(Dob26e),(Nak26)}, signature symmetry. In an axial nucleus, the former corresponds to orientation perpendicular to the axis of axial symmetry, while the latter relates to angular momentum aligned with that axis. Although detailed studies of orientation dependence are needed, there are two strong reasons in favour of the latter. First, for the broken signature, the magnitude of the angular-momentum polarisation is much larger than for the conserved signature, meaning the polarisation strength of the time-odd mean fields can act fully. Second, the perpendicular orientation of the angular momentum breaks the axial symmetry; thus, its study immediately involves complex triaxial calculations, as discussed above.

The final essential aspect of the methodology used in Reference~\cite{(Sas22c)} concerns the exact conservation of angular momentum. Although it was not necessary in studies conducted within the laboratory reference frame, References~\cite{(Co15),(Li18)}, and its restoration was applied in several earlier DFT studies, References~\cite{(Bal14b),(Bor16),(Bor17),(Bal22a),(Bal23)}; it was only in Reference~\cite{(Bon23c)} that its vital significance was explicitly demonstrated. As it turns out, the large-axial-deformation approximation, discussed in Section~\ref{Symmetry}, works well for the electric quadrupole moments, achieving accuracy within 1-5\% in well-deformed nuclei. Still, it can unpredictably worsen to 30\% when quadrupole moments are small. This indicates that calculations made solely in the intrinsic frame have limited applicability (References~\cite{(Rob12),(Bon23c)}). As shown in References~\cite{(Bon23c),(Dob26e)}, the situation is even more difficult for magnetic dipole moments, where values determined in the intrinsic frame fail to account for coupling to the c{\oe}r's quadrupole deformation and the large-axial-deformation approximation is not applicable.

\subsection{Nuclei near doubly magic systems}

\begin{figure}[h]
    \centering\includegraphics[width=\textwidth]{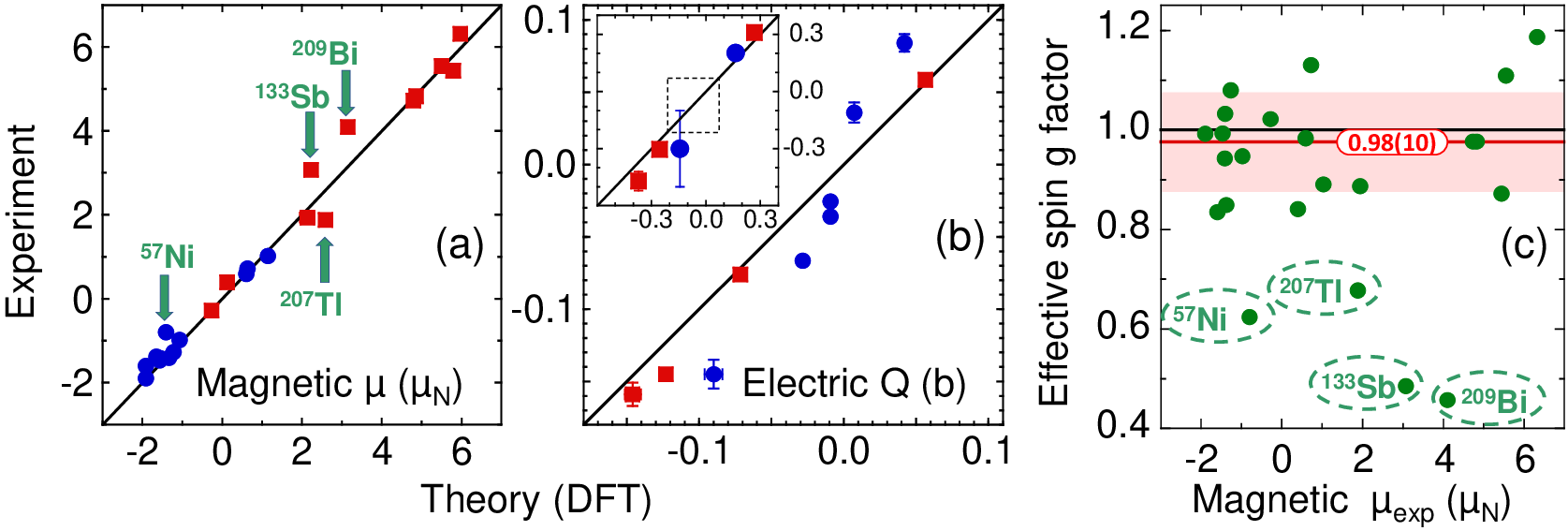}
\caption{
Calculated magnetic dipole moments $\mu$ (a) compared with 23 experimentally measured values (the arrows mark the outlier cases discussed in the text). Full circles (squares) show results obtained for odd-$N$ (odd-$Z$) nuclei. Panel (b) shows the electric quadrupole moments $Q$ compared with 15 experimentally
measured values (the inset shows values that are outside the area of the main
plot, as visualised by the dashed-line square drawn inside). Panel (c) shows the
effective spin $g$ factors described in the text, with ovals marking the outliers shown in panel (a).
Figure adapted from Reference~\cite{(Sas22c)}.
\label{fig:UNEDF1_JD-M-46-geff}}
\end{figure}

Reference~\cite{(Sas22c)} tested the novel approach by determining nuclear magnetic dipole and electric quadrupole moments in the 32 unpaired one-particle and one-hole neighbours of eight doubly magic nuclei. The results of this analysis are summarised in {\bf Figure}~\ref{fig:UNEDF1_JD-M-46-geff}. For the three studied Skyrme functionals, the isovector spin-spin interaction, defined by the Landau parameter $g_0'$, see References~\cite{(Ben02d),(Idi17)}, was adjusted to 23 known experimental values of magnetic dipole moments. As shown in panels~\ref{fig:UNEDF1_JD-M-46-geff}(a) and (c), for four studied nuclides, these moments significantly deviate from the data and can be qualified as outliers of the approach. With those excluded, the obtained average and RMS deviations from the data are 0.032\,$\mu_{\text{N}}$ and 0.178\,$\mu_{\text{N}}$, respectively. The corresponding deviations from 15 known experimental values of electric quadrupole moments, which are almost independent of $g_0'$, are 0.022\,b and 0.057\,b.

A significant result reported in Reference~\cite{(Sas22c)} concerns the so-called effective $g$ factors, which are adjustable constants that multiply the free spin and orbital $g$ factors (see Section~\ref{one-body}) to align calculated magnetic dipole moments with experimental data. In the past, numerous diverse physical effects have been invoked to justify the use of such numerical constants. These have included a limited size of the single-particle phase space, inadequate shape and/or angular-momentum polarisations, restrictions on conserved or broken symmetries or absence of angular-momentum restoration, missing terms or deficiencies in interactions or functionals, triaxial and/or octupole deformability, and contributions from two-body meson-exchange currents (Section~\ref{MEC}). Specifically, approaches that utilise a limited-size valence single-particle space, such as the shell model, always resort to effective $g$ factors.

\begin{marginnote}[]
\entry{Effective $g$ factors and effective charges}{Phenomenological corrections employed to account for missing elements in describing nuclear moments}
\end{marginnote}
{\bf Figure}~\ref{fig:UNEDF1_JD-M-46-geff}(c) displays the values of the hypothetical effective spin $g$ factors, $g_{\text{eff}}^{(n)}=g_{\text{eff}}^{(p)}$, that would have been necessary to adjust the calculated DFT magnetic dipole moments $\mu$ to match the 23 experimentally measured values individually.  As shown, except for the four outliers, the remaining 19 values cluster within a narrow range around $g_{\text{eff}}=0.98(10)$. This clearly indicates that using a sufficiently large single-particle space, as in DFT, eliminates the need for effective $g$ factors. Similarly, {\bf Figure}~\ref{fig:UNEDF1_JD-M-46-geff}(b) demonstrates that, for the same reason, the effective charges are then unnecessary.

\subsection{Open-shell nuclei}

Long chains of isotopes along the proton nearly magic nuclei were analysed in References~\cite{(Ver22b),(Nie23)} (indium),~\cite{(Gra23),(Wib25d)} (tin),~\cite{(deG24)} (silver), and antimony~\cite{(Wib25d)}. These studies did not reveal any specific pattern to motivate more extensive studies beyond the methodology used so far. In particular, the need to use unpaired rather than paired states to describe the magnetic moments of heavy indium isotopes (Reference~\cite{(Ver22b)}) and a poor description of proton spin-antialigned states in antimony (Reference~\cite{(Wib25d)}) remain subjects for further study.

\begin{figure}[h]
\centering\includegraphics[width=\textwidth]{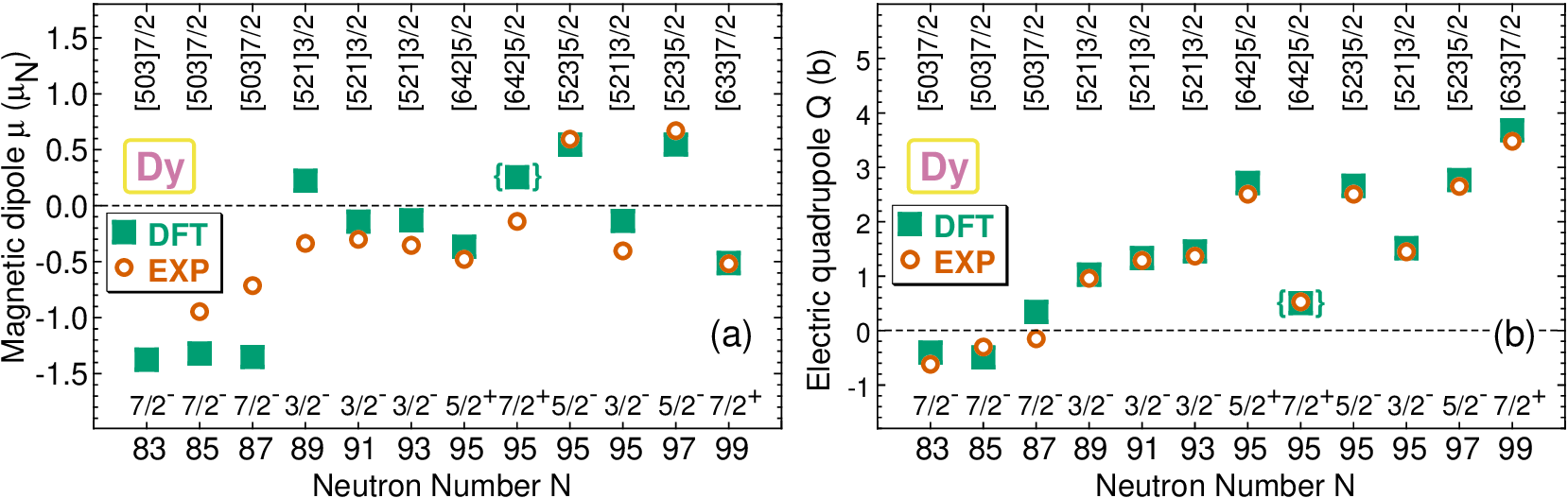}
\caption{Experimental and theoretical magnetic dipole (a) and electric quadrupole (b) moments in dysprosium isotopes ($Z = 66$). Braces denote the values obtained for the higher-spin members of the rotational bands. Figure adapted from Reference~\protect\cite{(Dob26e)}.
\label{fig:dyxxx-exp-the}}
\end{figure}

\begin{figure}[h]
\includegraphics[width=0.98\textwidth]{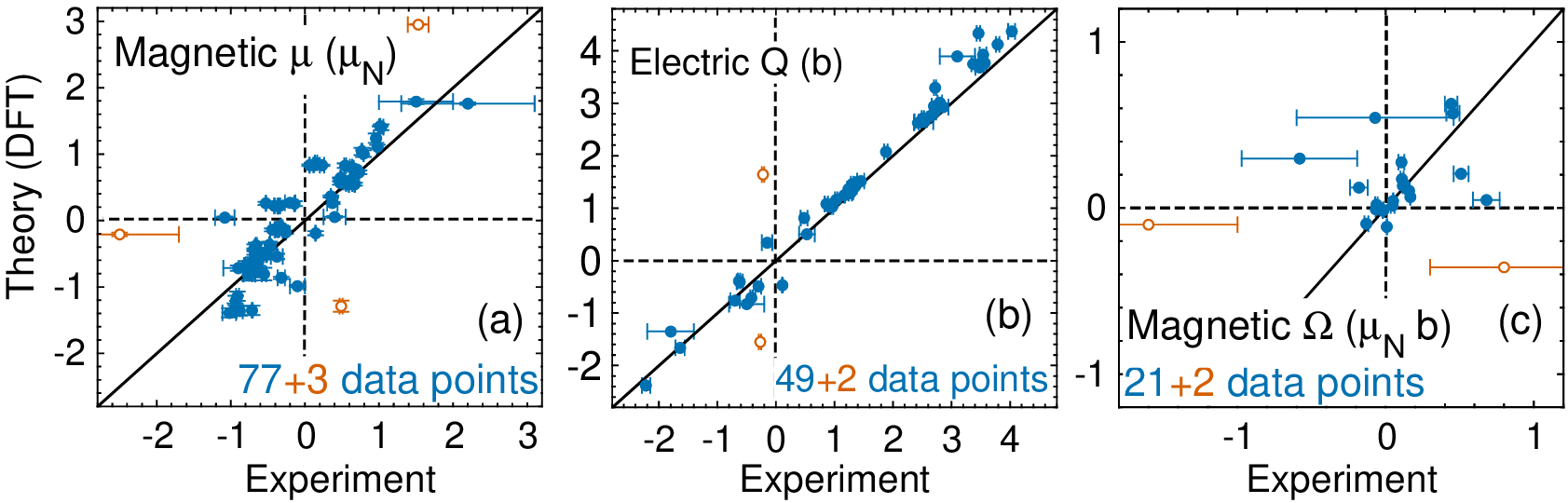}
\hspace*{2mm}\includegraphics[width=0.95\textwidth]{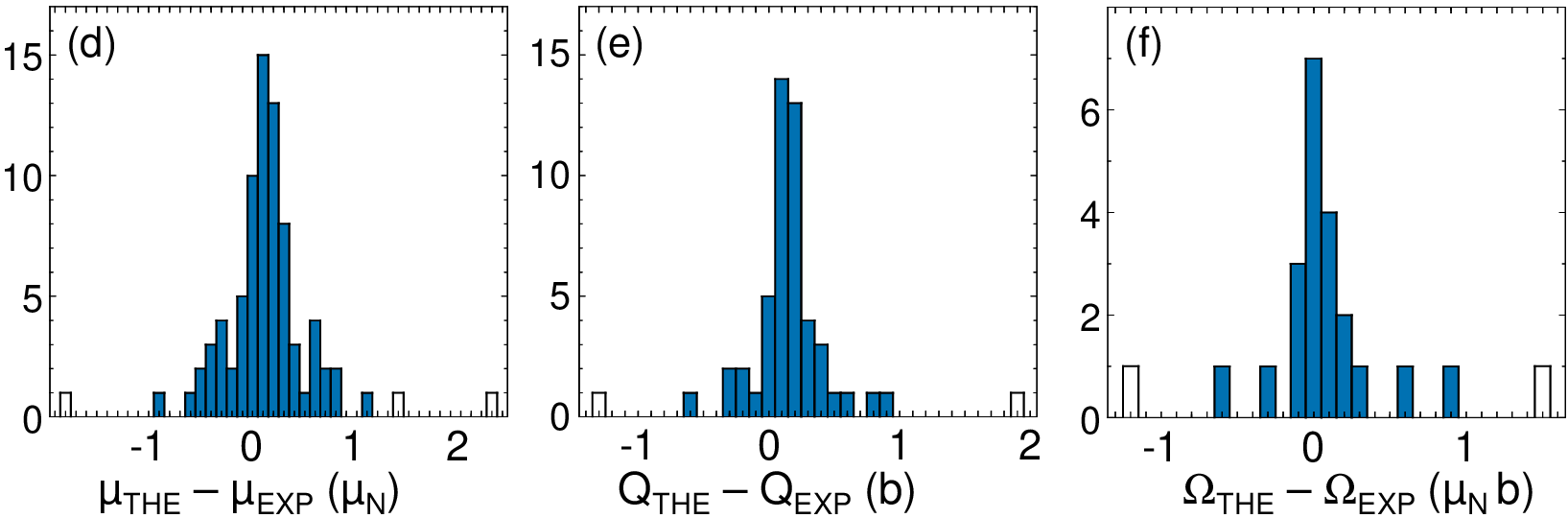}
\caption{Summary comparison of the experimental and theoretical DFT magnetic dipole moments $\mu$, (a) and (d), electric quadrupole moments $Q$, (b) and (e), in odd-$N$, even-$Z$ isotopes between gadolinium and osmium  (panels adapted from Reference~\protect\cite{(Dob26e)}). Panels (c) and (f) show the analogous comparison of the magnetic octupole moments $\Omega$ across the mass chart. Open symbols and bars denote the outliers, which for $\Omega$ are $^{155}$Gd and $^{165}$Ho.\label{fig:yyxxx-ttt-v67-170-HFT-N16-siq-UDF1.318j}}
\end{figure}

In the most recent study (Reference~\cite{(Dob26e)}), the spectroscopic magnetic dipole and electric quadrupole moments were calculated in various quasiparticle configurations of odd-$N$, even-$Z$, $83\leq{}N\leq125$ nuclei ranging from gadolinium to osmium. The tagging mechanism (Section~\ref{Odd-Pairing}) enabled efficient calculations of 22 prolate and 22 oblate states for each of the 154 nuclei and their tracking across the entire major neutron shell. This extensive set of theoretical results was compared with experimental data available for 82 states in the region. A detailed analysis of the pattern of agreement and disagreement between theory and experiment in individual nuclei was conducted.

In {\bf Figure}~\ref{fig:dyxxx-exp-the}, we display the comparison with experimental data of the results obtained for the isotopes of dysprosium. Here, an excellent description of the electric quadrupole moments is complemented by a reasonable description of the magnetic dipole moments, with discrepancies more evident for specific Nilsson configurations. As summarised in {\bf Figure}~\ref{fig:yyxxx-ttt-v67-170-HFT-N16-siq-UDF1.318j}, for all studied elements, the agreement of the magnetic dipole moments with data is characterised by an average and RMS deviations of 0.11\,$\mu_{\text{N}}$ and 0.35\,$\mu_{\text{N}}$, respectively. For the electric quadrupole moments, a good agreement of 0.16\,b and 0.29\,b was observed. The overall correspondence between the magnetic dipole moments and data is therefore better in near doubly magic nuclei than in open-shell systems. Since the values of the electric quadrupole moments in open-shell systems are an order of magnitude larger than in near doubly magic nuclei, an analogous overall comparison of the relative values is thus reasonable.

\subsection{Magnetic octupole moments\label{DFT_Magnetic_octupole}}

DFT studies of magnetic octupole moments constitute a virgin area of research, with only preliminary results available in Reference~\cite{(deG26)}. In {\bf Figures}~\ref{fig:yyxxx-ttt-v67-170-HFT-N16-siq-UDF1.318j}(c) and (f), we present the comparison between calculations and a few available experimental data, see Supplemental Material (Reference~\cite{ARNPS-suppl}, Section~\ref{Magnetic_octupole}). The calculated results were obtained using the exact same methodology as that employed for the other moments discussed in this section, with no parameter adjustments. Given the significant experimental uncertainties or the lack thereof, it is probably premature to draw definitive conclusions from this comparison, and further experimental and theoretical studies are warranted.

\begin{summary}[SUMMARY POINTS]
\begin{enumerate}
\item Electromagnetic moments carry information about the charge and current distributions inside the nucleus, making them crucial observables for testing nuclear structure theories.

\item The advanced nuclear DFT approach provides a unique opportunity to describe experimental electromagnetic properties of nuclei through collective degrees of freedom and core polarisation effects that go well beyond single-particle estimates.

\item Symmetry breaking and symmetry restoration of nuclear states with angular-momentum projections aligned with the axial symmetry axis are crucial elements in describing nuclear magnetic moments.

\item The parity-breaking exotic moments that are relevant for our understanding of fundamental symmetries of nature can be computed within the DFT framework, along with relevant symmetry-breaking two-body interactions.

\end{enumerate}
\end{summary}

\begin{issues}[FUTURE ISSUES]
\begin{enumerate}

\item Using the existing methodology as a baseline, identify missing terms or deficiencies in the interactions or functionals, study triaxial and/or octupole deformability and collectivity, investigate contributions from two-body meson-exchange currents, and implement configuration interaction, $K$-mixing, and/or cranking effects.


\item Prepare for upcoming experimental efforts by performing systematic calculations of parity-breaking exotic moments, which will allow for a better interpretation and understanding of fundamental interactions in nature.
\end{enumerate}
\end{issues}

\section*{DISCLOSURE STATEMENT}
The authors are not aware of any affiliations, memberships, funding, or financial holdings that
might be perceived as affecting the objectivity of this review.

\section*{ACKNOWLEDGMENTS}
We thank Wick Haxton for sharing his expertise on nuclear fundamental interactions.
This work was partially supported by the STFC Grant
Nos.~ST/P003885/1 and~ST/V001035/1, and
by a Leverhulme Trust Research Project Grant. COLFUTURO financially supported ARG.
We thank the CSC-IT Center for Science Ltd., Finland, and the IFT Computer Center at the University of Warsaw, Poland, for the allocation of computational resources.
This project was partly undertaken on the Viking Cluster,
which is a high performance compute facility provided by the
University of York. We are grateful for computational support
from the University of York High Performance Computing
service, Viking and the Research Computing team.
We thank Grammarly\textsuperscript{\textregistered} for its support with English writing.

%

\bibliographystyle{ar-style5}

\clearpage



\section{SUPPLEMENTAL MATERIAL}

In the Supplemental Material, we provide detailed derivations of the results presented in the main text, which are omitted there for the sake of clarity and to avoid unnecessary disruptions in the narrative. We also reiterate standard, well-known derivations, which may help some readers engage more smoothly with the subject matter.

\subsection{Multipole radiation\label{Radiation}}

In the first-order perturbation approximation, the electromagnetic transition probability between the nuclear states $|\Psi_i\rangle$ and $|\Psi_f\rangle$ is
\begin{equation}
    T_{fi} = \frac{2\pi}{\hbar}\left|\int \rmd^3\bm{r}\langle \Psi_f|\mathcal{H}_{\textrm{int}}|\Psi_i\rangle\right|^2 g(E_f),
\end{equation}
where $g(E_f)$ is the density of final states, and the interaction Hamiltonian density $\mathcal{H}_{\textrm{int}}=-j^\nu A_{\nu}$. For the photon with the quantum number $(\alpha,k,\lambda, \mu)$, the transition probability is
\begin{equation}
    T_{fi}(\alpha,k\lambda \mu) = \frac{8\pi(\lambda+1)}{\hbar \lambda((2\lambda+1)!!)^2}\left(\frac{E_{\gamma}}{\hbar c}\right)^{2\lambda+1}
    \big|\langle f|\hat{\mathcal{M}}(\alpha k \lambda \mu)|i\rangle\big|^2
\end{equation}
where the energy of the emitted photon is $E_{\gamma}=E_f - E_i=\hbar c k$, and the general multipole transition operator reads (Reference~\cite{(Rin80)}),
\begin{equation}\label{eq:gen_mult_op}
    \begin{aligned}
        \hat{\mathcal{M}}(E,k \lambda \mu) &= \frac{(2\lambda+1)!!}{k^\lambda(\lambda+1)}\int \rmd^3\bfr \left[\rho Y_{\lambda \mu}(\theta,\phi)\frac{\partial rj_\lambda(kr)}{\partial r} + i\frac{k}{c}\bm{j}\cdot\bm{r} j_\lambda(kr)Y_{\lambda \mu}(\theta,\phi)\right],\\
        \hat{\mathcal{M}}(M,k \lambda \mu) &= \frac{-(2\lambda+1)!!}{ck^\lambda(\lambda+1)}\int \rmd^3\bfr \bm{j}\cdot(\bm{r}\times\nabla)[j_\lambda(kr)Y_{\lambda \mu}(\theta,\phi)].
    \end{aligned}
\end{equation}

In the long-wave limit $kR_0\ll1$ with $R_0$ the nucleus radius, using the approximation $j_\lambda(kr)\approx (kr)^\lambda/(2\lambda+1)!!$ for small $kr$, Equation~\ref{eq:gen_mult_op} turns to the multipole operator
\begin{equation}
    \begin{aligned}
        \hat{Q}_{\lambda \mu} = & \int\rmd^3\bfr \rho \,r^\lambda Y_{\lambda \mu}(\theta,\phi),\\
        \hat{M}_{\lambda \mu} = & \frac{1}{\lambda+1}\int \rmd^3\bfr (\bm{r}\times\bm{j})\cdot\nabla\left[r^\lambda Y_{\lambda \mu}(\theta,\phi)\right].\\
    \end{aligned}
\end{equation}
The total transition probability for a certain $\lambda$, summing over all final states $|I_fM_f\rangle$ and averaging over all initial states $|I_iM_i\rangle$, is
\begin{equation}
    \begin{aligned}
        T_{fi}(\alpha,\lambda)
        =&\frac{1}{2I_i+1}\sum_{M_i,M_f,\mu}T_{fi}(\alpha,k\lambda\mu)
        =\frac{8\pi(\lambda+1)}{\hbar \lambda((2\lambda+1)!!)^2}(\frac{E_{\gamma}}{\hbar c})^{2\lambda+1}B(\alpha\lambda,I_i\rightarrow I_f)\\
    \end{aligned}
\end{equation}
where the reduced transition probability is
\begin{equation}
    \begin{aligned}
        &B(E \lambda,I_i\rightarrow I_f)=\frac{1}{2I_i+1}|\langle f\|\hat{Q}_\lambda\| i\rangle|^2,\\
        &B(M \lambda,I_i\rightarrow I_f)=\frac{1}{2I_i+1}|\langle f\|\hat{M}_\lambda\| i\rangle|^2.\\
    \end{aligned}
\end{equation}

\subsection{One-body magnetic operator\label{One-body_magnetic}}

This subsection is dedicated to presenting the derivation of the magnetic multipole operator in Equation~\ref{eq:magnetic}, originating from the definition of the magnetic multipole moment $M_{\lambda\mu}$ in Equation~\ref{Magnetic}.

Using vector analysis, the magnetic multipole moment, Equation \ref{Magnetic}, can be modified into
\begin{equation}
    {M}_{\lambda\mu}=-\frac{1}{c}\frac{1}{\lambda+1}\int \rmd^3\bm{r}\;{\bm{j}}_{\text{tot}}(\bm{r})\cdot\left[\bm{r}\times\bm{\nabla}\right]\left[r^{\lambda}Y_{\lambda\mu}(\theta,\phi)\right],\label{eq: modified magnetic multipole moment operator}
\end{equation}
where the total current density ${\bm{j}}_{\text{tot}}(\bm{r})$ is the sum of two parts, see Equations~\ref{density1} and~\ref{density2}, that is,
\begin{equation}
\bm{j}_{\text{tot}}(\bm{r})={\bm{j}}_L(\bm{r})+{\bm{j}}_s(\bm{r})=\bm{j}_L(\bm{r})+e\kappa_n\bm{\nabla}\times\bfs_n(\bm{r})+e(1+\kappa_p)\bm{\nabla}\times\bfs_p(\bm{r}).
\end{equation}
and the spin part can be represented as follows,
\begin{equation}
    {\bm{j}}_s(\bm{r})=c\bm{\nabla}\times{\bm{\mu}}_s(\bm{r}),\;\;\;{\bm{\mu}}_s(\bm{r})=\mu_{\text{N}}\left(g^{(n)}_{s}\;{\bm{s}}_{n}+g^{(p)}_{s}\;{\bm{s}}_{p}\right).\label{eq: spin part of the current density operator}
\end{equation}
The insertion of the first part of Equation \ref{eq: spin part of the current density operator} into Equation \ref{eq: modified magnetic multipole moment operator}, after substituting ${\bm{j}}_{\text{tot}}(\bm{r})$ with ${\bm{j}}_s(\bm{r})$, results in the spin part of the magnetic multipole moment,
\begin{align}
    {M}^{\text{s}}_{\lambda\mu}&=-\frac{1}{\lambda+1}\int \rmd^3\bm{r}\;{\bm{\mu}}_s(\bm{r})\cdot\left\{\bm{\nabla}\times\left[\bm{r}\times\bm{\nabla}\right]\left[r^{\lambda}Y_{\lambda\mu}(\theta,\phi)\right]\right\}\nonumber\\
    &=\int \rmd^3\bm{r}\;{\bm{\mu}}_s(\bm{r})\cdot\bm{\nabla}\left[r^{\lambda}Y_{\lambda\mu}(\theta,\phi)\right].
\end{align}
The last line is obtained by utilising the following identity:
\begin{equation}
    \bm{\nabla}\times\left[\bm{r}\times\bm{\nabla}\right]\left[r^{\lambda}Y_{\lambda\mu}(\theta,\phi)\right]=-(\lambda+1)\bm{\nabla}\left[r^{\lambda}Y_{\lambda\mu}(\theta,\phi)\right].
\end{equation}
This identity can be derived in two stages. Firstly, the identity involving a single summation over repeated index $k$ of the product of two Levi-Cività symbols, i.e., $\epsilon_{kij}\epsilon_{kmn}=\delta_{im}\delta_{jn}-\delta_{in}\delta_{jm}$, can be employed to obtain
\begin{align}
    \left\{\bm{\nabla}\times\left[\bm{r}\times\bm{\nabla}\right]\right\}_{i}Q_{\lambda\mu}(\bm{r})&=\epsilon_{ijk}\nabla_{j}\left[\bm{r}\times\bm{\nabla}\right]_{k}Q_{\lambda\mu}(\bm{r})=\epsilon_{kij}\epsilon_{kmn}\nabla_{j}\left[r_{m}\nabla_{n}Q_{\lambda\mu}(\bm{r})\right]\nonumber\\
    &=(\delta_{im}\delta_{jn}-\delta_{in}\delta_{jm})\nabla_{j}\left[r_{m}\nabla_{n}Q_{\lambda\mu}(\bm{r})\right]\nonumber\\
    &=\nabla_{n}\left[r_{i}\nabla_{n}Q_{\lambda\mu}(\bm{r})\right]-\nabla_{m}\left[r_{m}\nabla_{i}Q_{\lambda\mu}(\bm{r})\right]\nonumber\\
    &=\delta_{ni}\nabla_{n}Q_{\lambda\mu}(\bm{r})+r_{i}\nabla^{2}_{n}Q_{\lambda\mu}(\bm{r})-\delta_{mm}\nabla_{i}Q_{\lambda\mu}(\bm{r})-r_{m}\nabla_{m}\nabla_{i}Q_{\lambda\mu}(\bm{r})\nonumber\\
    &=\nabla_{i}Q_{\lambda\mu}(\bm{r})+r_{i}\nabla^{2}Q_{\lambda\mu}(\bm{r})-3\nabla_{i}Q_{\lambda\mu}(\bm{r})-(\bm{r}\cdot\bm{\nabla})\nabla_{i}Q_{\lambda\mu}(\bm{r})\nonumber\\
    &=r_{i}\nabla^{2}Q_{\lambda\mu}(\bm{r})-2\nabla_{i}Q_{\lambda\mu}(\bm{r})-(\bm{r}\cdot\bm{\nabla})\nabla_{i}Q_{\lambda\mu}(\bm{r}).
\end{align}
The last term can be manipulated further as follows:
\begin{align}
    \nabla_{i}\left[r_{\ell}\nabla_{\ell}Q_{\lambda\mu}(\bm{r})\right]&=\delta_{i\ell}\nabla_{\ell}Q_{\lambda\mu}(\bm{r})+r_{\ell}\nabla_{i}\nabla_{\ell}Q_{\lambda\mu}(\bm{r})=\nabla_{i}Q_{\lambda\mu}(\bm{r})+(\bm{r}\cdot\bm{\nabla})\nabla_{i}Q_{\lambda\mu}(\bm{r}),
\end{align}
which allows us to write
\begin{align}
    \left\{\bm{\nabla}\times\left[\bm{r}\times\bm{\nabla}\right]\right\}_{i}Q_{\lambda\mu}(\bm{r})&=r_{i}\nabla^{2}Q_{\lambda\mu}(\bm{r})-2\nabla_{i}Q_{\lambda\mu}(\bm{r})-\nabla_{i}\left[(\bm{r}\cdot\bm{\nabla})Q_{\lambda\mu}(\bm{r})\right]+\nabla_{i}Q_{\lambda\mu}(\bm{r})\nonumber\\
    &=r_{i}\nabla^{2}Q_{\lambda\mu}(\bm{r})-\nabla_{i}(1+\bm{r}\cdot\bm{\nabla})Q_{\lambda\mu}(\bm{r}),\nonumber
\end{align}
or
\begin{equation}
    \bm{\nabla}\times\left[\bm{r}\times\bm{\nabla}\right]\left[r^{\lambda}Y_{\lambda\mu}(\theta,\phi)\right]=\left\{\bm{r}\nabla^{2}-\bm{\nabla}\left(1+\bm{r}\cdot\bm{\nabla}\right)\right\}\left[r^{\lambda}Y_{\lambda\mu}(\theta,\phi)\right].
\end{equation}
Secondly, the last equation can be worked out in the spherical coordinate to give
\begin{align}
    \bm{\nabla}\times\left[\bm{r}\times\bm{\nabla}\right]\left[r^{\lambda}Y_{\lambda\mu}(\theta,\phi)\right]&=\bm{r}\left\{\frac{1}{r^{2}}\frac{\partial}{\partial r}\left(r^{2}\frac{\partial}{\partial r}\left[r^{\lambda}Y_{\lambda\mu}(\theta,\phi)\right]\right)-\frac{\hat{L}^{2}}{r^{2}}r^{\lambda}Y_{\lambda\mu}(\theta,\phi)\right\}\nonumber\\
    &-\bm{\nabla}\left\{1+r\frac{\partial}{\partial r}\right\}r^{\lambda}Y_{\lambda\mu}(\theta,\phi)\nonumber\\
    &=\bm{r}\left\{\frac{1}{r^{2}}\frac{\partial}{\partial r}\left[\lambda r^{\lambda+1}\right]-\frac{\lambda(\lambda+1)}{r^{2}}r^{\lambda}\right\}Y_{\lambda\mu}(\theta,\phi)\nonumber\\
    &-(\lambda+1)\bm{\nabla}r^{\lambda}Y_{\lambda\mu}(\theta,\phi)\nonumber\\
    &=\bm{r}\left\{\frac{\lambda(\lambda+1)}{r^{2}}-\frac{\lambda(\lambda+1)}{r^{2}}\right\}r^{\lambda}Y_{\lambda\mu}(\theta,\phi)\nonumber\\
    &-(\lambda+1)\bm{\nabla}r^{\lambda}Y_{\lambda\mu}(\theta,\phi)\nonumber\\
    &=-(\lambda+1)\bm{\nabla}r^{\lambda}Y_{\lambda\mu}(\theta,\phi),
\end{align}
where we have used the eigenvalue equation: $\hat{L}^{2}Y_{\lambda\mu}(\theta,\phi)=\lambda(\lambda+1)Y_{\lambda\mu}(\theta,\phi)$.
The insertion of the second part of Equation \ref{eq: spin part of the current density operator} into ${M}^{\text{s}}_{\lambda\mu}$ gives us the spin part of Equation \ref{eq:magnetic}.

The orbital part of Equation \ref{eq:magnetic} can be directly obtained by substituting ${\bm{j}}_{\text{tot}}(\bm{r})$ with ${\bm{j}}_L(\bm{r})$ in Equation \ref{eq: modified magnetic multipole moment operator}, that is,
\begin{equation}
\begin{aligned}
    {M}^L_{\lambda\mu}&=-\frac{1}{c}\frac{1}{\lambda+1}\int \rmd^3\bm{r}\;{\bm{j}}_L(\bm{r})\cdot\left[\bm{r}\times\bm{\nabla}\right]\left[r^{\lambda}Y_{\lambda\mu}(\theta,\phi)\right] \\
    &=-\frac{1}{c}\frac{1}{\lambda+1}\int \rmd^3\bm{r}\;e\frac{\bm{p}}{m}
    \cdot\left[\bm{r}\times\bm{\nabla}\right]\left[r^{\lambda}Y_{\lambda\mu}(\theta,\phi)\right] \\
    &=-\frac{1}{c}\frac{1}{\lambda+1}\int \rmd^3\bm{r}\;\frac{e}{m} \bm{\nabla}\left[r^{\lambda}Y_{\lambda\mu}(\theta,\phi)\right]
    \cdot(\bm{p}\times\bm{r}) \\
    &=\frac{1}{c}\frac{1}{\lambda+1}\int \rmd^3\bm{r}\;\frac{e}{m} \bm{\nabla}\left[r^{\lambda}Y_{\lambda\mu}(\theta,\phi)\right]
    \cdot\bm{L} \\
    &=\frac{2\mu_{\text{N}}}{\lambda+1}\int \rmd^3\bm{r}\;\bm{\nabla}\left[r^{\lambda}Y_{\lambda\mu}(\theta,\phi)\right]
    \cdot\bm{L}~,
\end{aligned}
\end{equation}
where the nuclear magneton, $\mu_{\text{N}}=\frac{e\hbar}{2mc}$, and the momentum density is defined as
\begin{equation}
    \bm{p}(\bm{r}) = \langle \Psi| \frac{1}{2}\left[\hat{\bm{p}}_i \delta(\bm{r} - \bm{r}_i) + \delta(\bm{r} - \bm{r}_i) \hat{\bm{p}}_i\right] |\Psi\rangle.
\end{equation}
The magnetic multipole moment, therefore, is
\begin{equation}
\begin{aligned}
    {M}_{\lambda\mu} &= {M}^L_{\lambda\mu} + {M}^s_{\lambda\mu}
    = \int \rmd^3\bm{r}\;\bm{\nabla}\left[r^{\lambda}Y_{\lambda\mu}(\theta,\phi)\right]
    \cdot\left( \tfrac{2}{\lambda+1}g_l^{(p)}\bm{L}_p + g_s^{(n)} \bm{S}_n + g_s^{(p)} \bm{S}_p \right)\mu_{\text{N}},
\end{aligned}
\end{equation}
which is the expectation value of Equation \ref{eq:magnetic}.

\subsection{Gradient of the solid harmonics\label{Gradient}}
Using the concept of the vector spherical harmonics (Reference~\cite{(Var88)}, Equations~5.8.3(13) and~7.3.2(12)), the gradient of the solid harmonics can be expressed as follows,
\begin{equation}
    \begin{aligned}
    \nabla_{\nu}[r^{\lambda}Y_{\lambda\mu}(\theta,\phi)]
    &=\sqrt{\lambda(2\lambda+1)} r^{\lambda -1} \left[Y_{\lambda\mu}^{\lambda -1}(\theta,\phi)\right]^{\nu} \\
    &=\sqrt{\lambda(2\lambda+1)} r^{\lambda -1} C_{\lambda-1 \mu-\nu, 1\nu}^{\lambda\mu}Y_{\lambda-1,\mu-\nu}(\theta,\phi),
    \end{aligned}
    \label{eq:grad_Y}
\end{equation}
where $\nu = -1, 0, 1$. For $\nu=0$ and $\mu=0$, and with the Clebsch-Gordan coefficient,
\begin{equation}
C_{\lambda-1\,0,10}^{\lambda 0} = \sqrt{\frac{\lambda}{2\lambda -1}},
\end{equation}
the $z$-derivative of the axial solid harmonics is given by,
\begin{equation}
\begin{aligned}
	\nabla_0[r^\lambda Y_{\lambda 0}(\theta,\phi)]
	&=\sqrt{\lambda(2\lambda+1)}C_{\lambda-1\,0,10}^{\lambda 0} r^{\lambda -1}Y_{\lambda-1,0}(\theta,\phi)\\
	&=\sqrt{\lambda(2\lambda+1)}\sqrt{\frac{\lambda}{2\lambda-1}} r^{\lambda -1} Y_{\lambda-1,0}(\theta,\phi)
    =\lambda\sqrt{\frac{2\lambda+1}{2\lambda-1}}r^{\lambda -1}Y_{\lambda-1,0}(\theta,\phi).
    \label{eq:grad_z}
\end{aligned}
\end{equation}

In general, for a rank-1 tensor $\bm{\mu}$, Equation~\ref{eq:grad_Y} suggests
\begin{equation}\label{eq:m-rank-lambda}
\begin{aligned}
    \bm{\mu}\cdot{\bm{\nabla}}[r^{\lambda}Y_{\lambda\mu}] & = \sum_{\nu}{\mu}_{1,\nu}\nabla_{\nu}[r^{\lambda}Y_{\lambda\mu}]\\
    &= \sum_{\nu}\sqrt{\lambda(2\lambda+1)} r^{\lambda -1} C_{\lambda-1 \mu-\nu, 1\nu}^{\lambda\mu}Y_{\lambda-1,\mu-\nu}{\mu}_{1,\nu} \\
    &=\sqrt{\lambda(2\lambda+1)}  \left[r^{\lambda -1}Y_{\lambda-1}\otimes {\mu}_{1}\right]_{\lambda\mu}.
\end{aligned}
\end{equation}
Therefore, the magnetic multipole moment in unit of $\mu_{\text{N}}$ can be expressed as
\begin{equation}
\begin{aligned}
    M_{\lambda\mu} &=\int\rmd^3\bm{r}\left(g_s\bm{S}+\tfrac{2}{\lambda+1}g_l\bm{L}\right)\cdot\bm{\nabla}[r^{\lambda}Y_{\lambda\mu}] \\
    &= \sqrt{\lambda(2\lambda+1)}\int\rmd^3\bm{r}\left[Q_{\lambda-1}\otimes \left(g_s{S}_1+\tfrac{2}{\lambda+1}g_l{L}_1\right)\right]_{\lambda\mu},
\end{aligned}
\end{equation}
and the magnetic multipole operator is as in Equation~\ref{eq:magnetic2}.

\subsection{Schmidt moments\label{Schmidt}} The expectation value of Equation~\ref{eq:mm_definition} with respect to the single-particle state with orbital ($l$) and spin ($\tfrac{1}{2}$) angular momenta coupled to $j,m=j$, $|nl\tfrac{1}{2};jj\rangle$, can be calculated using $\hat{\bm{J}}=\hat{\bm{L}}+\hat{\bm{S}}$ as follows
\begin{equation}
\begin{aligned}
\mu^{(\tau)}_{s.p.}&=\mu_{\text{N}}\left\langle nl\tfrac{1}{2};jj\left|\hspace{1mm}g^{(\tau)}_{l}\hat{L}_z+g^{(\tau)}_{s}\hat{S}_z \hspace{1mm}\right|nl\tfrac{1}{2};jj\right\rangle\\
    &=\mu_{\text{N}}\left\langle nl\tfrac{1}{2};jj\left|\hspace{1mm}g^{(\tau)}_{l}\hat{J}_z+\left(g^{(\tau)}_{s}-g^{(\tau)}_{l}\right)\hat{S}_z \hspace{1mm}\right|nl\tfrac{1}{2};jj\right\rangle\\
    &=\mu_{\text{N}}\left[g^{(\tau)}_{l}j+ \left(g^{(\tau)}_{s}-g^{(\tau)}_{l}\right)
    \left\langle nl\tfrac{1}{2};jj\left|\hspace{1mm}\hat{S}_z \hspace{1mm}\right|nl\tfrac{1}{2};jj\right\rangle\right].
    \label{eq:dm_step1}
\end{aligned}
\end{equation}
Since the basis chosen is in coupled total angular momentum, we need to use Wigner-Eckart and reduction theorems for the matrix element of $\hat{\bm{S}}=\tfrac{1}{2}\bm{\hat{\sigma}}$ as follows
\begin{equation}
\begin{aligned}
\mu^{(\tau)}_{s.p.}&=\mu_{\text{N}}\left[g^{(\tau)}_{l}j+ \left(g^{(\tau)}_{s}-g^{(\tau)}_{l}\right)\frac{C_{jj10}^{jj}}{2\sqrt{2j+1}}
    \left\langle nl\tfrac{1}{2};j\|\hspace{1mm}\hat{\bm{\sigma}} \hspace{1mm}\|nl\tfrac{1}{2};j\right\rangle\right]\\
    &=\mu_{\text{N}}\left[g^{(\tau)}_{l}j+ \left(g^{(\tau)}_{s}-g^{(\tau)}_{l}\right)\sqrt{\frac{j}{4(2j+1)(j+1)}}
    \left\langle nl\tfrac{1}{2};j\|\hspace{1mm}\hat{\bm{\sigma}} \hspace{1mm}\|nl\tfrac{1}{2};j\right\rangle\right],
    \label{eq:dm_step2}
\end{aligned}
\end{equation}
where the reduced matrix element involves a 6-$j$ symbol as follows,
\begin{equation}
\begin{aligned}
    \left\langle nl\tfrac{1}{2};j\|\hspace{1mm}\hat{\bm{\sigma}} \hspace{1mm}\|nl\tfrac{1}{2};j\right\rangle&=\sqrt{6}(2j+1)(-1)^{l+j+\frac{3}{2}}\begin{Bmatrix}
\tfrac{1}{2} & \tfrac{1}{2} & 1\\
j & j & l
\end{Bmatrix}\\&=\sqrt{\frac{2j+1}{j(j+1)}}\left[j(j+1)-l(l+1)+\tfrac{3}{4}\right].
    \label{eq:dm_step2a}
\end{aligned}
\end{equation}
Inserting this result into Equation~\ref{eq:dm_step2}, we obtain the final expressions as follows,
\begin{equation}
\begin{aligned}
    \mu^{(\tau)}_{s.p.}&=\mu_{\text{N}}\left[g^{(\tau)}_{l}j+ \left(g^{(\tau)}_{s}-g^{(\tau)}_{l}\right)\frac{j(j+1)-l(l+1)+\tfrac{3}{4}}{2(j+1)}
    \right]\\&= \begin{cases}
       \left[\Big(j-\tfrac{1}{2}\Big)g^{(\tau)}_l+\tfrac{1}{2}g_s^{(\tau)}\right]\mu_{\text{N}}, \hspace{17mm}j = l+\tfrac{1}{2},\\
       \frac{j}{j+1}\left[ \Big(j+\frac{3}{2}\Big)g^{(\tau)}_l-\tfrac{1}{2}g_s^{(\tau)} \right]\mu_{\text{N}}, \hspace{11mm}j = l-\tfrac{1}{2}.
     \end{cases}
    \label{eq:dm_step2b}
\end{aligned}
\end{equation}

\subsection{Single particle quadrupole moments\label{Single_particle_quadrupole}}
The expectation value we need to calculate can be divided into the radial and angular components as follows,
\begin{equation}
\begin{aligned}
    Q_{\text{s.p.}} &= \sqrt{\frac{16\pi}{5}}e\left\langle nl\tfrac{1}{2};jj\left|\hspace{1mm}r^2Y_{20}(\theta,\phi) \hspace{1mm}\right|nl\tfrac{1}{2};jj\right\rangle\\&=\sqrt{\frac{16\pi}{5}}e\langle nl|r^2|nl\rangle\left\langle l\tfrac{1}{2};jj\left|\hspace{1mm}Y_{20}(\theta,\phi) \hspace{1mm}\right|l\tfrac{1}{2};jj\right\rangle\\
    &=\sqrt{\frac{16\pi}{5}}e\langle r^2\rangle\frac{C_{jj20}^{jj}}{\sqrt{2j+1}}
    \left\langle l\tfrac{1}{2};j\left|\left|\hspace{1mm}Y_{2}(\theta,\phi) \hspace{1mm}\right|\right|l\tfrac{1}{2};j\right\rangle\\&=
    \sqrt{\frac{16\pi}{5}}e\langle r^2\rangle\sqrt{\frac{j(2j-1)}{(2j+3)(2j+1)(j+1)}}
    \left\langle l\tfrac{1}{2};j\left|\left|\hspace{1mm}Y_{2}(\theta,\phi) \hspace{1mm}\right|\right|l\tfrac{1}{2};j\right\rangle.
    \label{eq:qm_step1}
\end{aligned}
\end{equation}
Using the Wigner-Eckart and reduction theorems for the matrix element of $Y_2(\theta,\phi)$, we obtain
\begin{equation}
\begin{aligned}
    \left\langle l\tfrac{1}{2};j\left|\left|\hspace{1mm}Y_{2}(\theta,\phi) \hspace{1mm}\right|\right|l\tfrac{1}{2};j\right\rangle&=\sqrt{\frac{5}{4\pi}}(-1)^{j-\tfrac{1}{2}}(2j+1)\begin{pmatrix}
j & j & 2\\
\tfrac{1}{2} & -\tfrac{1}{2} & 0
\end{pmatrix}\\&=-\frac{1}{4}\sqrt{\frac{5}{4\pi}}\sqrt{\frac{(2j+3)(2j+1)(2j-1)}{j(j+1)}}.
    \label{eq:qm_step2}
\end{aligned}
\end{equation}
Inserting Equation~\ref{eq:qm_step2} into Equation~\ref{eq:qm_step1}, we obtain the final result of Equation~\ref{eq:q_sp}.

\subsection{Schwartz moments\label{Schwartz}}
Following a similar approach as for the magnetic dipole moment, we can separate Equation~\ref{eq:octupole_moment} into two terms corresponding to the orbital and spin angular momenta,
\begin{equation}
\begin{aligned}
\Omega^{(\tau)}_{s.p.}&=-\sqrt{\frac{4\pi}{7}}\mu_{\text{N}}\Bigg[\frac{g^{(\tau)}_{l}}{2}\left\langle nl\tfrac{1}{2};jj\left|\hspace{1mm}\nabla\Big(r^3Y_{30}(\theta,\phi)\Big)\cdot\hat{\bm{J}}\hspace{1mm}\right|nl\tfrac{1}{2};jj\right\rangle
\\&\hspace{18mm}+
\bigg(g^{(\tau)}_{s}-\frac{g^{(\tau)}_{l}}{2}\bigg)\left\langle nl\tfrac{1}{2};jj\left|\hspace{1mm}\nabla\Big(r^3Y_{30}(\theta,\phi)\Big)\cdot\hat{\bm{S}}\hspace{1mm}\right|nl\tfrac{1}{2};jj\right\rangle\Bigg].\label{eq:om_step1_1}
\end{aligned}
\end{equation}
Using Equation~\ref{eq:grad_Y} and the Wigner-Eckart and reduction theorems, for the spin term we obtain,
\begin{equation}
\begin{aligned}
&\bigg(g^{(\tau)}_{s}-\frac{g^{(\tau)}_{l}}{2}\bigg)\left\langle nl\tfrac{1}{2};jj\left|\hspace{1mm}\nabla\Big(r^3Y_{30}(\theta,\phi)\Big)\cdot\hat{\bm{S}}\hspace{1mm}\right|nl\tfrac{1}{2};jj\right\rangle\\&=\sqrt{7\cdot3}
\bigg(g^{(\tau)}_{s}-\frac{g^{(\tau)}_{l}}{2}\bigg)\langle r^2\rangle
\left\langle l\tfrac{1}{2};jj\left|\hspace{1mm}\Big[Y_{2}(\theta,\phi)\otimes\hat{\bm{S}}\Big]^{3}_0\hspace{1mm}\right|l\tfrac{1}{2};jj\right\rangle\\
&=\sqrt{7\cdot3}
\bigg(g^{(\tau)}_{s}-\frac{g^{(\tau)}_{l}}{2}\bigg)\langle r^2\rangle\sqrt{7}\sqrt{2j+1}C_{jj30}^{jj}
\begin{Bmatrix}
l & \tfrac{1}{2} & j\\
l & \tfrac{1}{2} & j\\
2 & 1 & 3
\end{Bmatrix}\left\langle l\left|\left|\hspace{1mm}Y_{2}(\theta,\phi)\hspace{1mm}\right|\right|l\right\rangle
\left\langle\tfrac{1}{2}\left|\left|\hspace{1mm}\hat{\bm{S}}\hspace{1mm}\right|\right|\tfrac{1}{2}\right\rangle,
\label{eq:om_step2}
\end{aligned}
\end{equation}
where the reduced matrix elements, 9-$j$ coefficient (taken from Table~10.3 of Reference~\cite{(Var88)}) and Clebsch-Gordan coefficient are given by
\begin{equation}
\begin{aligned}
\left\langle\tfrac{1}{2}\left|\left|\hspace{1mm}\hat{\bm{S}}\hspace{1mm}\right|\right|\tfrac{1}{2}\right\rangle&=\sqrt{\frac{3}{2}},\\
\left\langle l\left|\left|\hspace{0.5mm}Y_{2}(\theta,\phi)\hspace{0.5mm}\right|\right|l\right\rangle&=-\sqrt{\frac{5}{4\pi}}\sqrt{\frac{l(l+1)(2l+1)}{(2l-1)(2l+3)}}\\&=\begin{cases}
    -\sqrt{\frac{5}{4\pi}}\sqrt{\frac{j(2j+1)(2j-1)}{8(j+1)(j-1)}}, \hspace{10mm}j=l+\tfrac{1}{2}\\[2pt]
    -\sqrt{\frac{5}{4\pi}}\sqrt{\frac{(2j+1)(2j+3)(j+1)}{8j(j+2)}}, \hspace{4mm}j=l-\tfrac{1}{2}\
\end{cases}\\
\begin{Bmatrix}
l & \tfrac{1}{2} & j\\
l & \tfrac{1}{2} & j\\
2 & 1 & 3
\end{Bmatrix}&=\begin{cases}
    \frac{\sqrt{(2j+3)(j+2)}}{2j(2j+1)\sqrt{7\cdot 5}}, \hspace{21mm}j=l+\tfrac{1}{2}\\[2pt]
    -\frac{\sqrt{(j-1)(2j-1)}}{2(2j+1)(j+1)\sqrt{7\cdot 5}}, \hspace{13mm}j=l-\tfrac{1}{2}\
\end{cases}\\
C_{jj30}^{jj}&=\sqrt{\frac{j(j-1)(2j-1)}{(j+1)(j+2)(2j+3)}}.
\label{eq:om_step3}
\end{aligned}
\end{equation}
This gives the spin term as follows,
\begin{equation}
\begin{aligned}
&\bigg(g^{(\tau)}_{s}-\frac{g^{(\tau)}_{l}}{2}\bigg)\left\langle nl\tfrac{1}{2};jj\left|\hspace{1mm}\nabla\Big(r^3Y_{30}(\theta,\phi)\Big)\cdot\hat{\bm{S}}\hspace{1mm}\right|nl\tfrac{1}{2};jj\right\rangle\\
&=\frac{3}{8}\sqrt{\frac{7}{4\pi}}
\bigg(g^{(\tau)}_{s}-\frac{g^{(\tau)}_{l}}{2}\bigg)\langle r^2\rangle\times\begin{cases}
    -\frac{2j-1}{j+1}, \hspace{19mm}j=l+\tfrac{1}{2}\\[2pt]
    \frac{(j-1)(2j-1)}{(j+1)(j+2)}, \hspace{12mm}j=l-\tfrac{1}{2}
\end{cases},
\label{eq:om_om_final}
\end{aligned}
\end{equation}
For the orbital term, we use Equation~\ref{eq:grad_Y} to obtain
\begin{equation}
\begin{aligned}
&\frac{g^{(\tau)}_{l}}{2}\left\langle nl\tfrac{1}{2};jj\left|\hspace{1mm}\nabla\Big(r^3Y_{30}(\theta,\phi)\Big)\cdot\hat{\bm{J}}\hspace{1mm}\right|nl\tfrac{1}{2};jj\right\rangle\\&=\frac{g^{(\tau)}_{l}}{2}  \sum_{\nu=-1}^1\sqrt{21}C_{2-\nu1\nu}^{30}\langle r^2\rangle\left\langle nl\tfrac{1}{2};jj\left|
Y_{2-\nu}(\theta,\phi)
\hat{J}_{\nu}\hspace{1mm}\right|nl\tfrac{1}{2};jj\right\rangle\\&=\frac{g^{(\tau)}_{l}}{2}\langle r^2\rangle\Bigg[\left\langle nl\tfrac{1}{2};jj\left|3\sqrt{\tfrac{7}{5}}Y_{20}(\theta,\phi)
\hat{J}_{0}\hspace{1mm}\right|nl\tfrac{1}{2};jj\right\rangle\\&\hspace{20mm}-\left\langle nl\tfrac{1}{2};jj\bigg|\sqrt{\tfrac{21}{5}}Y_{21}(\theta,\phi)
\hat{J}_{-1}\hspace{1mm}\bigg|nl\tfrac{1}{2};jj\right\rangle\Bigg],
\label{eq:om_step5}
\end{aligned}
\end{equation}
where components of $\hat{J}_{\nu}$ act as follows,
\begin{equation}
\begin{aligned}
\hat{J}_{+1}\hspace{1mm}|nl\tfrac{1}{2};jj\rangle=0,\hspace{6mm}
\hat{J}_{0}\hspace{1mm}\left|nl\tfrac{1}{2};jj\right\rangle=j\hspace{1mm}\left|nl\tfrac{1}{2};jj\right\rangle,\hspace{6mm}
\hat{J}_{-1}\hspace{1mm}\left|nl\tfrac{1}{2};jj\right\rangle=\sqrt{j}\hspace{1mm}\left|nl\tfrac{1}{2};jj-1\right\rangle.
\label{eq:om_step6}
\end{aligned}
\end{equation}
This, together with the Wigner-Eckart theorem and Equation~\ref{eq:qm_step2} gives
\begin{equation}
\begin{aligned}
&=\frac{\sqrt{21}g^{(\tau)}_{l}}{2}\Bigg[\sqrt{\frac{3}{5}}\frac{jC_{jj20}^{jj}}{\sqrt{2j+1}}-\sqrt{\frac{1}{5}}\frac{\sqrt{j}C_{jj-1,21}^{jj}}{\sqrt{2j+1}}
\Bigg]\langle r^2\rangle\left\langle l\tfrac{1}{2};j\left|\left|
 Y_{2}(\theta,\phi)
\hspace{1mm}\right|\right|l\tfrac{1}{2};j\right\rangle
\\&=-\frac{\sqrt{21}g^{(\tau)}_{l}}{8}\langle r^2\rangle\sqrt{\frac{3}{5}}(j-1)\sqrt{\frac{j(2j-1)}{(j+1)(2j+1)(2j+3)}}\sqrt{\frac{5}{4\pi}}\sqrt{\frac{(2j+3)(2j+1)(2j-1)}{j(j+1)}}\\&
=-\frac{3}{8}g^{(\tau)}_l\langle r^2\rangle\sqrt{\frac{7}{4\pi}}\frac{(j-1)(2j-1)}{j+1}.
\label{eq:om_step7}
\end{aligned}
\end{equation}
Inserting Equations~\ref{eq:om_om_final} and~\ref{eq:om_step7} into Equation~\ref{eq:om_step1_1}, we obtain the final result as follows,
\begin{equation}
\begin{aligned}
\Omega^{(\tau)}_{s.p.}&=\frac{3}{8}\mu_{\text{N}}\langle r^2\rangle\Bigg[\hspace{1mm}g^{(\tau)}_l\frac{(j-1)(2j-1)}{j+1}-
\bigg(g^{(\tau)}_{s}-\frac{g^{(\tau)}_{l}}{2}\bigg)\times\begin{cases}
    -\frac{2j-1}{j+1}, \hspace{10mm}j=l+\tfrac{1}{2}\\[2pt]
    \frac{(j-1)(2j-1)}{(j+1)(j+2)}, \hspace{4mm}j=l-\tfrac{1}{2}
    \end{cases}\Bigg]
    \\&=\frac{3}{8}\mu_{\text{N}}\langle r^2\rangle\times\begin{cases}
    \frac{2j-1}{j+1}g^{(\tau)}_{s}+\frac{(2j-1)(2j-3)}{2(j+1)}g^{(\tau)}_{l}, \hspace{30mm}j=l+\tfrac{1}{2}\\[2pt]
    -\frac{(j-1)(2j-1)}{(j+1)(j+2)}g^{(\tau)}_{s}+\frac{(j-1)(2j-1)(2j+5)}{2(j+1)(j+2)}g^{(\tau)}_{l}, \hspace{11mm}j=l-\tfrac{1}{2}
    \end{cases}
    \\&=\frac{3}{2}\mu_{\text{N}}\langle r^2\rangle\frac{2j-1}{(2j+4)(2j+2)}\times\begin{cases}
    (j+2)\Big[(j-\frac{3}{2})g^{(\tau)}_{l}+g^{(\tau)}_{s}\Big], \hspace{10mm}j=l+\tfrac{1}{2}\\[4pt]
    (j-1)\Big[(j+\frac{5}{2})g^{(\tau)}_{l}-g^{(\tau)}_{s}\Big], \hspace{10mm}j=l-\tfrac{1}{2}.
    \end{cases}
\label{eq:om_step8}
\end{aligned}
\end{equation}

\subsection{Bohr deformation parameters\label{Bohr_deformation}}

Assuming the density is uniformly distributed inside the nucleus with a sharp surface $R(\theta,\phi)$ as defined in Equation~\ref{Bohr}, the density is simply $\rho_0=N/V$ with $N$ the number of particles and $V=4\pi R_0^3/3$ the volume of the nucleus, where $R_0$ is the radius of a sphere that has the same volume as the nucleus. The multipole moments defined from this uniform density distribution are as follows,
\begin{equation}
    \begin{aligned}
    \label{Moment}
        Q_{\lambda\mu} &= \int_V\rmd^3\bfr \frac{3N}{4\pi R_0^3}r^{\lambda} Y_{\lambda\mu}^*(\theta,\phi)
        =\frac{3N}{4\pi R_0^3}\int_V r^{\lambda+2} Y_{\lambda\mu}^*(\theta,\phi) \,\rmd\Omega\,\rmd r \\
        &=\frac{3N}{4\pi R_0^3}\int \rmd\Omega\, Y_{\lambda\mu}^*(\theta,\phi)\int_0^{R(\theta,\phi)} r^{\lambda+2} \rmd r
        =\frac{3N}{4\pi R_0^3}\int \rmd\Omega\, Y_{\lambda\mu}^*(\theta,\phi)\frac{1}{\lambda+3} \left[R(\theta,\phi)\right]^{\lambda+3},
    \end{aligned}
\end{equation}
where $\rmd\Omega=\sin\theta\,\rmd\theta\,\rmd\phi$ is the volume element for integration over the spherical angles.
For small $\beta_{\lambda\mu}$, keeping the first order of Taylor's expansion, we obtain,
\begin{equation}
\begin{aligned}
    \left[R(\theta,\phi)\right]^{\lambda+3}
		&\approx R_0^{\lambda+3} \left[1+(\lambda+3) \sum_{\lambda=1}^{\lambda_{\text{max}}}\sum_{\mu=-\lambda}^{\lambda} \beta_{\lambda\mu} Y_{\lambda\mu}(\theta,\phi)\right].
\end{aligned}
\end{equation}
This allows us to perform the integral over the spherical angles in Equation~\ref{Moment},
\begin{equation}
\begin{aligned}
    &\int \rmd \Omega  \left[1 + (\lambda+3) \sum_{\lambda'=1}^{\lambda'_{\text{max}}}\sum_{\mu'=-\lambda'}^{\lambda'} \beta_{\lambda'\mu'} Y_{\lambda'\mu'}(\theta,\phi)\right] Y^*_{\lambda\mu}(\theta,\phi) \\
    =&\int\rmd\Omega Y^*_{\lambda\mu}(\theta,\phi) + (\lambda+3)\sum_{\lambda'=1}^{\lambda'_{\text{max}}}\sum_{\mu'=-\lambda'}^{\lambda'} \beta_{\lambda'\mu'}\int \rmd \Omega Y_{\lambda'\mu'}(\theta,\phi)Y^*_{\lambda\mu}(\theta,\phi)\\
    =&\delta_{\lambda0}\sqrt{4\pi} + (\lambda+3)\beta_{\lambda\mu}.
\end{aligned}
\end{equation}
Finally, for $\lambda>0$, the first-order relation between the multipole moments and Bohr parameters, quoted in Section~\ref{Shapes}, reads as follows,
\begin{equation}
    Q_{\lambda\mu} = \frac{3N}{4\pi R_0^3}\frac{1}{\lambda+3}R_0^{\lambda+3}(\lambda+3)\beta_{\lambda\mu} = \frac{3NR_0^{\lambda}}{4\pi } \beta_{\lambda\mu},
\end{equation}
that is,
\begin{equation}
\label{first-order-Bohr_parameters}
    \beta_{\lambda\mu} = \frac{4\pi }{3NR_0^{\lambda}}Q_{\lambda\mu}.
\end{equation}
We observe that extending Taylor's terms to higher orders allows us to derive higher-order formulas, with moments $Q_{\lambda\mu}$ expressed as polynomials of order greater than 1, $P_{\lambda\mu}\left(\{\beta_{\lambda'\mu'}\}\right)$, of all the Bohr parameters $\beta_{\lambda'\mu'}$, that is,
\begin{equation}
\label{Bohr_parameters}
    Q_{\lambda\mu} = \frac{3NR_0^{\lambda}}{4\pi } \beta_{\lambda\mu}+P_{\lambda\mu}\left(\{\beta_{\lambda'\mu'}\}\right),\quad \mbox{for~} \lambda,\lambda'>0.
\end{equation}

An analytic inversion of these expressions, $\beta_{\lambda\mu}(\{Q_{\lambda'\mu'}\})$, is not generally possible; however, as implemented in Reference~\cite{(Dob09g)}, Equation~\ref{Bohr_parameters} can be formulated as an iterative fixed-point algorithm,
\begin{equation}
\label{Bohr_parameters2}
    \beta^{(n+1)}_{\lambda\mu} = \frac{4\pi }{3NR_0^{\lambda}}\Bigg(Q_{\lambda\mu}-P_{\lambda\mu}\left(\{\beta^{(n)}_{\lambda'\mu'}\}\right)\Bigg),\quad \mbox{for~}  \lambda,\lambda'>0,
\end{equation}
which starts from $\beta^{(0)}_{\lambda\mu}=0$, gives Equation~\ref{first-order-Bohr_parameters} for $\beta^{(1)}_{\lambda\mu}$, and with increasing $n$ rapidly converges to the exact solution $\beta_{\lambda\mu}(\{Q_{\lambda'\mu'}\})$.

\newcommand{\ph}{\phantom{$-$}}
\begin{table}[htb]
    \centering
    \caption{Known experimental values of the magnetic octupole moments $\Omega$ in~$\mu_{\text{N}}$\,b (taken from References~\protect\cite{(deG22b)} and~\protect\nocite{(Bro80a),(ger09),(lew13),(deG21),(Jia26)}\protect\citemany{(Bro80a)}{(Jia26)}) compared with the results of DFT calculations (preliminary results of Reference~\protect\cite{(deG26)}). For $^{209}$Bi, $^{133}$Cs, and $^{113,115}$In, recently calculated atomic structure matrix elements, References~\protect\nocite{(li22),(Li23),(Li24)}\protect\citemany{(li22)}{(Li24)}, were used to recompute the values of $\Omega$. When the sign of the experimental moment is given in parentheses, it indicates that no measured sign is provided in the literature; in such cases, the sign of the Schwarz prediction is used. Note that sometimes, no experimental uncertainties are available in the literature.\label{tab:MO}}
\vspace*{5mm}
\begin{tabular}{lrrrlll}
\hline
Nuclide     &  $Z$ &   $N$  &  $I^\pi$~~~~ &   Nilsson  &~~$\Omega_{\text{EXP}}$ & ~~$\Omega_{\text{DFT}}$ \\
\hline
$^{25}$Mg   &  12  &    13  &      5/2$^+$ &  [202]5/2  &  $(-)$0.042(9)       &  no conv      \\
$^{27}$Al   &  13  &    14  &      5/2$^+$ &  [202]5/2  &  $(+)$0.159(4)       & \ph0.1049     \\
$^{35}$Cl   &  17  &    18  &      3/2$^+$ &  [202]3/2  &  $-$0.0188           & $-$0.0160     \\
$^{37}$Cl   &  17  &    20  &      3/2$^+$ &  [202]3/2  &  $-$0.0146           & $-$0.0122     \\
$^{39}$K    &  19  &    20  &      3/2$^+$ &  [202]3/2  &   $(-)$0.066         & $-$0.0084     \\
$^{45}$Sc   &  21  &    24  &      7/2$^-$ &  [303]7/2  &  $-$0.07(53)         & \ph0.5440     \\
$^{69}$Ga   &  31  &    38  &      3/2$^-$ &  [301]3/2  &  +0.107(20)          & \ph0.2745     \\
$^{71}$Ga   &  31  &    40  &      3/2$^-$ &  [301]3/2  &  +0.146(20)          &  no conv      \\
$^{79}$Br   &  35  &    44  &      3/2$^-$ &  [301]3/2  &  +0.116              & \ph0.1341     \\
$^{81}$Br   &  35  &    46  &      3/2$^-$ &  [301]3/2  &  +0.129              & \ph0.1389     \\
$^{83}$Kr   &  36  &    47  &      3/2$^+$ &  [411]3/2  &  $-$0.18(6)          & \ph0.1231     \\
$^{87}$Rb   &  37  &    50  &      3/2$^-$ &  [301]3/2  &  $-$0.58(39)         & \ph0.2987     \\
$^{113}$In  &  49  &    64  &      9/2$^+$ &  [404]9/2  &  +0.455(44)          & \ph0.5698     \\
$^{115}$In  &  49  &    66  &      9/2$^+$ &  [404]9/2  &  +0.443(42)          & \ph0.6272     \\
$^{127}$I   &  53  &    74  &      5/2$^+$ &  [402]5/2  &  +0.167              & \ph0.0661     \\
$^{131}$Xe  &  54  &    77  &      3/2$^+$ &  [402]3/2  &  +0.048(12)          & \ph0.0151     \\
$^{133}$Cs  &  55  &    78  &      7/2$^+$ &  [404]7/2  &  +0.68(9)            & \ph0.0479     \\
$^{137}$Ba  &  56  &    81  &      3/2$^+$ &  [402]3/2  &  +0.05061(56)        & \ph0.0416     \\
$^{155}$Gd  &  64  &    91  &      3/2$^-$ &  [521]3/2  &  $-$1.6(6)           & $-$0.0994     \\
$^{165}$Ho  &  67  &    98  &      7/2$^-$ &  [523]7/2  &  +0.8(5)             & $-$0.3566     \\
$^{173}$Yb  &  70  &   103  &      5/2$^-$ &  [512]5/2  &  $-$0.062(8)         & \ph0.0190     \\
$^{197}$Au  &  79  &   118  &      3/2$^+$ &  [431]3/2  &  +0.0098(7)          & $-$0.1144     \\
$^{201}$Hg  &  80  &   121  &      3/2$^-$ &  [541]3/2  &  $-$0.130(13)        & $-$0.0949     \\
$^{209}$Bi  &  83  &   126  &      9/2$^-$ &  [503]5/2  &  +0.51(5)            & \ph0.2058     \\
$^{207}$Po  &  84  &   123  &      5/2$^-$ &  [505]9/2  &  +0.11(1)            & \ph0.1742     \\
\hline
\end{tabular}
\end{table}

\subsection{Magnetic octupole moments\label{Magnetic_octupole}}

\begin{figure}
    \centering
     \includegraphics[width=\textwidth]{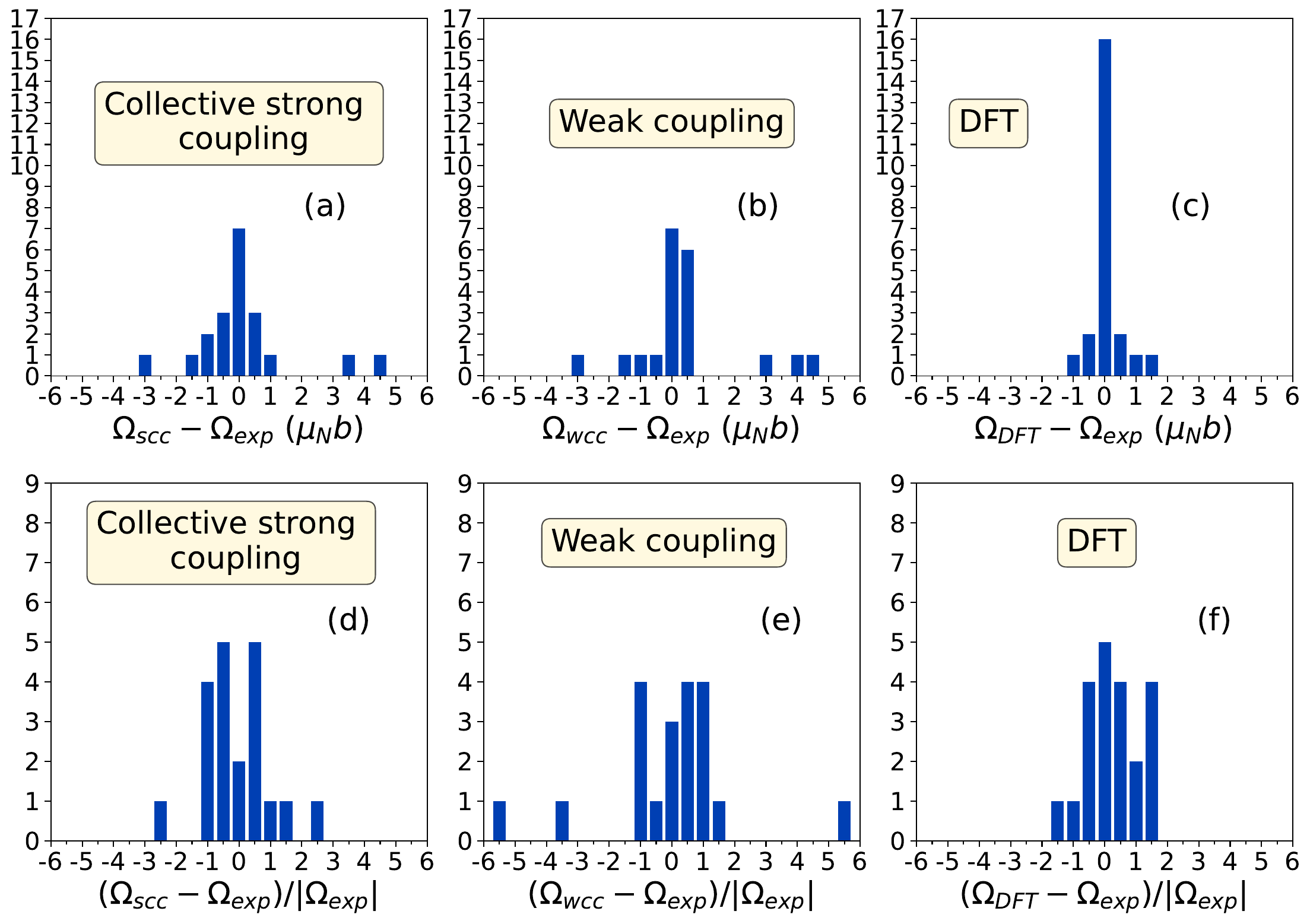}
    \caption{Residuals (upper panels), $\Omega_{\text{the}}-\Omega_{\text{exp}}$ (in $\mu_{\text{N}}b$),  or relative residuals, $(\Omega_{\text{the}}-\Omega_{\text{exp}})/|\Omega_{\text{exp}}|$, of magnetic octupole moments determined with respect to the collective strong coupling model (scc), left panels, weak coupling model (wcc), middle panels, and nuclear DFT, right panels. To facilitate comparison, the horizontal scales of the top and bottom panels are identical. For the left and middle panels, the model results and experimental data are taken from Reference~\cite{BOFOS2024101672}. For the right panels, the experimental and DFT data are shown in {\bf Table}~\ref{tab:MO} and those for $^{25}$Mg and $^{71}$Ga are omitted because nuclear-DFT results are unavailable. In panel (f), data for $^{45}$Sc and $^{197}$Au are not displayed because their values exceed the scale of the panel.}
    \label{fig: Residual Octupole moments}
\end{figure}

In the strong coupling case (scc) of the collective model (References~\cite{(Boh53a), (Sue57)}), the magnetic octupole moment $\Omega_{scc}$ in $I>3/2$ ground states of odd nuclei is calculated as
\begin{align}
    \Omega_{\text{scc}}=\frac{2I(2I-1)(2I-2)}{(2I+2)(2I+3)(2I+4)}\left[\Omega_{\text{s.p.}}+\Omega_{0}\right],
\end{align}
where the extreme single-particle estimate, $\Omega_{\text{s.p.}}$, is given by Equation \ref{eq:schwartz_limit}, and
\begin{equation}
    \Omega_{0}=-\frac{30}{49}\sqrt{\frac{2}{3}}g_{R}R^{2}_{0}.
\end{equation}
Here $g_{R}\simeq Z/A$ is the $g$-factor of the core and $R_{0}=1.35\;A^{1/3}$ fm (Reference~\cite{(Set58)}). For the case of $I=3/2$, the magnetic octupole moment, $\Omega_{\text{scc}}$ reads as follows,
\begin{equation}
    \Omega_{\text{scc}}\simeq\frac{1}{35}\left[19\Omega_{\text{s.p.}}+\Omega_{0}\right].
\end{equation}
In the weak coupling case (wcc) of the collective model, the magnetic octupole moment, $\Omega_{\text{wcc}}$, is decomposed into two terms, that is $\Omega_{\text{wcc}}=\Omega_{\text{p}}+\Omega_{\text{c}}$, where the core part, $\Omega_{\text{c}}$, takes the form as follows,
\begin{equation}
    \Omega_{\text{c}}=-\frac{45}{224\pi}g_{R}R^{2}_{0}\frac{k^{2}}{\hbar\omega_{2} C_{2}}\frac{(2I-1)(I-1)}{I(I+1)^{2}},
\end{equation}
whereas the particle part, $\Omega_{p}$, reads,
\begin{equation}
    \Omega_{\text{p}}=\Omega_{\text{s.p.}}\left[1+\frac{5}{32\pi}\frac{k^{2}}{\hbar\omega_{2} C_{2}}(F_{1}-F_{2})\right],
\end{equation}
where
\begin{equation}
    F_{1}=\frac{4I^{2}(I+1)^{2}-75I(I+1)+234}{4I^{2}(I+1)^{2}}
\end{equation}
and
\begin{equation}
    F_{2}=\frac{(2I-1)(2I+3)}{4I(I+1)}.
\end{equation}
Within the hydrodynamic model (Reference~\cite{(Boh53a)}), the mass parameter $B_{\lambda}$ is defined as
\begin{equation}
    B_{\lambda}=\frac{3}{4\pi}\frac{AMR^{2}_{0}}{\lambda},
\end{equation}
and the restoring force $C_{\lambda}$, originating from the surface tension, $S$, and Coulomb energy, reads
\begin{equation}
    C_{\lambda}=(\lambda-1)(\lambda+2)R^{2}_{0}S-\frac{3}{2\pi}\frac{\lambda-1}{2\lambda+1}\frac{Z^{2}\alpha\hbar c}{R_{0}}.
\end{equation}
Here the fine structure constant $\alpha\simeq 1/137$, $\hbar c = 197.327$ MeV$\cdot$fm, and $M=938\;\text{MeV}$. Following Reference~\cite{(Sue57)}, $k = 40$ MeV has been used. The surface tension, $S$, is estimated using $4\pi R_{0}^{2}S=15.4\cdot A^{2/3}$ MeV (Reference~\cite{(Leo48)}). The frequency $\omega_{\lambda}$ for a specific $\lambda$ is calculated via
\begin{equation}
    \hbar\omega_{\lambda}=\hbar c\sqrt{\frac{C_{\lambda}}{B_{\lambda}}}\;\;(\text{MeV}).
\end{equation}

\subsection{Restoration of rotational symmetry for non-axial states\label{3D}}

The projected wave function $|\Phi_{IM}\rangle$, projection operator $\hat{P}^I_{MK}$, Wigner function $D^I_{MK}(\alpha,\beta,\gamma)$, and the rotation operator $\hat{R}(\alpha,\beta,\gamma)$ are defined in Equations~\ref{projected}--\ref{Wigner}.
The normalisation of the projected wave function $|\Phi_{IM}\rangle$ is ensured by a proper normalisation condition of amplitudes $g_{IK}$, that is,
\begin{equation}
    \langle\Phi_{I'M'}|{\Phi}_{IM}\rangle=\delta_{II'}\delta_{MM'}\quad\Longrightarrow\quad
    \sum_{K,K'=-I}^{I}g_{IK'}^\ast\langle\Phi|\hat{P}_{K'K}^{I}|\Phi\rangle g_{IK}=1.
\label{normalisation}
\end{equation}
The matrix element of a spherical tensor operator $\hat{O}_{\lambda\mu}$ between two projected wave functions $\langle\Phi_{I'M'}|$ and $|\Phi_{IM}\rangle$ is given by
\begin{align}\label{eq: Matrix element of Qlambdamu 1}
    \langle\Phi_{I'M'}|\hat{O}_{\lambda\mu}|\Phi_{IM}\rangle&=\sum_{K,K'=-I}^{I}g^{\ast}_{I'K'}\langle\Phi|\hat{P}^{I'}_{K'M'}\hat{O}_{\lambda\mu}\hat{P}^{I}_{MK}|\Phi\rangle{}g_{IK}.
\end{align}

\subsubsection{Derivation of the identity in Equation~\protect\ref{Shift}\label{Derive_Shift}}

We proceed by following References~\cite{(Egido04),(Tan2023)}. The transformation property of the tensor operator $\hat{O}_{\lambda\mu}$ under rotation $\hat{R}(\mathbf{g})$ reads as follows (Reference~\cite{(Var88)}, Equation~3.1.3(11)),
\begin{equation}
    \hat{R}(\mathbf{g})\hat{O}_{\lambda\mu}\hat{R}^{+}(\mathbf{g})=\sum_{\mu'}D^{\lambda}_{\mu'\mu}(\mathbf{g})\hat{O}_{\lambda\mu'}.
    \label{spherical_tensor}
\end{equation}
For the reason that will become clear below, it is useful first to take the Hermitian conjugate of Equation~\ref{spherical_tensor} and apply the definition of a Hermitian conjugated tensor operator, $\hat{O}^{+}_{\lambda\mu}=(-1)^{\lambda+\mu}\hat{O}_{\lambda(-\mu)}$, to obtain
\begin{align}
    \hat{R}(\mathbf{g})\hat{O}^{+}_{\lambda\mu}\hat{R}^{+}(\mathbf{g})&=\sum_{\mu'}D^{\lambda\ast}_{\mu'\mu}(\mathbf{g})\hat{O}^{+}_{\lambda\mu'}\nonumber\\
    \hat{R}(\mathbf{g})(-1)^{\lambda+\mu}\hat{O}_{\lambda(-\mu)}\hat{R}^{+}(\mathbf{g})&=\sum_{\mu'}D^{\lambda}_{\mu\mu'}(\mathbf{g}^{-1})(-1)^{\lambda+\mu'}\hat{O}_{\lambda(-\mu')}\nonumber\\
    \hat{R}(\mathbf{g})\hat{O}_{\lambda(-\mu)}\hat{R}^{+}(\mathbf{g})&=\sum_{\mu'}(-1)^{\lambda+\mu'-\lambda-\mu}D^{\lambda}_{\mu\mu'}(\mathbf{g}^{-1})\hat{O}_{\lambda(-\mu')}\nonumber\\
    &=\sum_{\mu'}(-1)^{\mu'-\mu}D^{\lambda}_{\mu\mu'}(\mathbf{g}^{-1})\hat{O}_{\lambda(-\mu')}\nonumber\\
    \hat{R}(\mathbf{g})\hat{O}_{\lambda\mu}\hat{R}^{+}(\mathbf{g})&=\sum_{\mu'}(-1)^{\mu'+\mu}D^{\lambda}_{-\mu\mu'}(\mathbf{g}^{-1})\hat{O}_{\lambda(-\mu')}\nonumber\\
    &=\sum_{\mu'}(-1)^{\mu-\mu'}D^{\lambda}_{-\mu,-\mu'}(\mathbf{g}^{-1})\hat{O}_{\lambda\mu'}.
    \label{Hermitian}
\end{align}
Here $\mathbf{g}^{-1}$ stands for $(-\gamma,-\beta,-\alpha)$, which is defined as follows,
\begin{equation}
    \hat{R}^{+}(\mathbf{g})=e^{i\gamma\hat{I}_{z}}e^{i\beta\hat{I}_{y}}e^{i\alpha\hat{I}_{z}}=\hat{R}(-\gamma,-\beta,-\alpha)=\hat{R}(\mathbf{g}^{-1}).
\end{equation}
Multiplying both sides by $D^{I'\ast}_{K'M'}(\mathbf{g})=D^{I'}_{M'K'}(\mathbf{g}^{-1})$ (Reference~\cite{(Var88)}, Equations~4.1(4) and 4.1(13)) and applying the formula for the product of two Wigner functions (Reference~\cite{(Var88)}, Equation~4.6.1(1)),
\begin{align}
    D^{J_{1}}_{M_{1}N_{1}}(\mathbf{g})D^{J_{2}}_{M_{2}N_{2}}(\mathbf{g})=\sum_{J=|J_{1}-J_{2}|}^{J_{1}+J_{2}}\sum_{MN}C^{JM}_{J_{1}M_{1}J_{2}M_{2}}D^{J}_{MN}(\mathbf{g})C^{JN}_{J_{1}N_{1}J_{2}N_{2}},
\end{align}
we obtain
\begin{align}
    D^{I'\ast}_{K'M'}(\mathbf{g})\hat{R}(\mathbf{g})\hat{O}_{\lambda\mu}\hat{R}^{+}(\mathbf{g})&=\sum_{\mu'}(-1)^{\mu-\mu'}D^{I'\ast}_{K'M'}(\mathbf{g})D^{\lambda}_{-\mu,-\mu'}(\mathbf{g}^{-1})\hat{O}_{\lambda\mu'}\nonumber\\
    &=\sum_{\mu'}(-1)^{\mu-\mu'}D^{I'}_{M'K'}(\mathbf{g}^{-1})D^{\lambda}_{-\mu,-\mu'}(\mathbf{g}^{-1})\hat{O}_{\lambda\mu'}\nonumber\\
    &=\sum_{\mu'}(-1)^{\mu-\mu'}\sum_{I''}\sum_{\nu,\nu'}C^{I''\nu}_{I'M'\lambda(-\mu)}C^{I''\nu'}_{I'K'\lambda(-\mu')}D^{I''}_{\nu\nu'}(\mathbf{g}^{-1})\hat{O}_{\lambda\mu'}\nonumber\\
    &=\sum_{\mu'}(-1)^{\mu-\mu'}\sum_{I''}\sum_{\nu,\nu'}C^{I''\nu}_{I'M'\lambda(-\mu)}C^{I''\nu'}_{I'K'\lambda(-\mu')}D^{I''\ast}_{\nu'\nu}(\mathbf{g})\hat{O}_{\lambda\mu'}.
\end{align}
Multiplying both side of the last equation by $\hat{R}(\mathbf{g})$ from the right, and integrating over $\mathbf{g}$ with the inclusion of the normalization constant $(2I'+1)/8\pi^{2}$, we obtain
\begin{align}
    \frac{2I'+1}{8\pi^{2}}\int {\rm d}\mathbf{g}\;D^{I'\ast}_{K'M'}(\mathbf{g})\hat{R}(\mathbf{g})\hat{O}_{\lambda\mu}&=\sum_{\mu'}(-1)^{\mu-\mu'}\sum_{I''}\sum_{\nu,\nu'}C^{I''\nu}_{I'M'\lambda(-\mu)}C^{I''\nu'}_{I'K'\lambda(-\mu')}\hat{O}_{\lambda\mu'}\nonumber\\
    &\times\frac{2I'+1}{2I''+1}\frac{2I''+1}{8\pi^{2}}\int{\rm d}\mathbf{g}\;D^{I''\ast}_{\nu'\nu}(\mathbf{g})\hat{R}(\mathbf{g})\nonumber\\
    \hat{P}^{I'}_{K'M'}\hat{O}_{\lambda\mu}&=\sum_{\mu'}(-1)^{\mu-\mu'}\sum_{I''}\frac{2I'+1}{2I''+1}\sum_{\nu,\nu'}C^{I''\nu}_{I'M'\lambda(-\mu)}\nonumber\\
    &\times C^{I''\nu'}_{I'K'\lambda(-\mu')}\hat{O}_{\lambda\mu'}\hat{P}^{I''}_{\nu'\nu}.
\end{align}
Again, multiplying both sides from the right by projection operator $\hat{P}^{I}_{MK}$ and applying the property of projection operator, $\hat{P}^{I''}_{\nu'\nu}\hat{P}^{I}_{MK}=\delta_{I''I}\delta_{\nu M}\hat{P}^{I}_{\nu' K}$, we can write
\begin{align}
    \hat{P}^{I'}_{K'M'}\hat{O}_{\lambda\mu}\hat{P}^{I}_{MK}&=\frac{2I'+1}{2I+1}\sum_{\mu'}(-1)^{\mu-\mu'}C^{IM}_{I'M'\lambda(-\mu)}\sum_{\nu}C^{I\nu}_{I'K'\lambda(-\mu')}\hat{O}_{\lambda\mu'}\hat{P}^{I}_{\nu K}.\label{eq: POP version 2}
\end{align}
The reason for taking the Hermitian conjugate of the transformation property of the tensor operator $\hat{O}_{\lambda\mu}$, Equation~\ref{Hermitian}, becomes clear now, because applying the symmetry property of the Clebsch-Gordan coefficient (Reference~\cite{(Var88)}, Equation~8.4.3(10)),
\begin{equation}
    C^{j_{3}m_{3}}_{j_{1}m_{1}j_{2}m_{2}}=(-1)^{j_{1}-j_{3}+m_{2}}\sqrt{\frac{2j_{3}+1}{2j_{1}+1}}C^{j_{1}m_{1}}_{j_{3}m_{3}j_{2}(-m_{2})},
\end{equation}
yields
\begin{equation}
    C^{I\nu}_{I'K'\lambda(-\mu')}=(-1)^{I'-I-\mu'}\sqrt{\frac{2I+1}{2I'+1}}C^{I'K'}_{I\nu\lambda\mu'}
\end{equation}
and
\begin{equation}
    C^{IM}_{I'M'\lambda(-\mu)}=(-1)^{I'-I-\mu}\sqrt{\frac{2I+1}{2I'+1}}C^{I'M'}_{IM\lambda\mu}.
\end{equation}
Inserting these results into Equation~\ref{eq: POP version 2} results in
\begin{align}
    \hat{P}^{I'}_{K'M'}\hat{O}_{\lambda\mu}\hat{P}^{I}_{MK}&=\frac{2I'+1}{2I+1}\frac{2I+1}{2I'+1}C^{I'M'}_{IM\lambda\mu}\sum_{\mu'}\underbrace{(-1)^{2I'-2I-2\mu'}}_{=1}\sum_{\nu}C^{I'K'}_{I\nu\lambda\mu'}\hat{O}_{\lambda\mu'}\hat{P}^{I}_{\nu K}\nonumber\\
    &=C^{I'M'}_{IM\lambda\mu}\sum_{\nu\mu'}C^{I'K'}_{I\nu\lambda\mu'}\hat{O}_{\lambda\mu'}\hat{P}^{I}_{\nu K},
\end{align}
which is the identity of Equation~\ref{Shift} we wanted to prove.

\subsubsection{Derivation of the Wigner-Eckart theorem in Equation~\protect\ref{Wigner-Eckart}\label{Derive_Wigner-Eckart}}

Using the identity \ref{Shift}, Equation~\ref{eq: Matrix element of Qlambdamu 1} becomes:
\begin{equation}\label{eq: Matrix element of Qlambdamu 2}
    \langle\Phi_{I'M'}|\hat{O}_{\lambda\mu}|\Phi_{IM}\rangle=\sum_{K',K}g^{\ast}_{I'K'}g_{IK}C^{I'M'}_{IM\lambda\mu}\sum_{\nu\mu'}C^{I'K'}_{I\nu\lambda\mu'}\langle\Phi|\hat{O}_{\lambda\mu'}\hat{P}^{I}_{\nu K}|\Phi\rangle.
\end{equation}
To proceed, we need to apply the Wigner-Eckart theorem, which we now prove. Following Reference~\cite{(Sak11)}, the proof begins with the transformation of a spherical tensor operator $\hat{O}_{\lambda\mu}$, Equation~\ref{spherical_tensor}, specified to the infinitesimal rotation $\hat{R}(\delta g)$, that is,
\begin{equation}
    \hat{R}(\delta g)\hat{O}_{\lambda\mu}\hat{R}^{+}(\delta g)=\sum_{\mu'}D^{\lambda}_{\mu'\mu}(\delta g)\hat{O}_{\lambda\mu'},
\end{equation}
where $\delta g$ denotes the rotation angle about the rotation axis aligned with the unit vector $\hat{\bm{n}}$. This yields
\begin{align}
    \hat{R}(\delta g)\hat{O}_{\lambda\mu}\hat{R}^{+}(\delta g)&=\left(1-\frac{i\hat{\bm{I}}\cdot\hat{\bm{n}}\delta g}{\hbar}\right)\hat{O}_{\lambda\mu}\left(1+\frac{i\hat{\bm{I}}\cdot\hat{\bm{n}}\delta g}{\hbar}\right)\nonumber\\
    &=\hat{O}_{\lambda\mu}-i\frac{\delta g}{\hbar}\left(\hat{\bm{I}}\cdot\hat{\bm{n}}\;\hat{O}_{\lambda\mu}-\hat{O}_{\lambda\mu}\;\hat{\bm{I}}\cdot\hat{\bm{n}}\right)\nonumber\\
    &=\hat{O}_{\lambda\mu}-i\frac{\delta g}{\hbar}\left[\hat{\bm{I}}\cdot\hat{\bm{n}},\hat{O}_{\lambda\mu}\right]\nonumber\\
    \sum_{\mu'}D^{\lambda}_{\mu'\mu}(\delta g)\hat{O}_{\lambda\mu'}&=\sum_{\mu'}\langle\lambda\mu'|\left(1-\frac{i\hat{\bm{I}}\cdot\hat{\bm{n}}\delta g}{\hbar}\right)|\lambda\mu\rangle\;\hat{O}_{\lambda\mu'}\nonumber\\
    &=\sum_{\mu'}\delta_{\mu'\mu}\hat{O}_{\lambda\mu'}-i\frac{\delta g}{\hbar}\sum_{\mu'}\langle\lambda\mu'|\hat{\bm{I}}\cdot\hat{\bm{n}}|\lambda\mu\rangle\hat{O}_{\lambda\mu'}\nonumber\\
    &=\hat{O}_{\lambda\mu}-i\frac{\delta g}{\hbar}\sum_{\mu'}\langle\lambda\mu'|\hat{\bm{I}}\cdot\hat{\bm{n}}|\lambda\mu\rangle\hat{O}_{\lambda\mu'}\nonumber\\
    \left[\hat{\bm{I}}\cdot\hat{\bm{n}},\hat{O}_{\lambda\mu}\right]&=\sum_{\mu'}\hat{O}_{\lambda\mu'}\langle\lambda\mu'|\hat{\bm{I}}\cdot\hat{\bm{n}}|\lambda\mu\rangle.
\end{align}
Taking $\hat{\bm{n}}=\hat{\bm{z}}$, we obtain
\begin{align}
    \left[\hat{\text{I}}_{z},\hat{O}_{\lambda\mu}\right]&=\sum_{\mu'}\hat{O}_{\lambda\mu'}\langle\lambda\mu'|\hat{\text{I}}_{z}|\lambda\mu\rangle=\sum_{\mu'}\hat{O}_{\lambda\mu'}\;\hbar\mu\;\delta_{\mu'\mu}=\hbar\mu\;\hat{O}_{\lambda\mu}.\label{eq: Iz Olambdamu}
\end{align}
In turn, taking $\hat{\bm{n}}=\hat{\bm{x}}\pm i\hat{\bm{y}}$, we get
\begin{align}
    \left[\hat{\text{I}}_{x}\pm i\hat{\text{I}}_{y},\hat{O}_{\lambda\mu}\right]&\equiv\left[\hat{\text{I}}_{\pm},\hat{O}_{\lambda\mu}\right]=\sum_{\mu'}\hat{O}_{\lambda\mu'}\langle\lambda\mu'|\hat{\text{I}}_{\pm}|\lambda\mu\rangle\nonumber\\
    &=\sum_{\mu'}\hat{O}_{\lambda\mu'}\sqrt{(\lambda\mp\mu)(\lambda\pm\mu+1)}\hbar\underbrace{\langle\lambda\mu'|\lambda(\mu\pm 1)\rangle}_{=\delta_{\mu'(\mu\pm 1)}}\nonumber\\
    &=\sqrt{(\lambda\mp\mu)(\lambda\pm\mu+1)}\hbar\;\hat{O}_{\lambda(\mu\pm 1)}.\label{eq: Iplusminus Olambdamu}
\end{align}
From identity \ref{eq: Iz Olambdamu}, we can prove the following assertion:
\begin{equation}
    \langle\Phi_{I'M'}|\hat{O}_{\lambda\mu}|\Phi_{IM}\rangle = 0,\;\text{unless }M'=\mu+M.\label{eq: assertion}
\end{equation}
The proof involves taking the matrix element of Equation \ref{eq: Iz Olambdamu} between $|\Phi_{I'M'}\rangle$ and $|\Phi_{IM}\rangle$, which yields
\begin{align}
    0&=\left[\hat{\text{I}}_{z},\hat{O}_{\lambda\mu}\right]-\hbar\mu\hat{O}_{\lambda\mu}\nonumber\\
    &=\langle\Phi_{I'M'}|\left[\hat{\text{I}}_{z},\hat{O}_{\lambda\mu}\right]-\hbar\mu\hat{O}_{\lambda\mu}|\Phi_{IM}\rangle\nonumber\\
    &=\langle\Phi_{I'M'}|\hat{\text{I}}_{z}\hat{O}_{\lambda\mu}|\Phi_{IM}\rangle-\langle\Phi_{I'M'}|\hat{O}_{\lambda\mu}\hat{\text{I}}_{z}|\Phi_{IM}\rangle-\hbar\mu\langle\Phi_{I'M'}|\hat{O}_{\lambda\mu}|\Phi_{IM}\rangle\nonumber\\
    &=\hbar(M'-M-\mu)\langle\Phi_{I'M'}|\hat{O}_{\lambda\mu}|\Phi_{IM}\rangle.\label{eq: selection rule for matrix elements}
\end{align}
This result proves the assertion \ref{eq: assertion} and also suggests the proportionality of the matrix element $\langle\Phi_{I'M'}|\hat{O}_{\lambda\mu}|\Phi_{IM}\rangle$ to the Clebsch-Gordan coefficient, $C^{I'M'}_{IM\lambda\mu}$, since it also satisfies
\begin{equation}
    (M'-M-\mu)C^{I'M'}_{IM\lambda\mu}=0.\label{eq: selection rule for CG coefficients}
\end{equation}
This selection rule of the Clebsch-Gordan coefficient, $C^{I'M'}_{IM\lambda\mu}\equiv\langle IM\lambda\mu|I'M'\rangle$, can be obtained by taking the overlap between  $(\hat{\text{I}}'_{z}-\hat{\text{I}}_{z}-\hat{\lambda}_{z})|I'M'\rangle$ and the uncoupled state $|IM\lambda\mu\rangle$. If we take the matrix elements of Equation \ref{eq: Iplusminus Olambdamu} between $|\Phi_{I'M'}\rangle$ and $|\Phi_{IM}\rangle$, we obtain
\begin{align}
    \hbar\sqrt{(\lambda\mp\mu)(\lambda\pm\mu+1)}&\langle\Phi_{I'M'}|\hat{O}_{\lambda(\mu\pm 1)}|\Phi_{IM}\rangle\nonumber\\
    &=\langle\Phi_{I'M'}|\hat{\text{I}}_{\pm}\hat{O}_{\lambda\mu}|\Phi_{IM}\rangle-\langle\Phi_{I'M'}|\hat{O}_{\lambda\mu}\hat{\text{I}}_{\pm}|\Phi_{IM}\rangle\nonumber\\
    &=\hbar\sqrt{(I'\pm M')(I'\mp M'+1)}\langle\Phi_{I'(M'\mp 1)}|\hat{O}_{\lambda\mu}|\Phi_{IM}\rangle\nonumber\\
    &-\hbar\sqrt{(I\mp M)(I\pm M+1)}\langle\Phi_{I'M'}|\hat{O}_{\lambda\mu}|\Phi_{I(M\pm 1)}\rangle\nonumber\\
    \sqrt{(I'\pm M')(I'\mp M'+1)}&\langle\Phi_{I'(M'\mp 1)}|\hat{O}_{\lambda\mu}|\Phi_{IM}\rangle\nonumber\\
    &=\sqrt{(I\mp M)(I\pm M+1)}\langle\Phi_{I'M'}|\hat{O}_{\lambda\mu}|\Phi_{I(M\pm1)}\rangle\nonumber\\
    &+\sqrt{(\lambda\mp\mu)(\lambda\pm\mu+1)}\langle\Phi_{I'M'}|\hat{O}_{\lambda(\mu\pm 1)}|\Phi_{IM}\rangle.
    \label{Eq134}
\end{align}
A comparison between Equation~\ref{Eq134} and the recursion relation for the Clebsch-Gordan coefficients (Reference~\cite{(Sak11)}, Equation~3.8.49),
\begin{align}
    \sqrt{(I'\pm M')(I'\mp M'+1)}C^{I'(M'\mp1)}_{IM\lambda\mu}&=\sqrt{(I\mp M)(I\pm M+1)}C^{I'M'}_{I(M\pm 1)\lambda\mu}\nonumber\\
    &+\sqrt{(\lambda\mp\mu)(\lambda\pm\mu+1)}C^{I'M'}_{IM\lambda(\mu\pm 1)},
\end{align}
and between Equations \ref{eq: selection rule for matrix elements} and \ref{eq: selection rule for CG coefficients} suggests that the dependence of $\langle\Phi_{I'M'}|\hat{O}_{\lambda\mu}|\Phi_{IM}\rangle$ on $M'$, $\mu$, and $M$ is via Clebsch-Gordan coefficient. Therefore, we have arrived at the Wigner-Eckart theorem: For a spherical tensor operator $\hat{O}_{\lambda\mu}$, the matrix element $\langle\Phi_{I'M'}|\hat{O}_{\lambda\mu}|\Phi_{IM}\rangle$ can be written as
\begin{equation}
    \langle\Phi_{I'M'}|\hat{O}_{\lambda\mu}|\Phi_{IM}\rangle=c(I',I,\lambda)\;C^{I'M'}_{IM\lambda\mu},
    \label{Eq136}
\end{equation}
where the proportionality constant $c(I',I,\lambda)$ is independent of the magnetic quantum numbers $M'$, $M$, and $\mu$. Multiplying both sides of Equation~\ref{Eq136} with $C^{I'M'}_{IM\lambda\mu}$, summing up over $\mu$ and $M$, and using the orthogonality rule of the Clebsch-Gordan coefficients, Reference~\cite{(Var88)}, Equation~8.1.1(8),
\begin{equation}
    \sum_{\mu M}\;C^{I'M'}_{IM\lambda\mu}\;C^{I'M'}_{IM\lambda\mu}=1,
    \label{Eq136b}
\end{equation}
we finally obtain,
\begin{equation}
    \sum_{\mu M}\;C^{I'M'}_{IM\lambda\mu}\langle\Phi_{I'M'}|\hat{O}_{\lambda}|\Phi_{IM}\rangle=c(I',I,\lambda),
    \label{Eq136a}
\end{equation}
which proves the Wigner-Eckart theorem, Equation~\ref{Wigner-Eckart}, for
\begin{equation}
    c(I',I,\lambda)=\frac{1}{\sqrt{2I'+1}}\langle\Phi_{I'}||\hat{O}_{\lambda}||\Phi_{I}\rangle
\end{equation}
and the reduced matrix element $\langle\Phi_{I'}||\hat{O}_{\lambda}||\Phi_{I}\rangle$
defined in Equation~\ref{reduced_matrix_element}.

\subsubsection{Derivation of the unity resolution, Equations~\ref{complete} and~\ref{complete2}\label{unity_resolution}}

The completeness condition for two Wigner functions, $D^{I}_{K'K}(\alpha,\beta,\gamma)$ and $D^{I}_{K'K}(\alpha',\beta',\gamma')$, with integer and half-integer values of spin $I$ (Reference~\cite{(Var88)}, Equation 4.10(7)) is as follows,
\begin{align}
    \sum_{I=0,\tfrac{1}{2},1,\dots}^{\infty}\sum_{K'=-I}^{I}\sum_{K=-I}^{I}\frac{2I+1}{16\pi^{2}}D^{I\ast}_{K'K}&(\alpha,\beta,\gamma)D^{I}_{K'K}(\alpha',\beta',\gamma')\nonumber\\
    &=\delta(\alpha-\alpha')\delta(\cos\beta-\cos\beta')\delta(\gamma-\gamma'),
\end{align}
where the numerical prefactor corresponds to the Dirac $\delta$ distributions integrated over the ``doubled'' domain.
In particular, by setting $\alpha'=\beta'=\gamma'=0$ and using $D^{I}_{K'K}(0,0,0)=\delta_{K'K}$ (Reference~\cite{(Var88)}, Equation 4.16(1)), we have
\begin{align}
    \sum_{I=0,\tfrac{1}{2},1,...}^{\infty}\sum_{K=-I}^{I}\frac{2I+1}{16\pi^{2}}D^{I\ast}_{KK}(\alpha,\beta,\gamma) =\delta(\alpha)\delta(\cos\beta-1)\delta(\gamma).
\end{align}
We can now sum up both sides of Equation~\ref{projector} separately over the integer and half-integer angular momenta and use the Dirac $\delta$ distributions to perform the integrations over the ``single'' domains,
\begin{align}
    \left\{\sum_{I=0,1,2,\ldots}^{\infty}+\sum_{I=\tfrac{1}{2},\tfrac{3}{2},\ldots}^{\infty}\right\}\sum_{K=-I}^{I}\hat{P}_{KK}^I
    &=\left\{\sum_{I=0,1,2,\ldots}^{\infty}+\sum_{I=\tfrac{1}{2},\tfrac{3}{2},\ldots}^{\infty}\right\}\sum_{K=-I}^{I}\frac{2I+1}{8\pi^{2}} \nonumber\\
    &\times
    {{\int_0^{2\pi}} \rmd\alpha\int_0^{\pi}\sin\beta\,\rmd\beta\int_0^{2\pi}\rmd\gamma}\;D^{I\ast}_{KK}(\alpha,\beta,\gamma)\hat{R}(\alpha,\beta,\gamma)\nonumber\\&=\hat{R}_{\text{even}}(0,0,0)+\hat{R}_{\text{odd}}(0,0,0),
\end{align}
where $\hat{R}_{\text{even}}(0,0,0)$ and $\hat{R}_{\text{odd}}(0,0,0)$ are the identity rotation operators in the spaces of even ($I=0,1,2,\ldots)$ and odd $(I=\tfrac{1}{2},\tfrac{3}{2},\ldots)$ numbers of fermions. Since the nuclear states of even and odd numbers of nucleons are never mixed, we have the unity resolutions separately for each one, that is,
\begin{align}
\sum_{I=0,1,2,\ldots}^{\infty}\sum_{K=-I}^{I}\hat{P}_{KK}^I&=\hat{1}_{\text{even}},\\
\sum_{I=\tfrac{1}{2},\tfrac{3}{2},\ldots}^{\infty}\sum_{K=-I}^{I}\hat{P}_{KK}^I&=\hat{1}_{\text{odd}}.
\end{align}

\subsubsection{Derivation of the large-axial-deformation approximation, Equations~\ref{spectroscopic_final2} and~\ref{spectroscopic_final3}\label{large-axial-deformation approximation}}
The zero-order term of the Taylor series, Equation~\ref{reduced_kernel}, yields
\begin{equation}
    \langle\Phi_{\Omega}|\hat{O}_{\lambda\mu'}e^{-i\beta\hat{I}_{y}}|\Phi_{\Omega}\rangle=
    {\langle\Phi_{\Omega}|e^{-i\beta\hat{I}_{y}}|\Phi_{\Omega}\rangle}
    \delta_{\mu'0}\langle\Phi_{\Omega}|\hat{O}_{\lambda0}|\Phi_{\Omega}\rangle.
\end{equation}
Using the unity resolution of Equation~\ref{complete2} twice, we have
\begin{equation}
\begin{aligned}
    \langle\Phi_\Omega|e^{-i\beta\hat{J_{y}}}|\Phi_\Omega\rangle
    & =\sum_{I'I}  {\caN}_{I'\Omega}^{-1}{\caN}_{I\Omega}^{-1}\langle\widetilde{\Phi}_{I'\Omega\Omega}|e^{-i\beta\hat{J_{y}}}|\widetilde{\Phi}_{I\Omega\Omega}\rangle \\
    & =\sum_{I'} {\caN}_{I'\Omega}^{-2} \langle\widetilde{\Phi}_{I'\Omega\Omega}|e^{-i\beta\hat{J_{y}}}|\widetilde{\Phi}_{I'\Omega\Omega}\rangle
      =\sum_{I'} {\caN}_{I'\Omega}^{-2} d_{\Omega,\Omega}^{I'}(\beta),
\end{aligned}
\end{equation}
where in the second equality we used the fact that $\hat{J_{y}}$ conserves the total angular momentum, and in the third equality, we used the definition of the Wigner function $d_{\Omega,\Omega}^I(\beta)$, Reference~\cite{(Var88)}, Equation~4.3(1).
The spectroscopic multipole moment of Equation~\ref{reduced_final} then becomes
\begin{equation}
\begin{aligned}
&\langle\widetilde{\Phi}_{II\Omega}|\hat{O}_{\lambda0}|\widetilde{\Phi}_{II\Omega}\rangle\\
&=\tfrac{\caN_{I\Omega}^2(2I+1)}{2}C^{II}_{II,\lambda0}\sum_{\mu'}C^{I\Omega}_{I\Omega-\mu',\lambda\mu'}\int_0^{\pi}\sin\beta\rmd\beta\;d^{I}_{\Omega-\mu',\Omega}(\beta){\langle\Phi_{\Omega}|e^{-i\beta\hat{I}_{y}}|\Phi_{\Omega}\rangle}
    \delta_{\mu'0}\langle\Phi_{\Omega}|\hat{O}_{\lambda0}|\Phi_{\Omega}\rangle\\
&=
\tfrac{\caN_{I\Omega}^2(2I+1)}{2}C^{II}_{II,\lambda0}C^{I\Omega}_{I\Omega,\lambda 0}\langle\Phi_{\Omega}|\hat{O}_{\lambda0}|\Phi_{\Omega}\rangle\int_0^{\pi}\sin\beta\rmd\beta\;d^{I}_{\Omega,\Omega}(\beta)\sum_{I'} {\caN}_{I'\Omega}^{-2} d_{\Omega,\Omega}^{I'}(\beta). \\
\end{aligned}
\end{equation}
After applying the orthogonality condition of Wigner $d$-functions (Reference~\cite{(Var88)}, Equations~4.11.2(7)),
\begin{align}
    \int_{0}^{\pi}\sin\beta\,\rmd\beta\,d^{I}_{MM'}(\beta)d^{I'}_{MM'}(\beta)&=\tfrac{2}{2I+1}\delta_{II'},
\end{align}
we obtain the final results of Equations~\ref{spectroscopic_final2} and~\ref{spectroscopic_final3}.

\end{document}